\documentclass[aip,pof,12pt,reprint]{revtex4-2}
\usepackage{mathtools}
\usepackage[margin=2.0cm]{geometry}
\usepackage{color}
\usepackage{subfig}
\usepackage{diagbox}
\usepackage[colorlinks,
 linkcolor=blue,
 anchorcolor=blue,
 citecolor=red,
 ]{hyperref}

\usepackage{amsmath, amssymb, amsfonts}
\usepackage{tabularx, graphicx}
\usepackage{CJKutf8}
\usepackage[final]{changes}

\renewcommand\arraystretch{1.1}
\newcommand{\tmop}[1]{\ensuremath{\operatorname{#1}}}

\begin{document}
\begin{CJK*}{UTF8}{gbsn}
\title{Lyapunov exponents and Lagrangian chaos suppression in compressible homogeneous isotropic turbulence}



\author{Haijun Yu (于海军)}
\email{hyu@lsec.cc.ac.cn}\thanks{Corresponding author.}
\affiliation{NCMIS \& LSEC, Institute of Computational Mathematics and
  Scientific/Engineering Computing, Academy of Mathematics and Systems
  Science, Beijing 100190, People's Republic of China}
  \affiliation{School of Mathematical Sciences,
  University of Chinese Academy of Sciences, Beijing 100049, People's Republic of China}

\author{Itzhak Fouxon}
\email{itzhak8@gmail.com}\thanks{Corresponding author.}
\affiliation{Department of Mechanical Engineering, Ben-Gurion University of the Negev, 84105 Beer-Sheva, Israel}
\affiliation{Department of Computational Science and Engineering, Yonsei University, Seoul 03722, Korea}

\author{Jianchun Wang (王建春)}
\email{wangjc@sustech.edu.cn}
\affiliation{Department of Mechanics and Aerospace Engineering, Southern University of Science and Technology, Shenzhen 518055, People’s Republic of China}

\author{Xiangru Li (李相茹)}
\email{lixiangru@lsec.cc.ac.cn}
\affiliation{School of Mathematical Sciences,
  University of Chinese Academy of Sciences, Beijing 100049, People's Republic of China}
\affiliation{NCMIS \& LSEC, Institute of Computational Mathematics and
  Scientific/Engineering Computing, Academy of Mathematics and Systems
  Science, Beijing 100190, People's Republic of China}

\author{Li Yuan (袁礼)}
\email{lyuan@lsec.cc.ac.cn}
\author{Shipeng Mao (毛士鹏)}
\email{maosp@lsec.cc.ac.cn}
\affiliation{NCMIS \& LSEC, Institute of Computational Mathematics and
  Scientific/Engineering Computing, Academy of Mathematics and Systems
  Science, Beijing 100190, People's Republic of China}
\affiliation{School of Mathematical Sciences,
  University of Chinese Academy of Sciences, Beijing 100049, People's Republic of China}

\author{Michael Mond}
\email{mond@bgu.ac.il}\thanks{Corresponding author.}
\affiliation{Department of Mechanical Engineering, Ben-Gurion University of the Negev, 84105 Beer-Sheva, Israel}

\date{\today}

\begin{abstract}

We study Lyapunov exponents of tracers in compressible homogeneous isotropic turbulence at different turbulent Mach number $M_t$ and Taylor-scale Reynolds number $Re_\lambda$. We demonstrate that statistics of finite-time Lyapunov exponents have the same form as in incompressible flow due to density-velocity coupling. The modulus of the smallest Lyapunov exponent $\lambda_3$ provides the principal Lyapunov exponent of the time-reversed flow, which is usually wrong in a compressible flow. This exponent, along with the principal Lyapunov exponent $\lambda_1$, determines all the exponents due to vanishing of the sum of all Lyapunov exponents. Numerical results by high-order schemes for solving the Navier-Stokes equations and tracking particles verify these theoretical predictions. We found that:
1) The largest normalized Lyapunov exponent $\lambda_1 \tau_\eta$, where $\tau_\eta$ is the Kolmogorov timescale, is a decreasing function of $M_t$. Its dependence on $Re_\lambda$ is weak when the driving force is solenoidal, while it is an increasing function of $Re_\lambda$ when the solenoidal and compressible forces are comparable. Similar facts hold for $|\lambda_3|$, in contrast with well-studied short-correlated model;
2) The ratio of the first two Lyapunov exponents $\lambda_1/\lambda_2$ decreases with $Re_\lambda$, and is virtually independent of $M_t$ for $M_t \le 1$ in the case of solenoidal force but decreases as $M_t$ increases when solenoidal and compressible forces are comparable;
3) For purely solenoidal force, $\lambda_1 :\lambda_2 :\lambda_3 \approx 4:1:-5$ for $Re_\lambda > 80$, which is consistent with incompressible turbulence studies;
4) The ratio of dilation-to-vorticity is a more suitable parameter to characterize Lyapunov exponents than $M_t$.

 \end{abstract}


\maketitle 
\end{CJK*}

\section{Introduction}

Lyapunov exponents (LEs) provide a useful tool to characterize the stability of complex dynamical systems. It has been extensively used to identify chaos phenomena (e.g. the Lorenz 63 model \cite{lorenz_deterministic_1963},\cite{nagashima_csystemlike_1977}) and the so-called Lagrangian coherent structures in turbulence \cite{haller_lagrangian_2000, haller_distinguished_2001, hadjighasem_critical_2017},
{the later was widely adopted in engineering applications\cite{ma_eulerian_2020, cao_forced_2021, yuan_investigation_2022, wu_analysis_2023}}.
In the fluid mechanical studies of turbulence, the LEs are used both to characterize turbulent transport and turbulence itself, see e.g. \replaced{Ref~\onlinecite{balkovsky_universal_1999,falkovich_particles_2001}}{Balkovsky and Fouxon$^5$, Falkovich, Gawedzki, and Vergassola$^6$}.

For turbulence itself, the research has been so far mostly concentrated on the principal Lyapunov exponent of the three-dimensional incompressible turbulent flow fields $\lambda_1$, see e.g. \replaced{Ref.~\onlinecite{guido,mohan_scaling_2017,berera_chaotic_2018}.}{Ref. \cite{guido, mohan_scaling_2017}.} This exponent describes the asymptotic rate of divergence of two solutions of the Navier-Stokes equations that are initially almost equal. The small difference in the initial conditions grows exponentially with time where $\lambda_1>0$ provides the growth exponent. Thus $\lambda_1$ is a measure of the instability and complexity of the turbulence state. Recognizing that the fastest time-scale of turbulence is the Kolmogorov timescale $\tau_\eta$ (see e.g. Ref. \onlinecite{Frisch}), it was estimated in Ref. \onlinecite{ruelle} that $\lambda_1\sim \tau_{\eta}^{-1}$. This prediction, which disregards intermittency, was tested in \replaced{Ref.~\onlinecite{guido,mohan_scaling_2017,berera_chaotic_2018}}{ Ref.~\cite{guido,mohan_scaling_2017}}. To evolve in time the two solutions defining $\lambda_1$, \citet{mohan_scaling_2017} determined one of them by direct numerical simulation of the incompressible Navier-Stokes equations and the other by solving the linearized perturbation equation with a Fourier-Galerkin method coupled with a third-order Runge-Kutta method for time marching. For cases with Taylor-scale Reynolds numbers in the range $Re_\lambda \in [37,\ 211]$, they found that $\lambda_1$ increases with $Re_\lambda$ faster than the inverse of Kolmogorov timescale $\tau_\eta$, cf. similar results that have been reported in Ref.~\onlinecite{guido}{ and \onlinecite{berera_chaotic_2018}}. \citet{hassanaly_lyapunov_2019} made a significant step by determining a whole set of Lyapunov exponents as necessary to determine the Kaplan-Yorke (KY) dimension of the attractor of turbulent flow (such a set includes all positive exponents and some negative ones{, a comprehensive study of the 2-dimensional case was done by \citet{clark_chaos_2020}}). They used a second-order pressure-correction scheme with a staggered grid. They found that the KY dimension of the incompressible turbulence attractor $D_{KY}$ scales with the ratio of the domain size $L$, and the Kolmogorov scale $\eta$ as $D_{KY}\approx (L/\eta)^{2.8}$, and the Lyapunov eigenvectors (LVs) get more localized as $Re_\lambda$ increases. Localization, which was also observed in Ref.~\onlinecite{guido,mohan_scaling_2017}, signifies that the fast-growing perturbation has a spatial extent much smaller than $L$.

{We note that due to the quadratic computational cost to calculate $N$ Lyapunov exponents of turbulence itself with $N$ grows very fast with Reynolds number, earlier applications of LEs on turbulence are mostly based on reduced turbulence models. In fact, the Lorenz'63 system \cite{lorenz_deterministic_1963} can be regarded as a reduced turbulence model, in which only 3 Fourier modes are kept by the Galerkin method to approximate the Rayleigh-B{\'e}nard convection (RBC) problem. It can generate chaotic motions but has a phase diagram very different from the original RBC problem\cite{barrio_threeparametric_2007, yu_onsagernet_2021}.
A more realistic and capable reduced turbulence model is the shell model proposed by Gledzer \cite{gledzer_system_1973}.
The first attempts to establish the Lyapunov-Re dependence were made by Yamada and Ohkitani \cite{yamada_lyapunov_1987, ohkitani_temporal_1989,yamada_asymptotic_1998} based on shell models. This approach was later extended to elastic systems by \citet{ray_elastic_2016}. Recent updates in this direction are the Lyapunov analysis based on Fourier-decimated turbulence\cite{ray_nonintermittent_2018} and Galerkin-truncated inviscid turbulence\cite{murugan_manybody_2021}. A novel property of decimation models was found by \citet{ray_nonintermittent_2018}: it is possible to have nonintermittent, yet chaotic, turbulent flows, with an emergent time reversibility as the effective degrees of freedom are reduced through decimation. Conversely, the study on the Galerkin-truncated inviscid fluids by \citet{murugan_manybody_2021} reveals an interesting link between the thermodynamic variables $T$ and the maximal Lyapunov exponents, the measure of (many-body) chaos, by the relation $\lambda_1 = \sqrt{T}$.
}

We remark that probably, the most immediate quantity that remains to be studied in this direction is the backward-in-time principal Lyapunov exponent. That characterizes the divergence of solutions for backward in-time evolution, which can provide a measure of time-reversal symmetry breaking by turbulence. It would be also of interest to study how finite compressibility changes $\lambda_1$, by providing its dependence on the Mach number $Ma$ (the Mach number is the ratio of the typical flow velocity and speed of sound, a measure of compressibility that equals zero for incompressible flow). Indirectly, the study below provides information on this.

The number of the Lyapunov exponents that characterize the Navier-Stokes turbulence, which constitutes an infinite-dimensional dynamical system, is infinite. However, the effective number $N$ of intrinsic degrees of freedom of the turbulent flow, e.g. the KY dimension, is finite \cite{landauLFluid} (in standard phenomenology of turbulence $N^{4/9}$ scales linearly with the Reynolds number \cite{Frisch}). The solutions of the Navier-Stokes equations then determine trajectories in the $N-$dimensional space that are characterized by $N$ Lyapunov exponents that describe the evolution of distances between infinitesimally close trajectories. The number $N$ is influenced by intermittency and calls for further investigations.

In contrast, the LEs of small-scale turbulent transport describe the divergence of trajectories of infinitesimally close fluid particles in ordinary three-dimensional space. These trajectories are called Lagrangian trajectories and thus we occasionally refer to the corresponding exponents as Lagrangian LEs. There are three such exponents that describe the evolution of infinitesimal parcels of passive tracers (in practice, the largest linear size of the parcel must be much smaller than $\eta$). At large times the parcels attain the shape of an ellipsoid whose axes evolve in time exponentially with {the} LEs providing the growth (or decay) exponents. Thus, the {Lagrangian} LEs provide a robust characterization of the long-time effect of the transport of matter on the smallest scales of turbulence, below the Kolmogorov scale. The principal Lagrangian {LE} is positive, as that of the Navier-Stokes {equations}, so the tracers' motion is chaotic (so-called Lagrangian chaos).

There might be a connection between the principal Lyapunov exponent of turbulence {and that of} turbulent transport. Crudely, both are estimated as the inverse Kolmogorov time. Moreover, localization of the fastest-growing mode of perturbations of the Navier-Stokes equations might cause this mode to localize below $\eta$ in the limit of large $Re_{\lambda}$ and thus be determined by the small-scale turbulence similarly to  Lagrangian LEs. Still, it seems that the two exponents, after multiplication by $\tau_{\eta}$ have different Reynolds number dependence: the Lagrangian $\lambda_1\tau_{\eta}$ must decay with $Re_{\lambda}$ for incompressible flow, see detailed discussion in Ref. \onlinecite{fm} where the difference is explained and references therein. Further studies of the difference between the two principal exponents are necessary.

The Lagrangian LEs have been thoroughly studied for incompressible flows, see e.g. Ref. \onlinecite{johnson_largedeviation_2015}. Here the LEs were calculated based on particle tracking with a second-order predictor-correction time marching scheme and a {fourth}-order central finite difference for velocity gradients. \citet{johnson_largedeviation_2015} verified that $\lambda_1 : \lambda_2 : \lambda_3 \approx 4:1:-5$, the ratio that has been known for some time, see the Refs. \onlinecite{johnson_largedeviation_2015,falkovich_particles_2001}, and references therein. The exponents sum to zero so that the volume of the ellipsoid of passive particles is conserved. A remarkable property of the exponents is universality: at large $Re_{\lambda}$ the dimensionless exponents $\lambda_i\tau_{\eta}$ are functions of $Re_{\lambda}$ only, having no dependence on the details of the forcing. The large length of the inertial interval guarantees that statistics of the flow gradients, which is a small-scale turbulence property, are independent of the details of the mechanism that stirs the large-scale turbulence.

In the case of compressible flows much less is known and universality is more restricted. A distinction must be made here between two different types of compressible flows. The first type of compressible flow arises as an effective description of the motion of the passive particles. Probably the first example of such a flow was presented by \citet{maxey} who considered the motion of weakly inertial particles passively transported by incompressible turbulence. Due to inertia, a particle cannot fully follow the transporting flow and its velocity is slightly different from that of the turbulent flow at its position. The velocity difference is proportional to the local Lagrangian acceleration so that the particle's velocity is a linear combination of the flow velocity and acceleration at its position. Thus the particle velocity is determined uniquely by its spatial position and an effective transporting flow of particles can be introduced. This flow is compressible since the divergence of the Lagrangian accelerations' field is non-zero. Thus the solution to the continuity equation, obeyed by the particles' concentration, is a non-constant, time-dependent inhomogeneous field. This field is passive and it does not change the flow.

The flow's independence of the solutions to the continuity equation is the benchmark of the first type of compressible flow. Since Ref. \onlinecite{maxey}, many other effective flow field descriptions of particles' motion in turbulence appeared. These include particles with large inertia which sediment fast \cite{seul,int}, bubbles \cite{bubbles}, phytoplankton \cite{patches,fl} and phoretic particles \cite{phor}. The density of particles in all these cases has universal properties and allows a unified description \cite{unified}. Here, the compressible flows are generic, and hence their sum of Lyapunov exponents, providing the logarithmic rate of growth of infinitesimal volumes, is negative \cite{generic}. This implies that the KY dimension is smaller than the dimension of space and particles distribute over a random strange attractor with multifractal density support \cite{nature}. The density obeys the continuity equation, providing its singular (weak) solution. It is also observed that compressibility damps chaos: all Lagrangian LEs decrease with compressibility in a model random flow \cite{falkovich_particles_2001}, and cases with negative $\lambda_1$ could be envisioned \cite{negl}. Whereas the Lagrangian LEs describe the motion of particles below the Kolmogorov scale, quite similar phenomena occur in the inertial range. \citet{gawedzki_phase_2000} demonstrated for a model compressible flow that tracers explosively separate for low compressibility however {they} may collapse onto each other when compressibility is high.

The other type of compressible flow is where the transport of matter by the flow is coupled to the flow, i.e. solutions to the continuity equation actively change the flow. This includes the case of inertial particles with high concentration and also the compressible turbulence at a finite Mach number, the case with many applications on which we focus below. In compressible turbulence, the density changes must be such as to keep the pressure finite, thus avoiding infinite forces. This introduces the forbidding principle for isothermal (or any other barotropic) turbulence: Lagrangian transport may not result in infinite density or solutions to the continuity equation must remain finite and smooth \cite{falkovich_particles_2001,fouxon_density_2019}. In cases where both density and temperature are non-trivial, care is needed. For instance, if the ideal gas equation of state holds, then density can become infinite provided the temperature vanishes simultaneously so the product of density and temperature remains finite. In fact, this possibility is realized in fluid mechanics with cooling where density may blow up in finite time \cite{fouxon_2007a, Fouxon_2007b, Fouxon_2009}. In the case of conservative hydrodynamics, we conjecture that this type of singularity may not occur and density stays finite so that
\begin{eqnarray}&&
\lim_{t\to\infty} \frac{1}{t} \ln\left[\frac{\rho(\boldsymbol{q}(t), t)}{\rho(\boldsymbol{q}(0), 0)}\right]=0,
 \end{eqnarray}
where the density field $\rho(\boldsymbol{x}, t)$ is evaluated on the Lagrangian trajectory $\boldsymbol{q}(t)$. However, by mass conservation, the product of infinitesimal volume and the density is constant so the above implies that the sum of Lagrangian LEs must vanish for compressible turbulence \cite{falkovich_particles_2001}. The physical reasons for the vanishing are temporal anticorrelations of the flow divergence evaluated on $\boldsymbol{q}(t)$ as was formalized via several identities in the main text and in the Appendix of \citet{fouxon_density_2019}, see also below.
Thus the sum of the Lyapunov exponents obtained from numerical simulations must vanish, provided that the flow gradients are fully resolved, see discussion in Ref. \onlinecite{fouxon_density_2019}.

The conclusion on the vanishing of the sum of LEs is robust. However, \citet{schwarz_lyapunov_2010} observed the sum to be non-zero. These authors studied the LEs of passive particles in the frame of isothermal compressible Euler equations. They adopted the CWENO scheme \cite{kurganov_thirdorder_2000} for direct numerical simulation (DNS), and employed a {linear velocity interpolation} and a third-order strong-stability-preserving Runge-Kutta method \cite{shu_efficient_1988} for particle tracking. The LEs are calculated using a standard method proposed by \citet{benettin_lyapunov_1980}. They found that the sum of LEs is not equal to 0, which implies that the KY dimension is smaller than $3$ and the particles' density is a singular field supported on a multifractal. This contradiction requires an explanation and is one of the reasons for the current work which provides seemingly the only measurement of the sum of the LEs alternative to Ref. \onlinecite{schwarz_lyapunov_2010}.

\citet{schwarz_lyapunov_2010} simulated the motion of passive tracers whose density does not react on the flow. In this form, their result does not contradict the forbidding principle. However, the authors claimed that the density of tracers is equal to that of the fluid, so that the fluid density is singular, which does contradict the finiteness of forces in the fluid and, in fact, the continuum assumption itself. We observe that despite that both densities, that of the fluid and that of the tracer particles, obey the continuity equation with the same flow, these densities might differ as was explained in Ref.~\onlinecite{fouxon_density_2019}. This is because the continuity equation is not dissipative and a fine difference may persist. Yet, the numerical observation of the difference is highly delicate. The result of \citet{schwarz_lyapunov_2010} could be an observation of such a difference. Another possible explanation could be that the non-zero sum of the LEs arises in Ref.~\onlinecite{schwarz_lyapunov_2010} because they essentially measure the coarse-grained LEs in the inertial range and not the true LEs that demand the introduction of viscosity in the numerical scheme and, resolution of sub-Kolmogorov scales.

Lagrangian LEs of compressible turbulence, it seems, were studied in Ref.~\onlinecite{schwarz_lyapunov_2010} only. These authors did not study the dependence of the LEs on the Mach and Reynolds numbers and also their results, as explained, demand a further study. For starters, a consistent study of the LEs demands determining the needed resolution and which parameters determine the exponents uniquely. Several issues arise.

Compressible turbulence is characterized by a larger number of relevant spatial scales than its incompressible counterpart. 
The phenomenology of incompressible turbulence operates with the energy pumping scale $L$ and energy dissipation (Kolmogorov) scale $\eta$ only, see e.g. Ref.~\onlinecite{Frisch}. Velocity is determined mainly by the flow fluctuations at the scale $L$ and its gradients by those at the scale $\eta$. In compressible turbulence, there are two components to the flow, solenoidal and compressible, and their behavior can be different. Both components are excited at the scale $L$, either directly by forcing, or indirectly by coupling to the other component. However, the scales that determine the components' gradients might differ. Roughly speaking, gradients of the solenoidal component are determined by the size of the smallest vortices in the flow, whereas gradients of the compressible component (the flow divergence) are determined by the smallest width of the shocks. These scales might differ. In the current work, we consider Mach numbers smaller than one so that the Kolmogorov scaling might be anticipated to describe the solenoidal component fairly well. That would imply that the smallest scale associated with this component is the Kolmogorov scale. The latter scales as $Re^{-3/4}$, where $Re$ is the ordinary Reynolds number, and is the scale that determines the gradients of the solenoidal component of the flow (vorticity). On the other hand, the naive smallest scale of the shock, estimated from the Burgers-type solutions, scales proportionally to $Re^{-1}$. Hence the smallest scale of the flow is anticipated to behave as $L/Re$. Thus, in principle, accurate determination of the flow derivatives, which is necessary to determine the LEs, demands resolution below this scale.

The LEs derive from the gradients of the flow $\nabla\boldsymbol{u}$. Hence, in computing the exponents, it is significant to understand whether the statistics of the gradients of the compressible turbulent flow are universal and what they depend on. Both, the solenoidal and the compressible components of the flow, are relevant. The sum of the Lyapunov exponents is determined by the flow divergence and thus, mostly by the compressible component of the flow (it still has an implicit dependence on the solenoidal component via the Lagrangian trajectory on which the divergence is determined, see below). Other combinations of the exponents, generally speaking, depend on both components. It can be hoped that if the statistics of the gradients are universal, then the Lyapunov exponents, made dimensionless by multiplying with $\langle |\nabla \bm u|^2\rangle^{-1/2}$ would depend only on the degree of compressibility $C\equiv \langle (\nabla\cdot \bm u)^2\rangle/\langle |\nabla \bm u|^2\rangle$ and the Reynolds number. The ratio $C$ measures the compressibility of the small-scale turbulence, changing from zero for incompressible flow to one for potential flow.

The forcing that drives the turbulence imposes on the flow comparative weights of the solenoidal and compressible components at the scale $L$. In order to have universal statistics of the gradients, the inertial range must be as large as needed so that at the smallest scale of the flow the weights of the flow components attain intrinsic values that are independent of their values at scale $L$. Thus, unless the inertial range is large (larger than in the case of incompressible turbulence), the gradients and the LEs would not be universal. They would depend on $Re$, on $Ma$, and the type of force. Moreover, the dependence on $Ma$ is non-obvious since the local Mach number associated with the small-scale turbulence is anticipated to be small in many cases. That would make small-scale turbulence effectively incompressible with all its implications. It is seen from the above that compressible turbulence imposes many issues. Some of them are tackled below.

In this paper, we investigate LEs of passive particles in three-dimensional compressible isotropic turbulence by using high-order methods for direct numerical simulation and for tracking particles, with a particular emphasis on their dependence on Taylor Reynolds number and turbulent Mach number, as well as the driving force. We check whether the sum of the LEs vanishes and whether the LEs decay as a function of the Mach number. The remaining part of the paper is presented in 5 sections. We first describe the governing equations and the driving force in Section \ref{sec:2}. Definitions and properties of the Lyapunov exponents are presented in Section \ref{sec:3} while Section \ref{sec:4} describes the high-order numerical schemes for DNS, tracking particles, and the calculation of LEs. The major results are presented and discussed in Section \ref{sec:5}. Conclusions are given in Section \ref{sec:6}.

\section{Governing equations} \label{sec:2}

\subsection{Compressible Navier-Stokes equations and the driving force}
We study compressible homogeneous isotropic turbulence governed by the following three-dimensional Navier-Stokes equations in a dimensionless form:\cite{wang_hybrid_2010}
 \begin{equation}
 \frac{\partial \rho}{\partial t} + \frac{\partial (\rho u_j)}{\partial
 x_j} = 0,
 \end{equation}
 \begin{equation}
 \frac{\partial (\rho u_i)}{\partial t} + \frac{\partial [\rho u_i u_j +
 {p_m} \delta_{i j}]}{\partial x_j} =
 \frac{1}{{\tmop{Re}}} \frac{\partial \sigma_{ji}}{\partial
 x_j} +\mathcal{F}_i,
 \end{equation}
 \begin{equation}
 \frac{\partial \mathcal{E}}{\partial t} + \frac{\partial [(\mathcal{E}+
 p_m) u_j]}{\partial x_j} = \frac{1}{\alpha} \frac{\partial}{\partial x_j}
 \left( {\kappa} \frac{\partial T}{\partial x_j} \right) +
 \frac{1}{{\tmop{Re}}} \frac{\partial (\sigma_{ji}
 u_i)}{\partial x_j} - \Lambda +\mathcal{F}_j u_j,
 \end{equation}
 where $\rho$ is the fluid density, $u_i$ the $i$th component of fluid velocity,  viscous stress tensor $\sigma_{i j} = {\mu} (\frac{\partial u_i}{\partial x_j} + \frac{\partial u_j}{\partial x_i})
 - \frac{2}{3} {\mu} \theta \delta_{i j}$,
 dilatation $\theta = \partial u_i/\partial x_i$,
 scaled pressure $p_m = \frac{p}{\gamma {\tmop{Ma}}^2}$,
 pressure $p = \rho T$, and $\mathcal{E}= \frac{p}{(\gamma - 1) \gamma \tmop{Ma}^2} + \frac{1}{2} \rho (u_j u_j)$ the total energy per unit volume.
 $\mathcal{F}_i$ is a large-scale driving force, and $\Lambda$ a cooling function.
 Einstein's summation convention is adopted. For simplicity, Stokes' hypothesis is assumed to be true, i.e. the bulk viscosity is set to 0.
 Note that the equations are nondimensionalized such that the spatial averaged density $\langle \rho \rangle =1$, and the size of the computational box is a cube with side length $L=2\pi$.
 There are three reference dimensionless parameters: the reference Mach number $Ma = U_f / c_f$, the reference Reynolds number $Re = \rho_f U_f L_f /\mu_f$, and the reference Prandtl number $Pr = \mu_f C_p/\kappa_f$, where $U_f, c_f, L_f, \mu_f, \kappa_f, C_p$ are reference velocity, speed of sound, length, viscosity, thermal conductivity, and specific heat at constant pressure, respectively.
 The coefficient $\alpha$ is equal to $\Pr \tmop{Re}(\gamma-1)\tmop{Ma}^2$. Here $\gamma$ is the ratio of specific heat at constant pressure $C_p$ to that at constant volume $C_v$. In this study, we fix $\Pr = 0.7$ and $\gamma=1.4$.
 The scaled $\mu$ {and} $ \kappa$ satisfy the following Sutherland's formula\cite{sutherland1893}
 \[ \mu = \kappa = \frac{1.4042 T^{1.5}}{T + 0.40417} . \]
 The cooling function takes {the} form $\Lambda = a T^b$.
 Several studies have shown that the statistics are not sensitive to the choice of the cooling function\cite{wang_hybrid_2010, jagannathan_reynolds_2016, liu_hybrid_2019}. So we fix $b=1$ in this paper.

The decomposition of the flow 
into the sum of the solenoidal 
and compressible 
components is done in the Fourier space
\begin{eqnarray}&&
\boldsymbol{v}(\boldsymbol{k}, t)\equiv
\int_{[0,2\pi]^3} \exp(-i\boldsymbol{k}\cdot \boldsymbol{x})\boldsymbol{u}(\boldsymbol{x}, t)d\boldsymbol{x}; \quad
\boldsymbol{v} (\boldsymbol{k}, t) =\boldsymbol{v}^{s} (\boldsymbol{k},
 t) +\boldsymbol{v}^{c} (\boldsymbol{k}, t);\quad
\boldsymbol{v}^{c} =\frac{\boldsymbol{k}(\boldsymbol{k} \cdot
 \boldsymbol{v})}{|\boldsymbol{k}|^2},\ \ \boldsymbol{v}^{s}
 =\boldsymbol{v}-\boldsymbol{v}^{c}.
\end{eqnarray}
The solenoidal and compressible components determine the flow's curl (vorticity) and divergence, respectively.

It is known that the statistics in (compressible) turbulence are sensitive to the driving force. Different kinds of forces have been considered in the literature, see e.g. \citet{eswaran_examination_1988,kida_energy_1990,mininni_large-scale_2006,petersen_forcing_2010,wang_hybrid_2010,konstandin_statistical_2012, john_does_2021}. Here, we adopt a large-scale driving force that can lead to a statistical steady state quickly \cite{wang_hybrid_2010}.
 The large-scale forcing   is constructed in Fourier space by fixing $E
 (k)$ within the two lowest wave number shells: $k=1$ and $k=2$, where $E(k) = \sum_{k-0.5 < |\boldsymbol{k}| \leqslant k+0.5} |\boldsymbol{v}(\boldsymbol{k})|^2/2 $,
  to prescribed values that are consistent with the $k^{-5/3}$ kinetic energy
 spectrum.
More precisely, if a pure solenoidal force is adopted,  after the forcing, we let
\begin{equation} \label{eq:sforce1}
    \boldsymbol{v}_{\tmop{new}} (\boldsymbol{k}, t) = \alpha_{k}
 \boldsymbol{v}^{s} (\boldsymbol{k}, t) +\boldsymbol{v}^{c}
 (\boldsymbol{k}, t)
\end{equation}
 such that the velocity magnitude in $k=1$ and $k=2$ shells are
 equal to pre-specified values $E(k), k = 1, 2$.
 This is achieved by taking
 \begin{equation} \label{eq:sforce2}
     \alpha_{k} =
 \sqrt{\frac{E(k)-E^{c}(k)}{E^{s}(k)}},
 \qquad
 E^{\beta} (k) = \sum_{k-0.5<| \boldsymbol{k} | \leqslant
 k + 0.5} \frac{1}{2} | \boldsymbol{v}^{\beta} (\boldsymbol{k}, t) |^2,
 \quad
 \beta = c, s.
\end{equation}
For forces with nonzero compressible components, we modify the compressible component as well. To this end, we use a factor $\gamma_0$ to denote the ratio of the solenoidal part to the compressible part. The driving force is applied to obtain
\begin{equation}\label{eq:cforce1}
    \boldsymbol{v}_{\tmop{new}} (\boldsymbol{k}, t) = \alpha^s_{k}
 \boldsymbol{v}^{s} (\boldsymbol{k}, t) + \alpha^c_{k} \boldsymbol{v}^{c}
 (\boldsymbol{k}, t),
\end{equation}
 where
 \begin{equation}
    \label{eq:cforce2}
  \alpha^s_{k} =
 \sqrt{\frac{\gamma_0 \left(E(k)-E^{c}(k)\right)+E^s(k)}{ (1+\gamma_0) E^{s}(k)}},
 \qquad
 \alpha^c_{k} =
 \sqrt{\frac{ \left(E(k)-E^{s}(k)\right)+\gamma_0 E^c(k)}{ (1+\gamma_0) E^{c}(k)}}.
\end{equation}
In particular,
\begin{itemize}
    \item $\gamma_0 =\infty$ generates a purely solenoidal force, the case called below ST.
    \item $\gamma_0 =1$ generates a force with the solenoidal and compressible parts comparable, the case called below C1.
    \item $\gamma_0=0$ generates {a force with only the compressible part}.
\end{itemize}
Due to the limitation of computational resources, in this paper, we will only investigate the ST and C1 cases.

\subsection{Statistical quantities}

The following important statistical quantities are of interest in compressible turbulence.
The root mean square (r.m.s.) component fluctuation velocity $u'$ is defined and related to the kinetic energy spectrum $E (k)$ as
\begin{equation}
\frac{3}{2} (u')^2 \equiv \tfrac{1}{2} \langle \boldsymbol{u} (\boldsymbol{x},
 t) \cdot \boldsymbol{u} (\boldsymbol{x}, t) \rangle = \int_0^{\infty}
  {E (k)} \mathrm{d} k,
  \label{eq:rmsvelocity}
  \nonumber
  \end{equation}
where $\langle \cdot\rangle$ stands for the spatial average.
 The longitudinal integral length scale is
 \[ L_f = \frac{\pi}{{2 u'}^2} \int_0^{\infty} \frac{E (k)}{k}
 \mathrm{d} k . \]
The transverse Taylor microscale $\lambda$ and Taylor
 Reynolds number $\tmop{Re}_{\lambda}$ are defined as
 \[ \lambda = \frac{u'}{\langle (\partial u_i / \partial x_i )^2 /
 3 \rangle^{1 / 2}}, \quad \tmop{Re}_{\lambda} = \frac{u' \lambda
 \langle \rho \rangle}{\langle \mu \rangle} \text{Re} , \]
where the summation convention is not assumed in the equation. The viscous dissipation rate $\varepsilon$, Kolmogorov length scale
 $\eta$, and timescale $\tau_{\eta}$ are defined as
 \[ \varepsilon =  \langle {\sigma_{i
 j}} \frac{\partial u_i}{\partial x_j} \rangle,
 \quad \eta = Re^{-3/4}[\langle \mu^3 \rangle/ \langle {\rho}^2 \rangle /
 \varepsilon]^{1 / 4},
 \quad \tau_{\eta} = Re^{-1/2}(\langle \mu \rangle /
 \varepsilon)^{1 / 2} . \]
 The turbulent Mach number $M_t$ is defined as
 \[ M_t = \tmop{Ma} \langle u_1^2 + u_2^2 + u_3^2 \rangle^{1 / 2} / \langle
 \sqrt{T} \rangle . \]

\section{Definitions and properties of Lyapunov exponents of compressible turbulence}\label{sec:3}

\subsection{Definitions}

The trajectories of passive particles in a fluid are determined by
\begin{equation}
 \frac{\mathrm{d} \boldsymbol{q} (t)}{\mathrm{d} t} =\boldsymbol{u} (\boldsymbol{q} (t),
 t), \quad \boldsymbol{q} (t = 0) =\boldsymbol{q}_0, \quad \boldsymbol{q} (t) \in
 \mathbb{R}^3, \label{eq:pp}
\end{equation}
where $\boldsymbol{u}$ is the flow field obtained from DNS. An infinitesimal perturbation $\boldsymbol{r}_0$ in the initial condition $\boldsymbol{q} (t = 0) =\boldsymbol{q}_0+\boldsymbol{r}_0$ leads to a slightly different trajectory. The distance $\boldsymbol{r} (t)$ between the two resulting trajectories obeys the evolution equation
\begin{equation}
 \frac{\mathrm{d} \boldsymbol{r} (t)}{\mathrm{d} t} =(\boldsymbol{r} (t)\cdot\nabla) \boldsymbol{u}, \quad \boldsymbol{r} (t = 0)
 =\boldsymbol{r}_0, \quad \boldsymbol{r} (t) \in
 \mathbb{R}^3, \label{eq:pertub}
\end{equation}
where the flow gradients are evaluated on the trajectory $\boldsymbol{q} (t)$. The Lyapunov characteristic number (LCN) is defined as
 \begin{equation}
 \lambda (\boldsymbol{q}_0, \boldsymbol{r}_0) = \limsup_{t \rightarrow \infty}
 \frac{1}{t} \frac{| \boldsymbol{r} (t) |}{|\boldsymbol{r}_0 |}. \label{eq:LCN}
 \end{equation}
 It has the following basic property: there exist $d$ linear subspaces $Z_d \subset
 Z_{d - 1} \subset \cdots \subset Z_1 =\mathbb{R}^d$, such that
 \[ \lambda_i (\boldsymbol{q}_0) = \max_{\boldsymbol{r}_0 \in Z_i} \lambda
 (\boldsymbol{q}_0, \boldsymbol{r}_0) = \left\{ \lambda (\boldsymbol{q}_0,
 \boldsymbol{r}_0),\:\; \forall \, \boldsymbol{r}_0 \in Z_i \backslash
 Z_{i + 1} \right\} . \]
 Alternatively, we may define Lyapunov exponents using the Cauchy-Green tensor.
 Consider
 \begin{equation}
 \frac{\mathrm{d} D_{i j} (t)}{\mathrm{d} t} = \frac{\partial
 u_i}{\partial x_k} \left(\boldsymbol{q} (t), t\right) D_{k j} (t), \quad D_{i
 j} (t = 0) = \delta_{i j} . \label{auchy-Green}
 \end{equation}
 The finite-time Lyapunov exponents (FTLEs) are defined by using the singular
 values
 $\sigma_i (\boldsymbol{q}_0, t)$ of the deformation tensor $D_{i j}$,
 i.e. the square roots $\sigma_i$ of the eigenvalues of the symmetric positive Cauchy-Green tensor $C_{i
 j} = D_{i k} D_{j k}$.
 \[ \gamma_i (\boldsymbol{q}_0, t) := \frac{1}{t} \ln \sigma_i
 (\boldsymbol{q}_0, t) . \]

 The Lyapunov exponents (LEs) are defined as the infinite time limit of FTLEs and are equivalent to LCNs, i.e.
 \begin{equation}
 \lim_{t \rightarrow \infty} \gamma_i (\boldsymbol{q}_0, t) = \lambda_i
 (\boldsymbol{q}_0) .
 \end{equation}

\subsection{Representation of the Lyapunov exponents for three-dimensional flow}

We are concerned with three-dimensional flows where the spectrum of the Lyapunov exponents is fully described by $\lambda_1$, $\lambda_3$, and $\sum_{i=1}^3\lambda_i$. These three quantities allow a simplified description in any dimension. Considering $\boldsymbol{q} (t)$ as a vector function of the initial condition $\boldsymbol{q}_0$, we find from Eq.~(\ref{auchy-Green}) that the Jacobi determinant of this function, $J(t)=|\partial \textbf{q}(t)/\partial \textbf q_0|$, obeys a simple equation. We have by using the definition of $\lambda_i$
 \begin{equation}\label{eq:LEsum1}
\frac{\mathrm{d} \ln J(t)}{\mathrm{d} t} = \nabla\cdot \boldsymbol{u},\ \
    \sum \lambda_i \equiv \lim_{t \rightarrow \infty} \frac{\ln J(t)}{t}= \lim_{t \rightarrow \infty} \frac{1}{t}
    \int_0^t \langle \nabla \cdot \boldsymbol{u} (\boldsymbol{q}, t') \rangle_L
    \mathrm{d} t', 
 \end{equation}
  where $\langle \cdot \rangle_L$ denotes Lagrangian average. Ergodicity is assumed, which will be numerically verified in the numerical results section.
 Due to the fact that $\frac{\mathrm{d}}{\mathrm{d} t} \ln \rho (\boldsymbol{q},
 t) = - \nabla \cdot \boldsymbol{u}$, we also have
 \begin{equation}\label{eq:LEsumalt}
    \sum \lambda_i = \lim_{t \rightarrow \infty} \frac{1}{t} \int_0^t
    - \frac{\mathrm{d}}{\mathrm{d} t} \ln \rho (t') \mathrm{d} t' = \lim_{t
    \rightarrow \infty} \frac{1}{t} (\ln \rho (0) - \ln \rho (t)) .
 \end{equation}
 If we take uniform initial density, i.e. $\rho(0)\equiv 1$, then we have
 \begin{equation} \label{eq:LEsum2}
    \sum \lambda_i = \lim_{t \rightarrow \infty} -\ln \rho(t)/t   =
    \lim_{t \rightarrow \infty} -\langle \ln \rho(t) \rangle_L /t.
 \end{equation}
 Since $\langle \cdot \rangle_L$ denotes the Lagrangian average and the passive particles
 and the fluid density {have} the same probability distribution in a statistically steady turbulence state, so $\langle \ln \rho(t) \rangle_L = \langle \rho(t) \ln \rho(t) \rangle \ge \langle \rho(t)\rangle \ln \langle \rho(t)\rangle =0$ by Jensen inequality. Here, the assertion that Lagrangian averages are equivalent to density-weighted averages is assumed and will be justified by a numerical investigation given in the numerical results section. Thus, if the start-up density satisfies a uniform distribution, the sum of finite time LEs will take negative values and converge to $0$ from below at a speed $\mathcal{O}(1/t)$.

 For largest and smallest {{LEs}}, they can be evaluated using {the} following dynamics \cite{falkovich_particles_2001,balkovsky_universal_1999}
 \begin{align}
 \lambda_1 = \lim_{t \rightarrow \infty} \frac{1}{t} \int_0^t \sigma :
 \boldsymbol{r}_1 \boldsymbol{r}_1 \mathrm{d} t', & \qquad
 \frac{d \boldsymbol{r}_1}{dt} =
 \sigma \cdot \boldsymbol{r}_1 - (\sigma : \boldsymbol{r}_1
 \boldsymbol{r}_1) \boldsymbol{r}_1, \label{eq:LE1}\\
\lambda_3 = \lim_{t \rightarrow \infty} \frac{1}{t} \int_0^t \sigma :
 \boldsymbol{r}_3 \boldsymbol{r}_3 \mathrm{d} t', & \qquad
 \frac{d \boldsymbol{r}_3}{dt} =
 - \sigma^T \cdot \boldsymbol{r}_3 + (\sigma : \boldsymbol{r}_3
 \boldsymbol{r}_3) \boldsymbol{r}_3, \label{eq:LE3}
 \end{align}
 where $\sigma = \nabla_{\boldsymbol{x}}\boldsymbol{u}$ and $\boldsymbol{r}_1, \boldsymbol{r}_3$ are two randomly initialized vectors {in the respective linear subspaces(e.g. $\boldsymbol{r_1(0)} \in Z_1\backslash Z_2$)}.
 We note that the \eqref{eq:LE1} is mathematically equivalent to
 \eqref{eq:pertub}-\eqref{eq:LCN}, so we can use \eqref{eq:FTLEnum} below to estimate it. Equation \eqref{eq:LE3} is obtained in Appendix B of \citet{balkovsky_universal_1999}. Roughly speaking, the representation for $\lambda_3$ can be explained by observing that the equation on the Cauchy-Green tensor, which in matrix form reads $\dot{D}=\sigma D$, implies $\dot{D}^{-1, T}=-\sigma^T D^{-1, T}$. The singular values of $D^{-1, T}$ are $-\lambda_i$ which implies Eq.~(\ref{eq:LE3}). Since \eqref{eq:LE1} and \eqref{eq:LE3} have similar forms, we can use the standard numerical method for $\lambda_1$ to calculate $\lambda_3$ by considering the transformed backward dynamics.
\subsection{Incompressible-like large deviations theory of finite-time Lyapunov exponents}

We saw that $\sum\lambda_i$ of compressible turbulence equals zero as in the incompressible flow. We demonstrate here that fluctuations of the finite-time Lyapunov exponents also behave as in the incompressible flow. This has a remarkable consequence that $|\lambda_3|$ is the first Lyapunov exponent of the time-reversed flow. Generally speaking, an infinitesimal spherical parcel of the fluid is transformed by the flow into an ellipsoid, whose largest and smallest axes, asymptotically at large times, grow and shrink exponentially, with exponents $\lambda_1$ and $\lambda_3$, respectively. Reversing this evolution naively, we find that in time-reversed flow the stretching occurs with exponent $|\lambda_3|$ and shrinking with the exponent $-\lambda_1$. In fact, this consideration is generally true in incompressible flow only: compressibility introduces a non-trivial Jacobian (volume) factor which usually depends on time exponentially causing the principal exponent of the time-reversed flow to differ from $|\lambda_3|$, see the arXiv version of \citet{itzhak_1999}. In compressible turbulence, however, the Jacobian is stationary since infinitesimal volumes define stationary density. This makes the volume factor irrelevant to the exponential behavior of the ellipsoid's dimensions.

We provide a more formal description of the considerations above. We consider the joint probability density function (PDF) of
\begin{eqnarray}&&
\kappa(t)\equiv \ln\left(\frac{\rho(0)}{\rho(\bm q(t), t)}\right)=\int_0^t \nabla \cdot \boldsymbol{u} (\boldsymbol{q}, t')
    \mathrm{d} t',\ \ \rho_1(t)=\int_0^t \sigma :
 \boldsymbol{r}_1 \boldsymbol{r}_1 \mathrm{d} t',\ \ \rho_3(t)=\int_0^t \sigma :
 \boldsymbol{r}_3 \boldsymbol{r}_3 \mathrm{d} t',
\end{eqnarray}
where $\boldsymbol{r}_{1, 3}$ are defined in Eq.~(\ref{eq:LE3}).

At large times the distribution of $\kappa(t)$ becomes stationary since it becomes a difference of two independent stationary processes $\ln \rho(0)$ and $\ln \rho(\bm q(t), t)$. The time integral in the definition of $\kappa(t)$ does not accumulate with time due to the anticorrelations in the temporal behavior of $\nabla \cdot \boldsymbol{u} (\boldsymbol{q}, t)$ that manifests via the vanishing of $\int \langle \nabla \cdot \boldsymbol{u} (\boldsymbol{q}, t) \nabla \cdot \boldsymbol{u} (\boldsymbol{q}, t')\rangle dt'$. The vanishing of this integral in several forms (quasi-Eulerian and Lagrangian) was discussed at length in Appendix A of \citet{fouxon_density_2019}. Thus $\kappa(t)$ depends only on the values of the velocity divergence in the time vicinity of times $0$ and $t$ whose size is of the order of the correlation time of $\nabla \cdot \boldsymbol{u} (\boldsymbol{q}, t)$.

In contrast, $\rho_{1, 3}$ depend on the gradients during the whole time interval $(0, t)$. At large times, the contribution of time intervals near $0$ and $t$ that determine $\kappa(t)$ can be neglected in $\rho_{1, 3}(t)$. We infer from this that $\kappa(t)$ and $\rho_{1, 3}(t)$ become independent, see similar considerations in \citet{balkovsky_universal_1999}. We conclude that the joint PDF $P(\kappa, \rho_1, \rho_3, t)$ of $\kappa(t)$ and $\rho_{1, 3}(t)$ reads
\begin{eqnarray}&&
P(\kappa, \rho_1, \rho_3, t)\sim P_0(\kappa) \exp\left(-tS\left(\frac{\rho_1}{t}-\lambda_1, \frac{\rho_3}{t}-\lambda_3\right)\right), \label{for}
\end{eqnarray}
where the standard large deviations form is used for the PDF of $\rho_{1, 3}(t)$, see e.g. \citet{balkovsky_universal_1999}. The convex large deviations function $S(x_1, x_2)$ is non-negative and it has a unique minimum of zero at $x_1=x_2=0$. It is readily seen that at $t\to\infty$ the above distribution becomes a $\delta-$function which guarantees that the limits in Eqs.~(\ref{eq:LE1}) and (\ref{eq:LE3}) are non-random. The typical values of $\rho_{1, 3}(t)$ scale linearly with $t$ so that at large times bounded fluctuations of $\kappa$ become negligibly small, compared with the spread of the distribution. For most relevant observables, such as those considered in turbulent mixing \cite{balkovsky_universal_1999}, the results of the averaging with the PDF in Eq.~(\ref{for}) coincide with those obtained by using the effective PDF given by
\begin{eqnarray}&&
P(\kappa, \rho_1, \rho_3, t)\sim \delta(\kappa) \exp\left(-tS\left(\frac{\rho_1}{t}-\lambda_1, \frac{\rho_3}{t}-\lambda_3\right)\right). \label{fro}
\end{eqnarray}
This has the same form as in the incompressible flow, as explained at the beginning of the Section. The recalculation of this result into the statistics of the finite time Lyapunov exponents is straightforward, see Ref.~\onlinecite{fouxonkalda}. The form of the statistics is the same as in the incompressible flow, as derived in Ref.~\onlinecite{balkovsky_universal_1999}. We conclude that long-time statistics of the finite-time Lyapunov exponents have the same form as in the incompressible flow. For instance, this implies that the principal Lyapunov exponent of the time-reversed flow is $|\lambda_3|$, the fact that is non-true for a general compressible flow \cite{itzhak_1999}.

\subsection{Perturbation theory for Lyapunov exponents at small Mach number}

We study below the dependence of $\lambda_i$ on the Mach number $Ma$. In the limit of very small $Ma$ the results must be describable by the perturbation theory around the exponents of the incompressible turbulent flow that holds at $Ma\to 0$ \cite{Ristorcelli_1997, Zank_1991, Sarkar_1991, jagannathan_reynolds_2016, Wang_2017}. The corrections to the exponents of incompressible turbulence are of the same order in $Ma$ as the compressible component. However, this perturbation theory holds only at very small Mach numbers, empirically below $0.1$\cite{jagannathan_reynolds_2016, Wang_2017} and it is of no use below where larger $Ma$ are considered.

\section{Numerical methods\label{sec:4}}

\subsection{Numerical methods for direct numerical simulation of turbulence}
We use a hybrid WENO-central compact difference method proposed by \citet{liu_hybrid_2019}, which is an improved version of \citet{wang_hybrid_2010} for DNS of compressible isotropic turbulence. The DNS solver has the following properties:
\begin{itemize}
 \item An 8th-order compact central difference scheme
 ({{CCD8}}\cite{lele_compact_1992}) for smooth regions, a 7th-order WENO scheme
 ({{WENO7}}\cite{balsara_monotonicity_2000}) for shock regions is used to discretize the convection terms.

 \item Shock fronts are determined by $\theta < - R_{\theta} \theta'$,
 where $R_{\theta} ~= 3$, $\theta'$ is the r.m.s. of $\theta$.

 \item Viscous dissipation terms are handled by a {{non-compact 6th-order central difference scheme (CD6)}}.

 \item Thermal diffusion term in the energy equation is {discretized} using a {{non-hybrid CCD8 scheme}}.

 \item Strong-stability preserving third-order Runge-Kutta method (SSP-RK3\cite{shu_efficient_1988}) is used for time marching.

 \item {{Numerical hyperviscosity}} is applied to $\rho, u, v, w, T $ in every 5 steps (filtering) as
 \begin{equation}
    \frac{\partial f}{\partial t} = \nu_n [f_n'' - (f'_n)'_n],
 \end{equation}
 where {$f$ stands for any of $\rho, u, v, w$ and $T$}, and  $f_n'$, $f''_n$  stands for numerical discretizations for $f_x$ and $f_{xx}$ using CCD8 scheme respectively. We set $\nu_n = 2\times 10^{-2}$ in the numerical simulation.
 Note that \cite{wang_hybrid_2010}
 \[ \nu_n [f_n'' - (f'_n)'_n] \sim \nu_n C (0) (\Delta x)^8 (\nabla^2)^5
 f, \quad \quad C (0) \approx 2.88
 \times 10^{- 5}. \]
So $\nu_n C (0) \ll \mu / \tmop{Re}$. Thus, the numerical hyperviscosity only has a strong smoothing effect on high wave number parts. The smoothing effect on low-frequency
 part $\hat{f}_k, k< k_{max}/2$ is quite small compared to the physical diffusion $\tfrac{\mu}{Re} \nabla^2 f$ in the Navier-Stokes equation.
 \end{itemize}
 To get reliable results of Lyapunov exponents, we will only consider the cases where the shock scales can be resolved or nearly resolved numerically, which means that we focus on low Reynolds number cases. For these low Reynolds cases, we will also {present numerical results computed by using a 5th-order upwind compact scheme \citep{adams_high-resolution_1996} together with a compact central difference scheme with hyperviscosity for comparison}.

 \subsection{Numerical methods for passive particles and Lyapunov exponents}
 The standard method for numerically calculating all LEs\citep{benettin_lyapunov_1980} is described below.
 The method of computing all LEs needs first to choose random vectors $\boldsymbol{r}^0_1, \ldots,
 \boldsymbol{r}^0_m \in \mathbb{R}^d$ with $(\boldsymbol{r}^0_i, \boldsymbol{r}^0_j)
 = \delta_{i j}$, $m \leqslant d$, then evolve them via equation \eqref{eq:pertub}, then the LEs are formally given by
 \begin{equation}\label{eq:LEnum}
   \lambda_p = \lim_{k \rightarrow \infty} \frac{1}{k \tau} \sum_{i =
 1}^k \ln r_p^{(i)},
 \qquad r_p^{(i)} = |
 \tilde{\boldsymbol{r}}_p (\boldsymbol{r}^{(i-1)}_{p}, \tau) |,
 \qquad p = 1, \ldots, .
 \end{equation}
 Here ${\boldsymbol{r}}_p (\boldsymbol{r}^{(i-1)}_{p}, \tau)$ stands for the vector obtained by marching equation \eqref{eq:pertub} with initial value $\boldsymbol{r}^{(i-1)}_{p}$ for a time period $\tau$. $\tilde{\boldsymbol{r}}_p, p = 1, \ldots, d$ are vectors after
 applying Gram-Schmidt orthogonalization (but not normalized) to $\boldsymbol{r}_p$. More precisely,
 \[ \tilde{\boldsymbol{r}}_p =\boldsymbol{r}_p - \sum_{j = 1}^{p - 1}
 (\boldsymbol{r}_p, \boldsymbol{r}_j^{(i)})
 \boldsymbol{r}_j^{(i)}, \quad \boldsymbol{r}_p^{(i)}=\tilde{\boldsymbol{r}}_p/|\tilde{\boldsymbol{r}}_p|,
 \quad p = 1, \ldots, m. \]

 We note that there exist some improvements in the orthogonalization procedure
 (see, e.g. Ref.~\onlinecite{eckmann_ergodic_1985}), but the standard Gram-Schmidt
 orthonormalization is good enough for our purpose since we only need to compute three LE exponents.

 Numerically we can't solve the perturbation equation \eqref{eq:pertub} for an infinitely long time to obtain the $\lambda_p$ defined by \eqref{eq:LEnum}.
 Instead, we will use a relatively large terminal time $t$, and use the corresponding FTLEs to approximate LEs. The FTLEs are evaluated as
 \begin{equation}
    \label{eq:FTLEnum}
    \gamma_p = \frac{1}{K-K_0} \sum_{i=K_0}^{K-1} a_p^{(i)},\quad p=1,2,3,
 \end{equation}
where $a_p^{(i)} = \tfrac{1}{\tau}\ln r_p^{(i)}$, $\tau =t/K$.
$K_0$ is a parameter to be tuned to get faster convergence by discarding data from a transient region.

 To verify the numerical accuracy of the standard numerical methods for LEs, we also implemented alternative approaches \eqref{eq:LEsumalt} for the sum of all LEs and a different approach \eqref{eq:LE3} for the largest and smallest LE.

\section{Numerical results and discussions\label{sec:5}}

 \subsection{Verifying the numerical schemes}

 \subsubsection{Checking numerical resolutions of DNS}

\begin{figure}[!htb]
    \centering
    \includegraphics[width=1\textwidth]{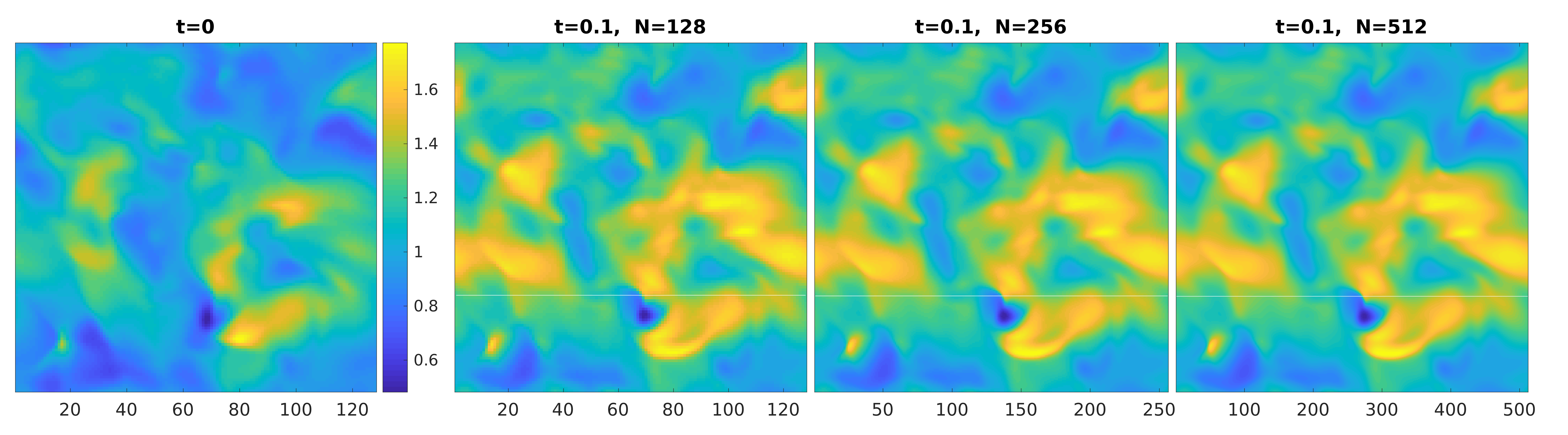}
    \includegraphics[width=1\textwidth]{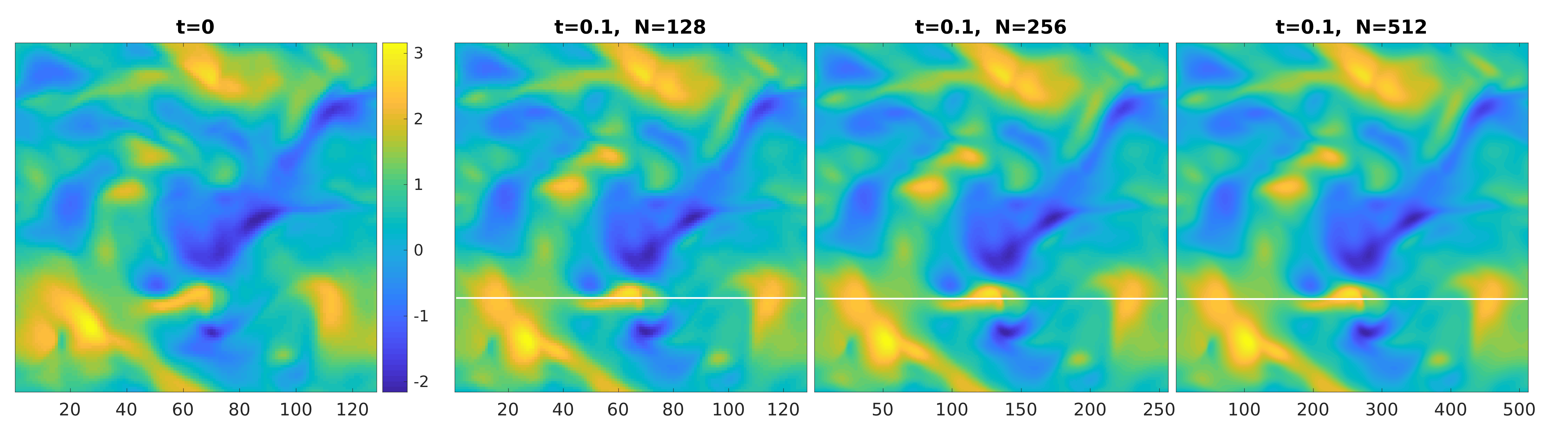}
    \includegraphics[width=1\textwidth]{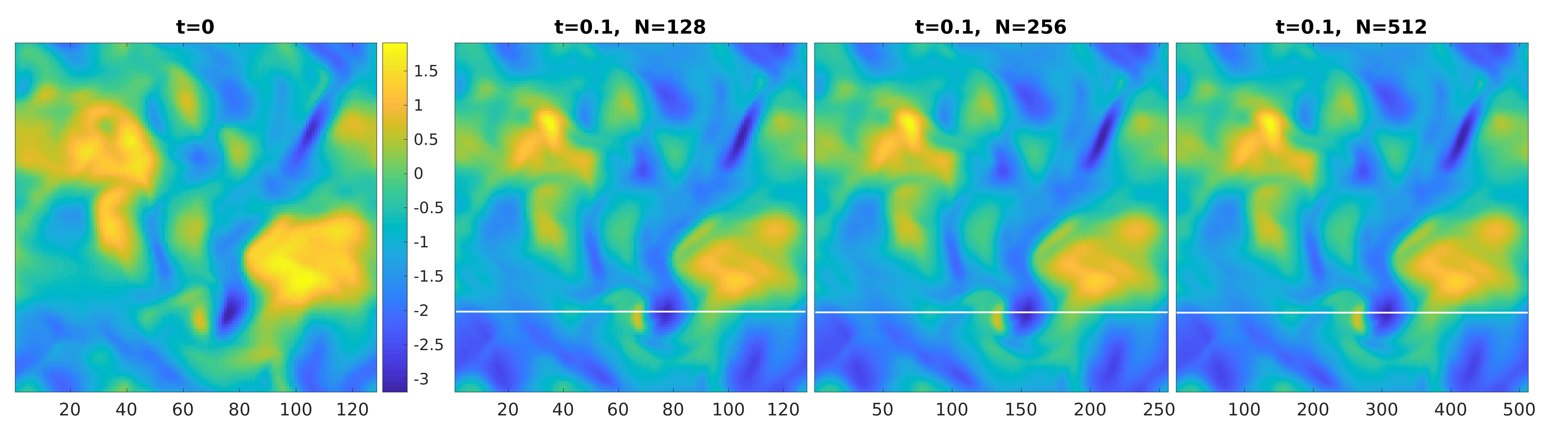}
    \includegraphics[width=0.49\textwidth]{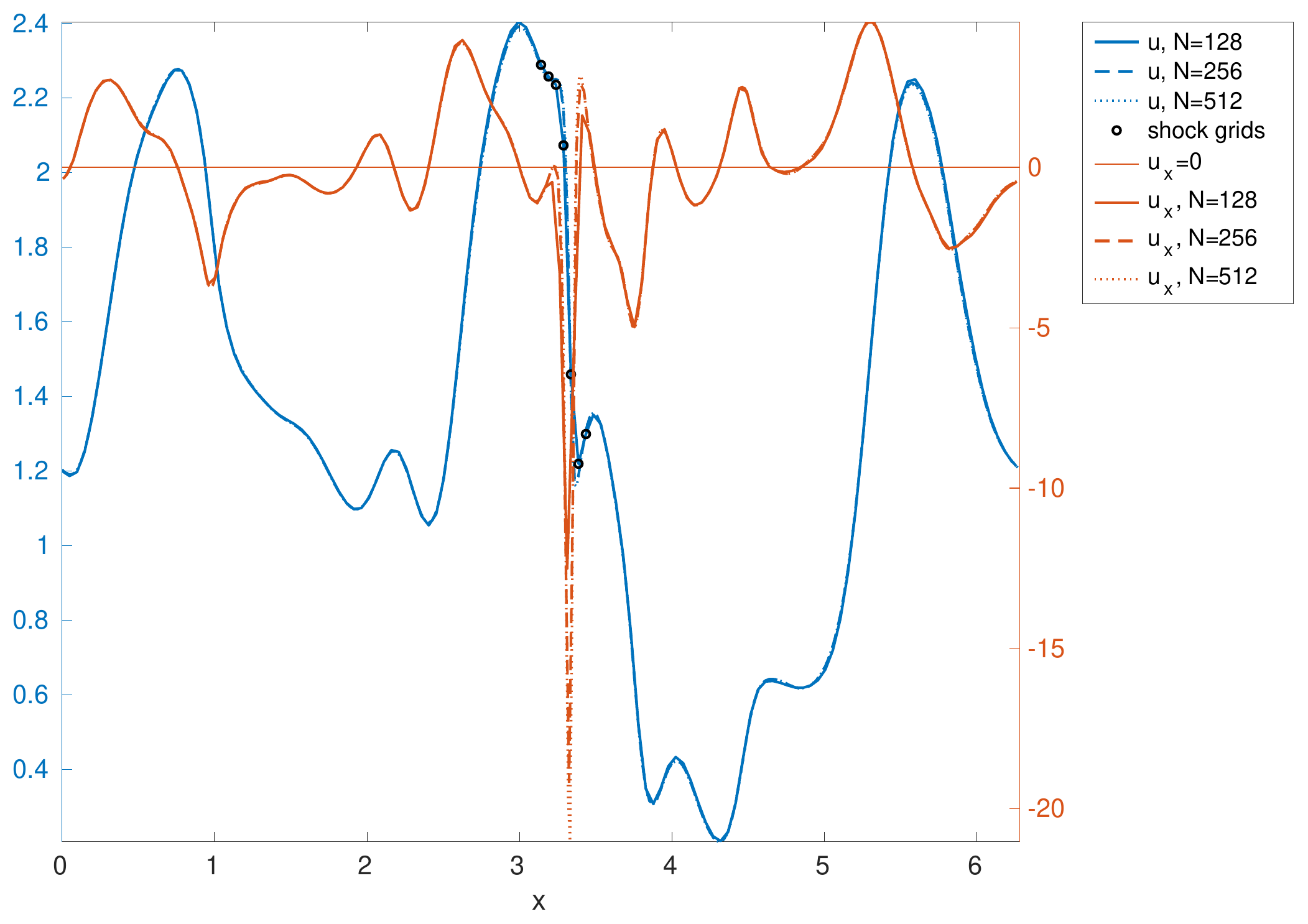}
    \includegraphics[width=0.49\textwidth]{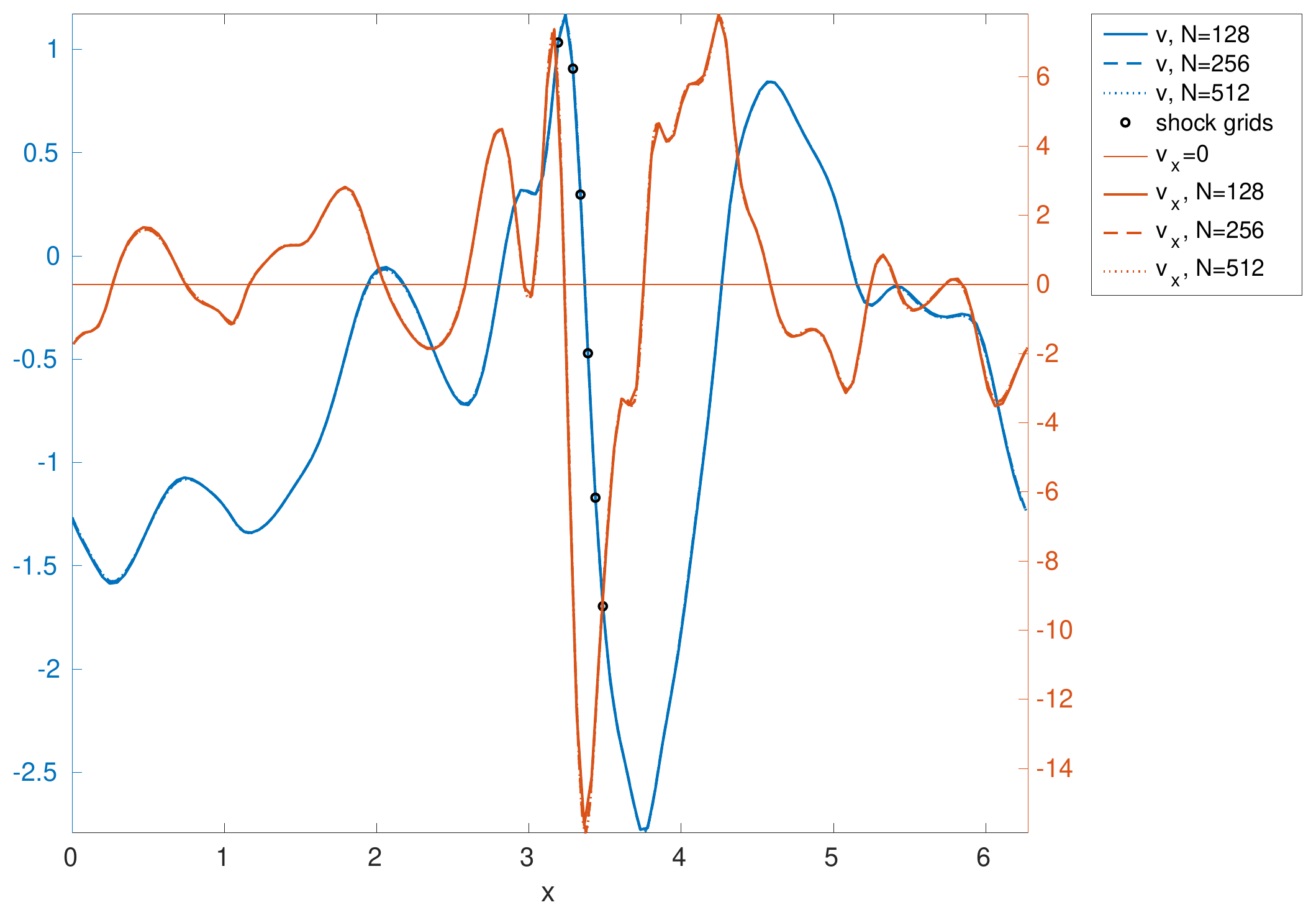}
    \caption{Density and velocity profiles using different resolutions (ST, $Re_\lambda = 80 , M_t = 0.8$, hybrid scheme). From top to bottom: snapshots of $\rho$, $u$, $v$ at a fixed $z$ slice and profiles along white lines in $u, v$ slices.}
    \label{fig:densityST}
    \end{figure}

\begin{figure}[!htb]
    \centering
    \includegraphics[width=1\textwidth]{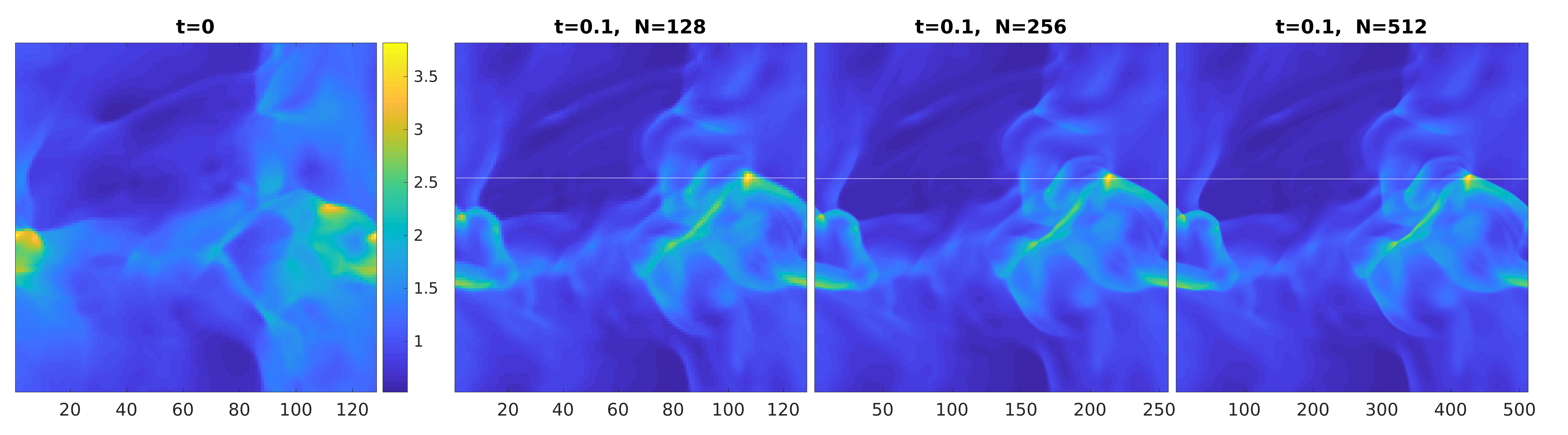}
    \includegraphics[width=1\textwidth]{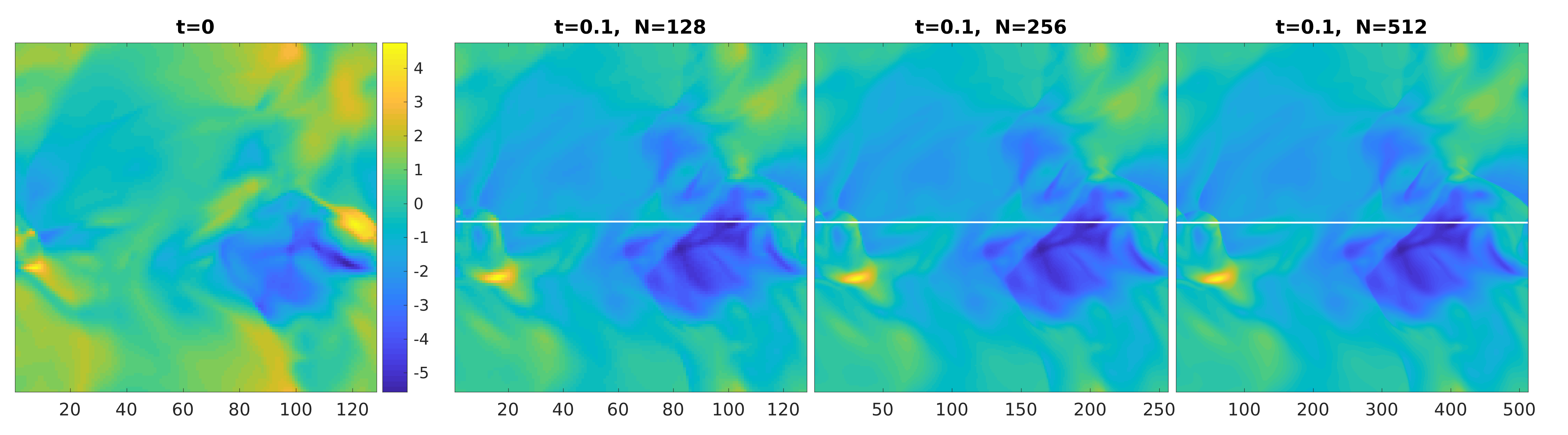}
    \includegraphics[width=1\textwidth]{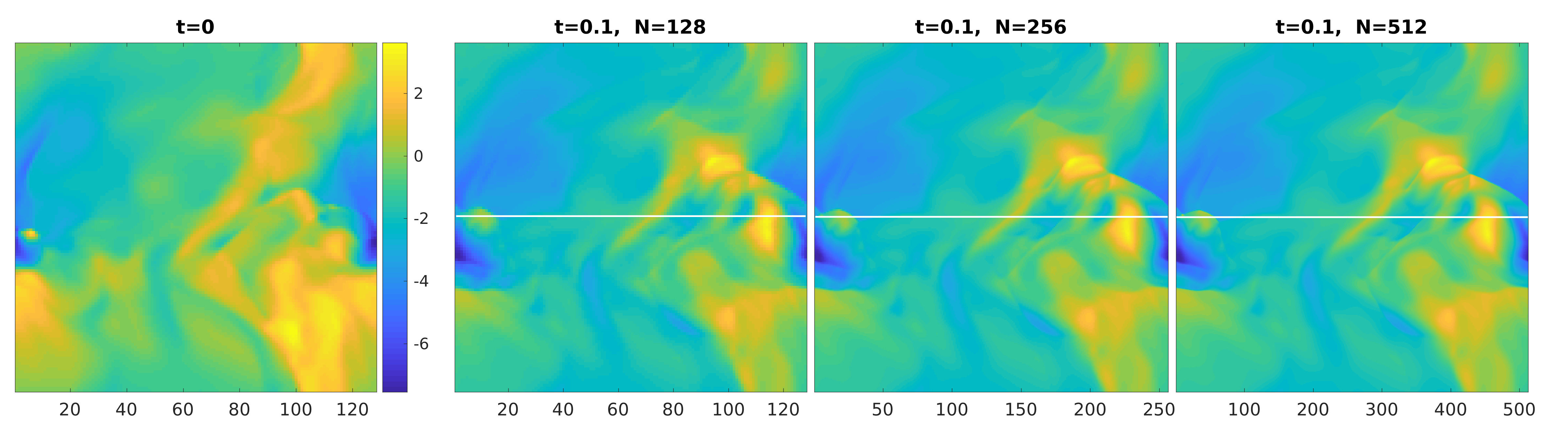}
    \includegraphics[width=0.49\textwidth]{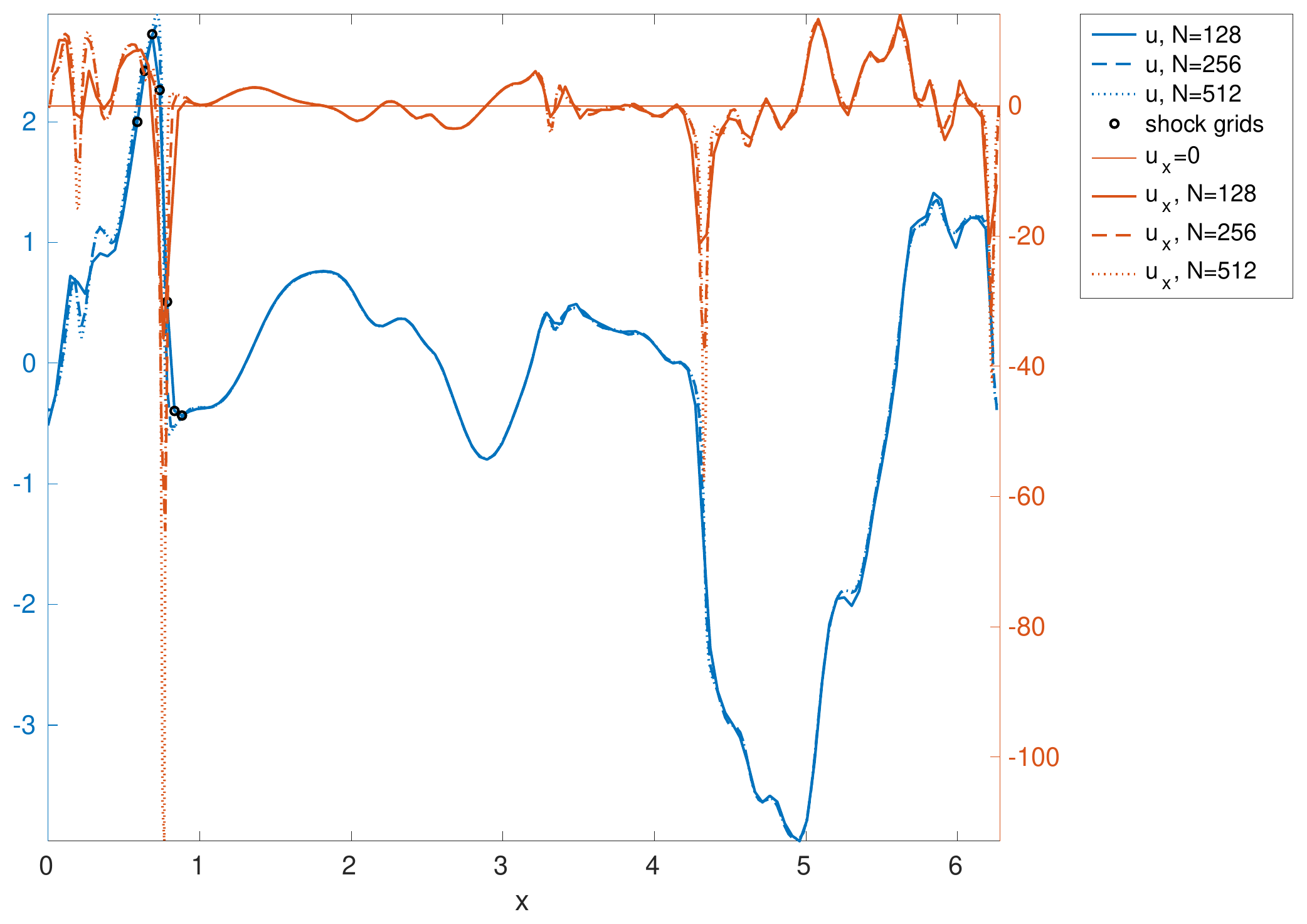}
    \includegraphics[width=0.49\textwidth]{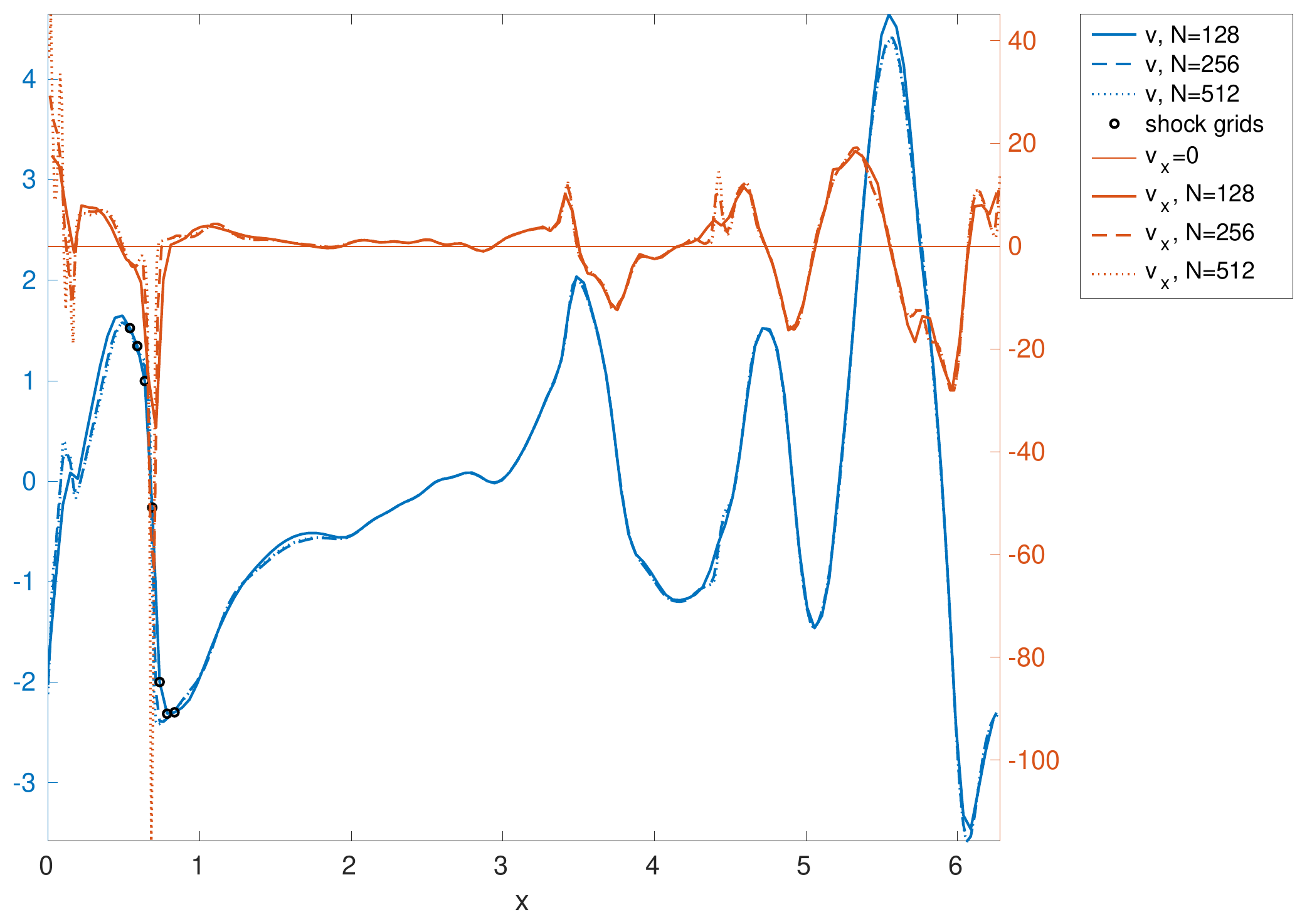}
    \caption{Density and velocity fields using different resolutions (C1, $Re_\lambda = 80 , M_t = 0.8$, hybrid scheme). From top to bottom: snapshots of $\rho, u, v$ and profiles along the white lines in $u, v$ slices.}
    \label{fig:densityCT}
    \end{figure}

 To check if the shocks are resolved by the numerical experiments, we first run a testing case using the hybrid scheme \cite{liu_hybrid_2019} with $128^3$ grids for 20
 time units. Then use the solution as the initial condition and run with resolutions using
 $128^3$, $256^3$, and $512^3$ grid points, for a short time, e.g. 0.1 time unit (which is approximately equal to 1 Kolmogorov timescale).
The results are reported in Fig.\ref{fig:densityST} and \ref{fig:densityCT}.
 These figures show that all the solutions are very accurate for the density and velocity fields except for some locations with huge gradients, which occupy only a small portion of the total computational domain.

\begin{figure}[!htb]
    \centering
    \includegraphics[width=0.49\textwidth]{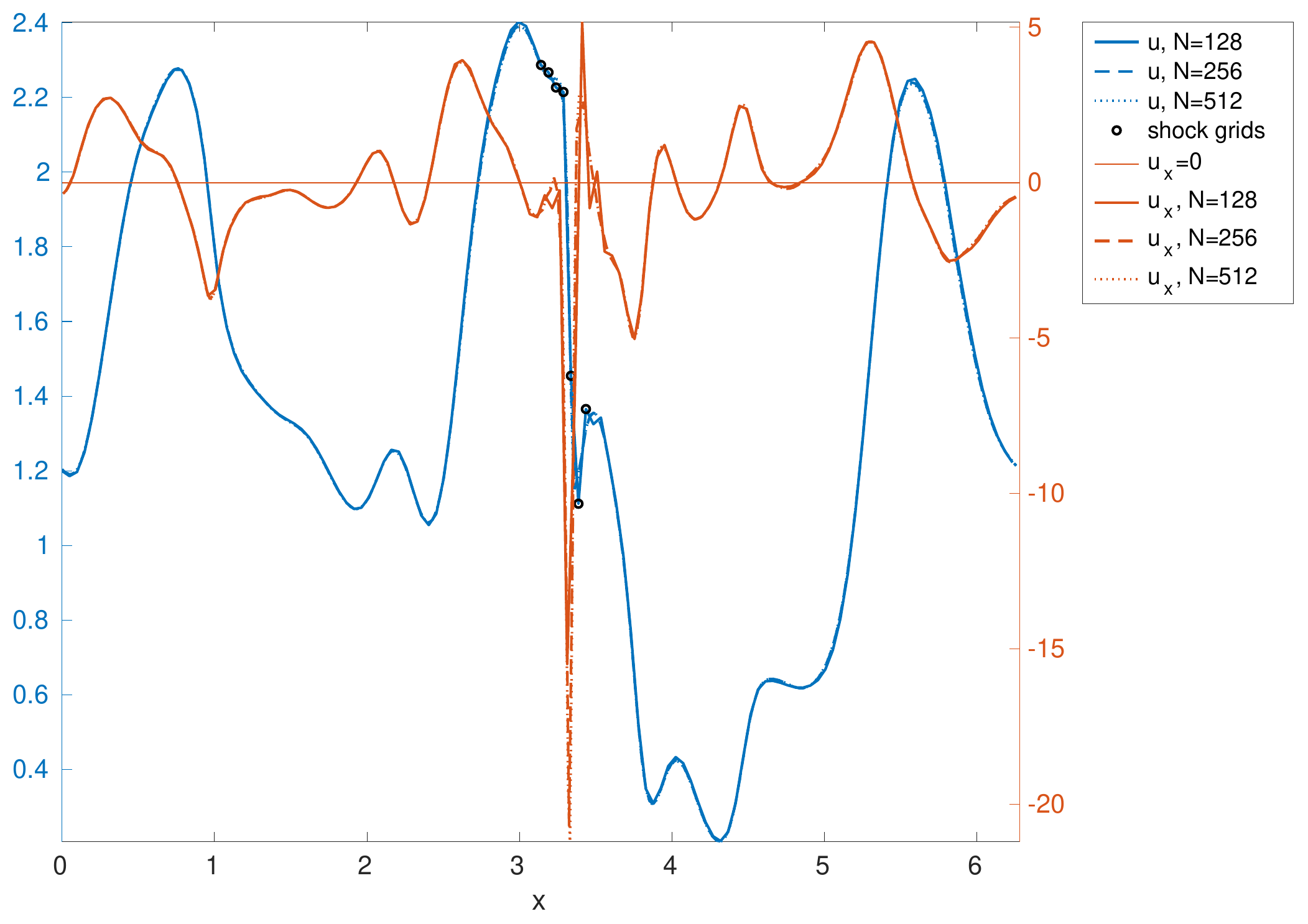}
    \includegraphics[width=0.49\textwidth]{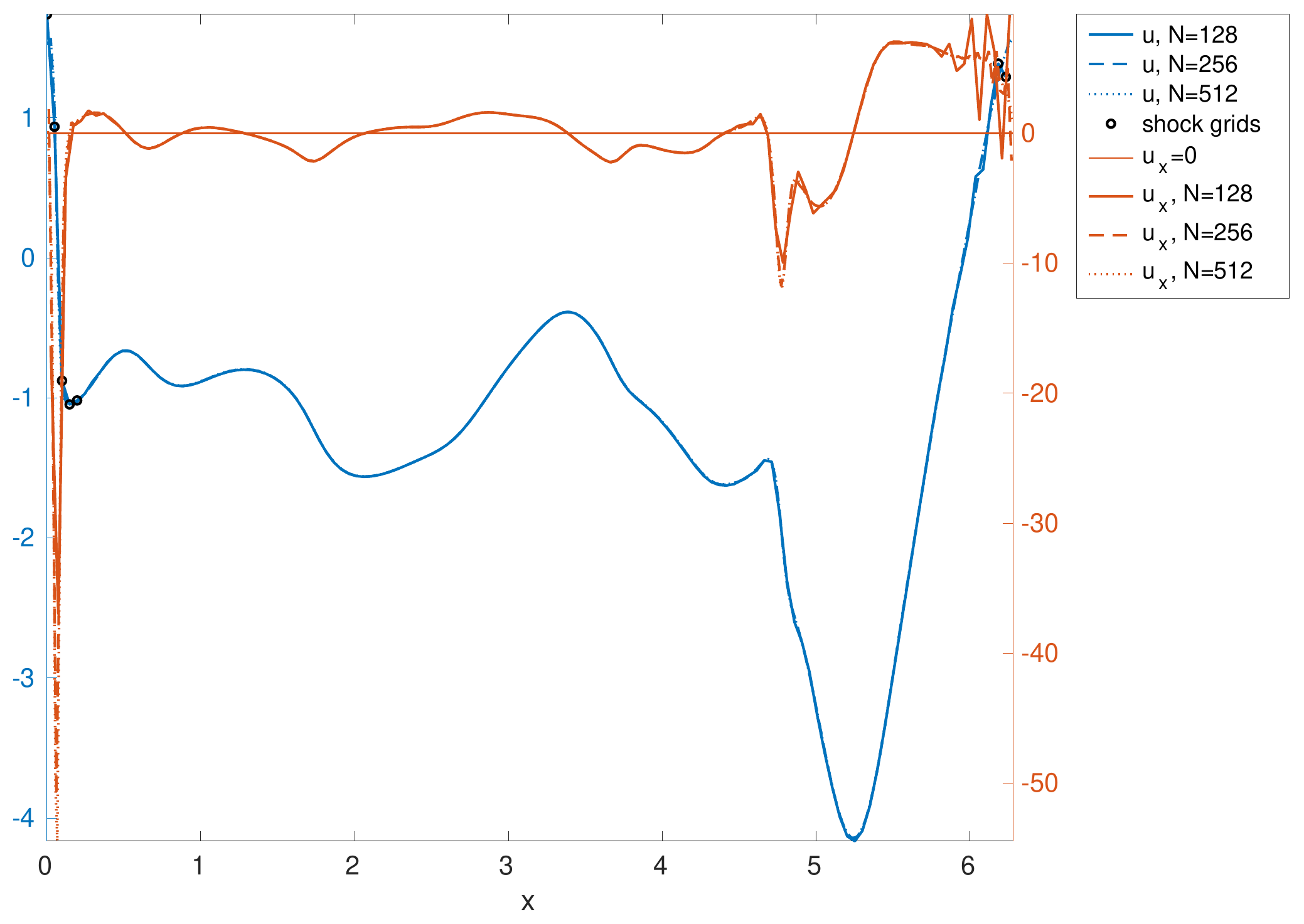}
    \caption{Resolution check ($u$ $x$-profile) using compact difference scheme. {\bf Left}: ST forcing, $Re_\lambda = 80 , M_t = 0.8$: Acceptable results; {\bf Right}: C1 forcing, $Re_\lambda = 40 , M_t = 0.4$: Oscillation happened for $u_x$ near $x=6$. }
    \label{fig:vecCompactScheme}
\end{figure}

 For the solenoidal forcing (ST) case (Fig \ref{fig:densityST}), we see that the $128^3$ resolution does a decent job, except for the place where $u_x$ is smaller than $-5$, which corresponds to a shock wave. Notice that the shock calculated by $128^3$ resolution is a little bit wider and less steep than that obtained using $256^3$ and $512^3$ resolutions.

 The result of the C1 forcing case (Fig \ref{fig:densityCT}) is similar to the ST case. However, we see larger errors in the $128^3$ grid case {compared to the reference solution obtained by using $512^3$ grid points, at places where gradients are large}. So, we use $256^3$ grid points for most of the numerical experiments.

 In Figure \ref{fig:vecCompactScheme}, we present the results of using a compact central difference scheme for a solenoidal forcing case and a mixed forcing case. We see that
 the compact difference scheme has better resolutions of shock thickness, but its solution
 has non-physical oscillations near the rear boundaries of the shocks. For this reason, we will not use this scheme to generate DNS data.

\subsubsection{High order schemes are necessary for calculating LEs}

There are several numerical methods that are used for evaluating LEs of passive particles in turbulence. To make a proper choice, we tested two such schemes with different spatial and temporal discretization orders.
For the low-order scheme (LS1, L stands for Lyapunov), we use a second-order central difference scheme to evaluate the velocity gradient tensor $\nabla \boldsymbol{u}$, then carry out linear interpolation to get values on Lagrangian points and use a second-order Crank-Nicolson scheme for marching the perturbation equation \eqref{eq:pertub}. For the high-order scheme (LS2), we use a 4th-order central difference for $\nabla \boldsymbol{u}$ and then use cubic interpolation to get values on Lagrangian points, and a third-order Runge-Kutta\cite{shu_efficient_1988} is adopted to march equation \eqref{eq:pertub}.

\begin{figure}[!ht]
 \centering
 \includegraphics[width=0.45\textwidth]{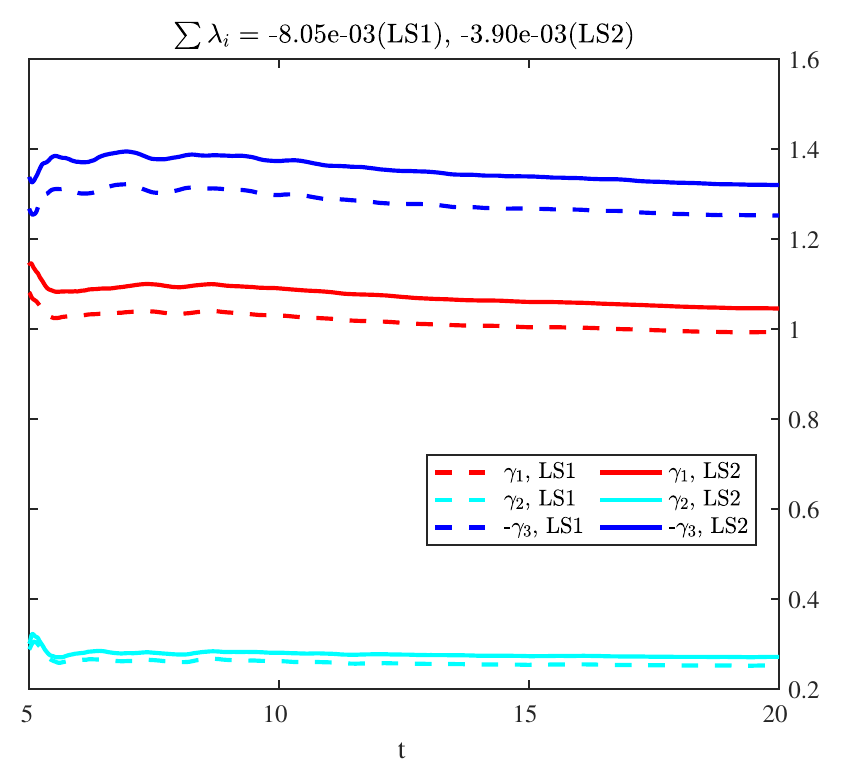}
  \includegraphics[width=0.45\textwidth]{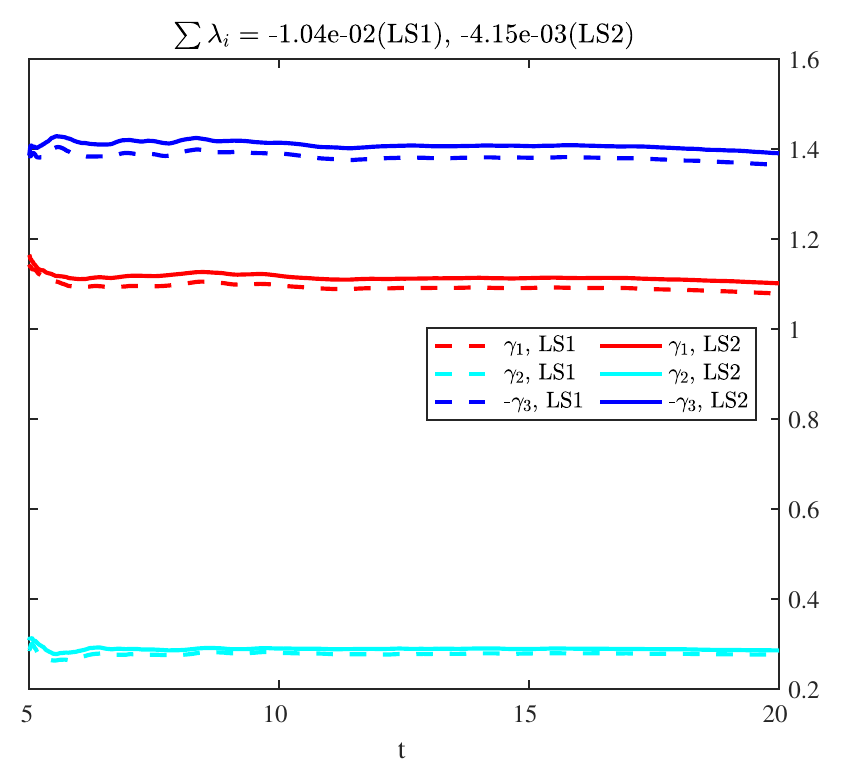}
 \caption{LE results obtained with different schemes for calculating
 $\sigma=\nabla_{\boldsymbol{x}} \boldsymbol{u}$ and
 for time marching of perturbing vectors (C1 forcing case with $M_t \approx 0.8, Re_\lambda \approx 100$).
 {\bf Left:} $128^3$ DNS grids;
 {\bf Right:} $256^3$ DNS grids.
 }
 \label{fig:scheme}
 \end{figure}

In Figure \ref{fig:scheme}, we present these results for a C1 forcing case with $M_t \approx 0.8, Re_\lambda \approx 100$, where we observe that the lower-order scheme and the high-order scheme generate results with observable differences in $128^3$ grid resolution\replaced{. By taking the computed LEs of the high-order scheme with $256^3$ grid points as reference solutions}{, by comparing the results with $256^3$ grid resolution}, we see that the lower order scheme lead to large numerical errors in the computation of LEs. We also observed that the high-order method (LS2) leads to {significantly} smaller values of the sum of LEs.
For this reason, we adopt the high-order scheme in the remaining part of this paper. {We also note that the sum of LEs in $256^3$ grid case is larger than the case with $128^3$ grid case. But this difference is insignificant compared to the differences in LEs themselves. As shown by \eqref{eq:LEsumalt} and \eqref{eq:LEsum2}, the sum of finite time LEs usually takes negative values and converges to 0 from below at a speed $O(1/t)$. The small difference is due to the fluctuations in the Lagrangian-averaged density at the start and terminal time.}

\subsubsection{Remove the dependence on initial conditions}

\begin{figure*}[!htb]
    \centering
    {\includegraphics[width=0.6\textwidth]{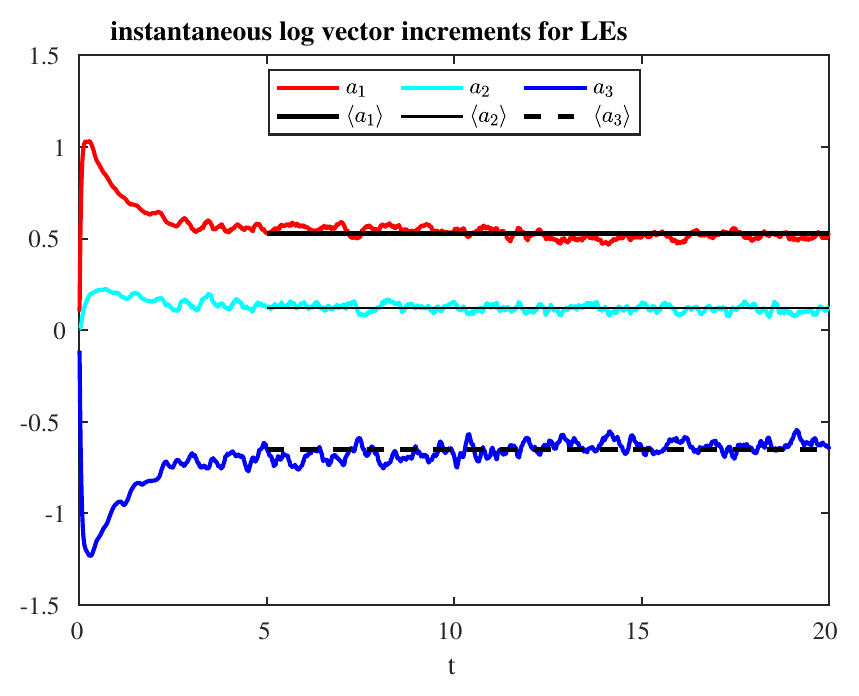}}
	 \caption{\label{fig:LEsHist-t5} Results of instantaneous log vector length increments $a_p^{(i)}$ used in \eqref{eq:FTLEnum} for calculating FTLEs in a mixed-type forcing (C1) case with $Re_\lambda=40$, $M_t=0.4$. In this figure, $\langle \cdot \rangle$ stands for the time average for $t>5$.
     }
\end{figure*}

\begin{figure*}[!htb]
    \centering
    \subfloat[$t_0=0$.]{\includegraphics[width=0.45\textwidth]{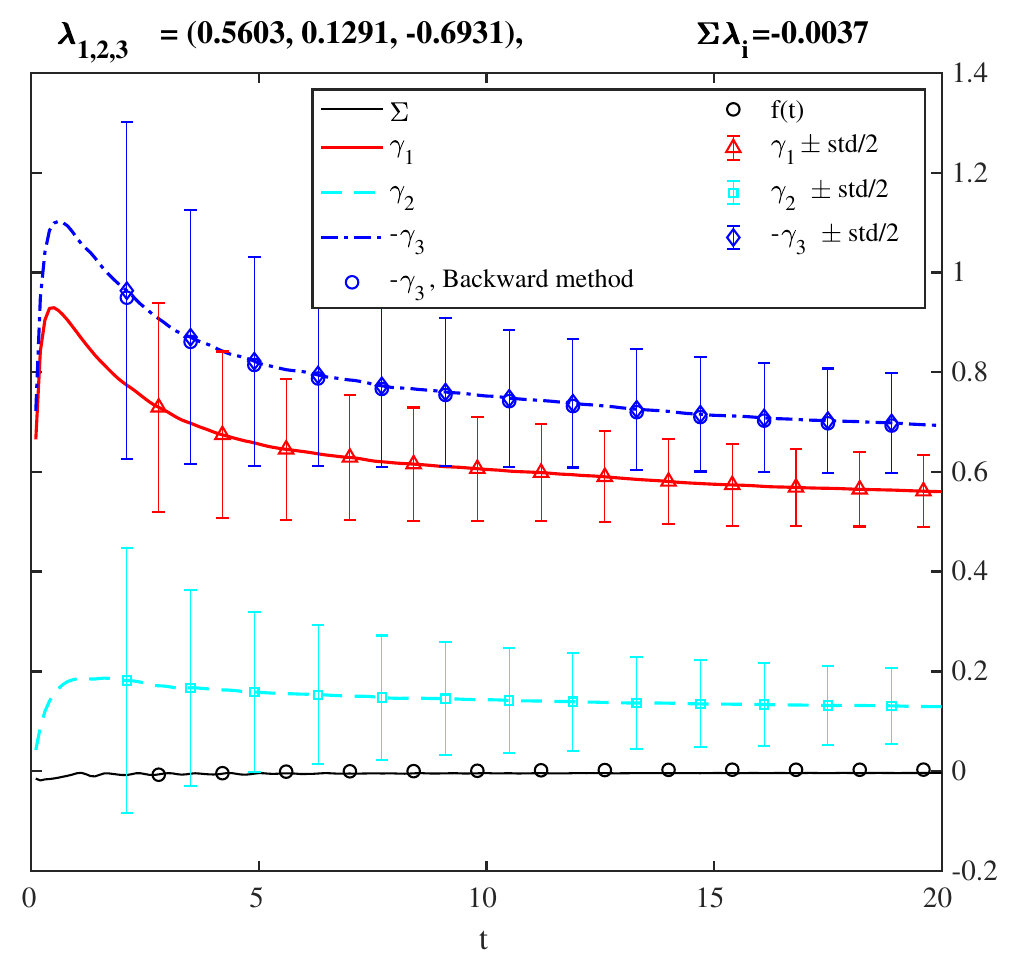}}
    \subfloat[$t_0=5$.]{\includegraphics[width=0.45\textwidth]{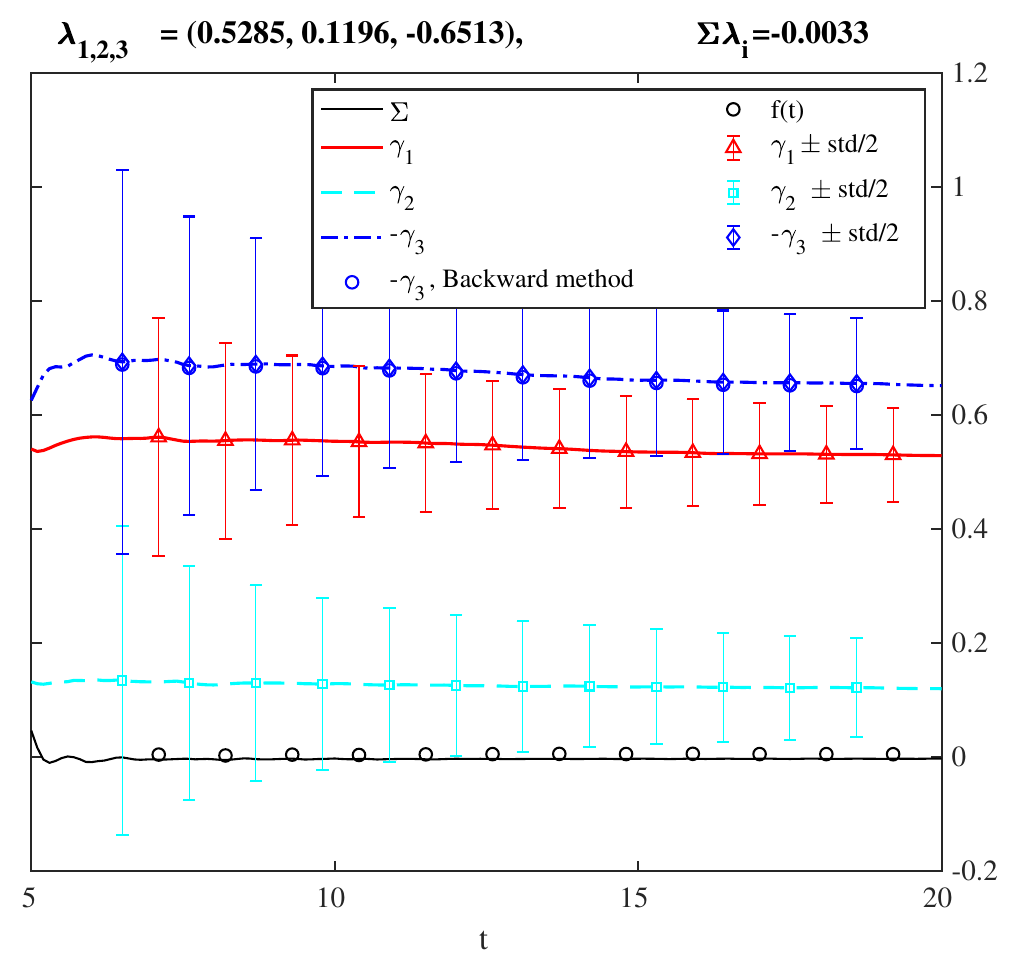}}
     \caption{\label{fig:LEst5} Results of FTLEs calculated by discarding or not discarding the relaxation period. The result for a typical mixed-type forcing (C1) case with $Re_\lambda=40$, $Mt=0.4$ is shown.
     }
\end{figure*}

 For the initial condition of the DNS, we use uniform density $\rho(x, 0)\equiv 1$,
 uniform temperature $T(x, 0)\equiv 1$, and a randomly initialized velocity field that is consistent with the $k^{-5/3}$ scaling law. We use a uniform initial distribution for the passive particles to calculate LEs. However, there is a noticeable transient region in the first several time units, as shown in Fig.~\ref{fig:LEsHist-t5}.
 Since we are interested in the properties of the statistical steady state, we expect to obtain more physical and accurate results by removing the initial transient region related to the initial conditions. To this end, we tested doing particle tracking from $t_0 \approx 3T_e$, where $T_e= L_f / u'$ is the large-eddy turnover time. However, the numerical results are similar to the ones without discarding the transient region of the flow. We also tried doing particle tracking from the very beginning but evaluating finite-time Lyapunov exponents from $t_0 \approx 3T_e$, which is equivalent to replacing the
 summation $i=1$ by $i=t_0/\tau$ in \eqref{eq:LEnum}, or setting $K_0=t_0/\tau$ in \eqref{eq:FTLEnum}. By doing this, the relaxation periods of both DNS and the LE dynamics are removed, and a faster convergence is observed, see Fig.~\ref{fig:LEst5}.
 In this figure, calculated FTLEs as functions of time using different methods are plotted. The FTLEs at the largest simulated time are taken as estimates of LEs.
 The red, green, and blue curves are three FTLEs calculated using three tracked vectors with Gram-Schmidt orthonormalization at each time step. The one marked by blue dots is the negative of $\gamma_3$ obtained using a method similar to the one to the largest FTLE, but with the evolution tensor $\sigma=\nabla\boldsymbol{u}$ replaced by $\sigma=(-\nabla\boldsymbol{u})^t$. The figure shows that the $\lambda_3$ calculated using two different methods agree with each other very well. The sum of FTLEs converges to 0, which asserts that the sum of LEs is 0, despite some minor oscillations existing in the initial transient region.

We further run some typical cases using $128^3$ and $256^3$ grid resolutions with different DNS schemes to check the effect of numerical resolutions and schemes of DNS on the calculation of LEs. The parameters and results are summarized in Table \ref{tbl:gird}, from which, we see the maximum relative differences of $\lambda_1/\lambda_2$ and $\lambda_3$ in difference resolutions/schemes are less than $2.2\%$.
We note that the grid size $\Delta x$ for $128^3$ and $256^3$ are close to $0.049$ and $0.024$, respectively, which are smaller than the standard $\Delta x \approx 2\eta$ used in existing DNS studies\cite{jagannathan_reynolds_2016, john_does_2021}.

\begin{table}[!htb]
    \centering
    \begin{tabular}{l|r|ccc|ccccc|c}
        \hline
        Force-Scheme & Grid & $Re_L$ & $Re_\lambda$ & $M_t$ & $\eta$ & $\tau_\eta$ &  $\lambda_1$ & $\lambda_2$ & $\lambda_3$ & $\lambda_1/\lambda_2$ \\
        \hline
ST-Hybrid & $128^3$ & 1395.43 &   79.81 &  0.7961 &  0.0353 &  0.1307 &  0.8620 &  0.2130 & -1.0761 & 4.047 \\
ST-Unwind & $128^3$ & 1402.75 &   78.36 &  0.8007 &  0.0349 &  0.1281 &  0.8801 &  0.2176 & -1.0997 & 4.045 \\
ST-Hybrid & $256^3$ &  1407.12 &   81.11 &  0.8001 &  0.0346 &  0.1293 &  0.8660 &  0.2151 & -1.0817 & 4.027 \\
ST-Upwind & $256^3$ &  1400.49 &   79.19 &  0.7988 &  0.0350 &  0.1286 &  0.8592 &  0.2154 & -1.0763 & 3.989 \\
\hline
C1-Hybrid & $128^3$ &   628.48 &   39.84 &  0.4042 &  0.0603 &  0.1842 &  0.5763 &  0.1328 & -0.7117 & 4.340 \\
C1-Upwind & $128^3$ &   627.46 &   38.04 &  0.4036 &  0.0595 &  0.1796 &  0.5673 &  0.1303 & -0.7007 & 4.354 \\
C1-Hybrid & $256^3$ &   627.73 &   37.92 &  0.4036 &  0.0594 &  0.1787 &  0.5635 &  0.1302 & -0.6976 & 4.328 \\
C1-Upwind & $256^3$ &   627.87 &   37.14 &  0.4037 &  0.0589 &  0.1762 &  0.5628 &  0.1318 & -0.6995 & 4.270 \\
    \hline
    \end{tabular}
    \caption{\label{tbl:gird} Summary of results using different schemes and grid sizes. ST stands for the solenoidal driving force. C1 stands for the force with comparable solenoidal and compressible parts ($\gamma_0=1$).}
\end{table}

\subsection{Comparing the average of velocity divergence weighted by fluid density and particle number density.}

\begin{figure}[!ht]
    \centering
    \subfloat[A ST forcing case with $Re_\lambda=80$, $M_t=0.8$]{\includegraphics[width=0.75\textwidth,height=0.3\textwidth]{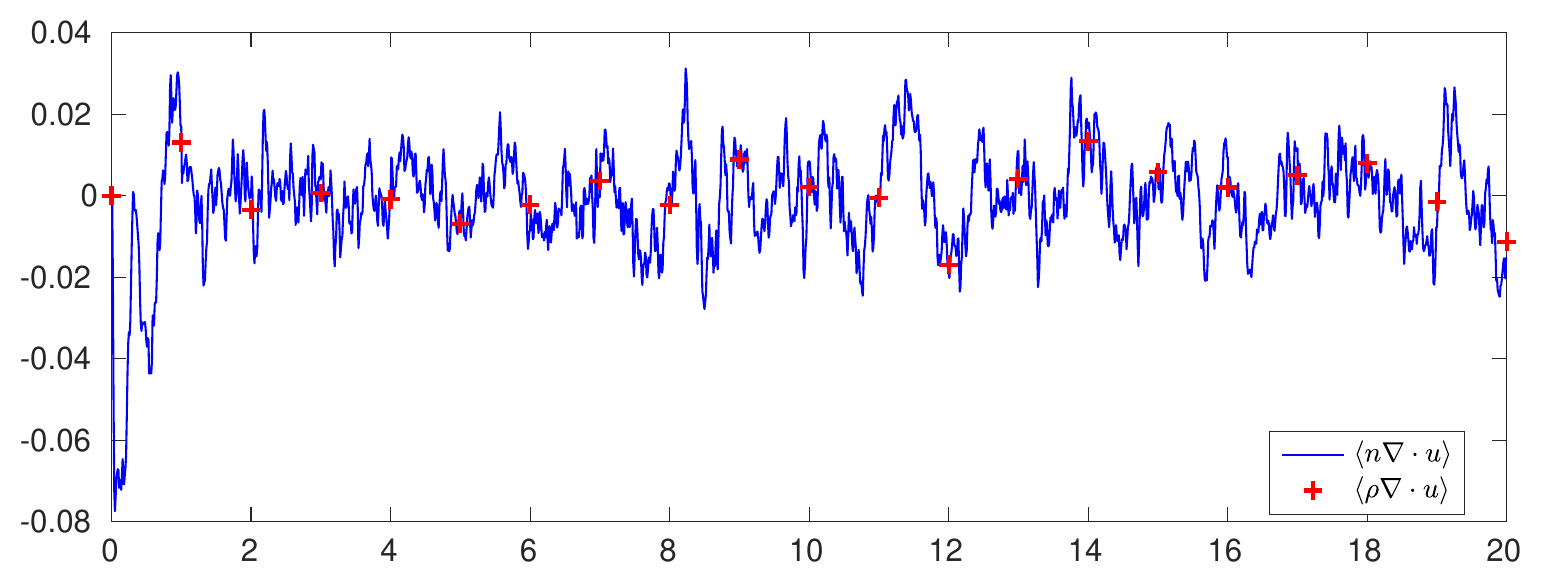}}
    \\
    \subfloat[A C1 forcing case with $Re_\lambda=80$, $M_t=0.8$]{\includegraphics[width=0.75\textwidth,,height=0.3\textwidth]{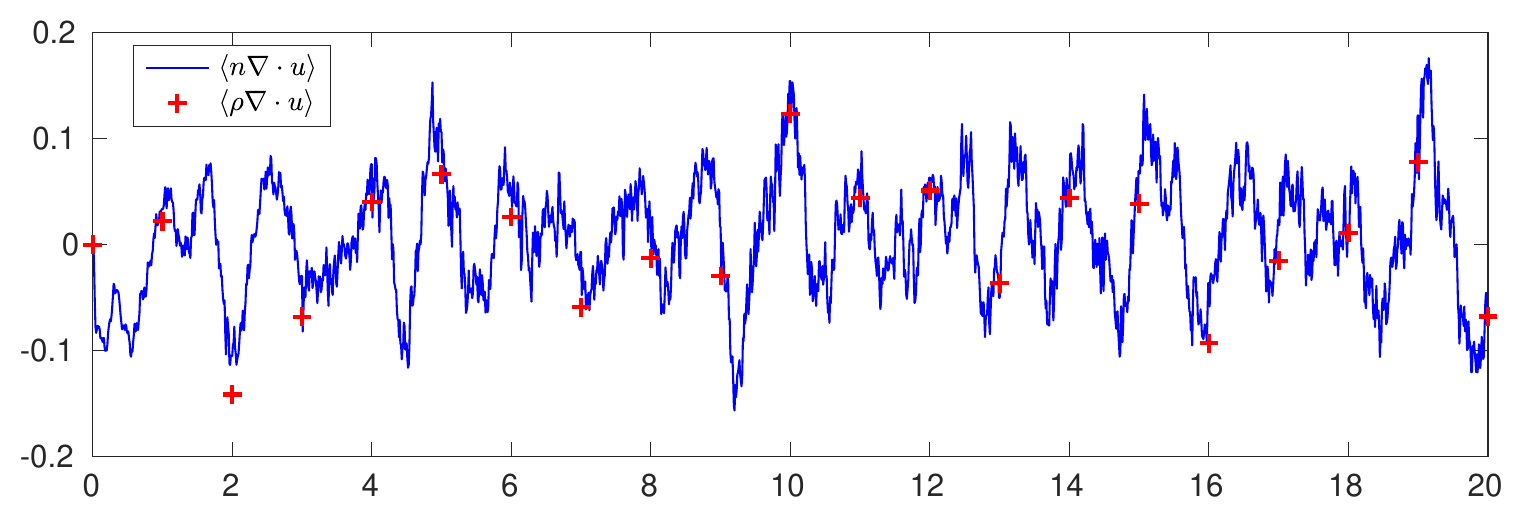}}
    \caption{Comparison of weighted dilation.\label{fig:divuwt}}
\end{figure}

The quantities of $\int_\Omega \rho \nabla\!\cdot\!\boldsymbol{u}\, d\boldsymbol{x}$ and
$\int_\Omega n(\boldsymbol{x}) \nabla\!\cdot\!\boldsymbol{u}\, d\boldsymbol{x}$ as functions of time $t$ are plotted in Fig.~\ref{fig:divuwt}, where $n(\boldsymbol{x})$ is the number density of passive particles.
We see that the two quantities agree with each other in both solenoidal and mixed forcing cases.

\subsection{A direct comparison of fluid density and particle number density.}

We next present a direct numerical study to check if the fluid density $\rho$ and particle number density $n$ have the same invariant distribution since they are related to the statistics of LEs.
We adopt the following strategy.

\begin{figure}[!ht]
    \centering
    \subfloat[T=5]{\includegraphics[width=0.43\textwidth]{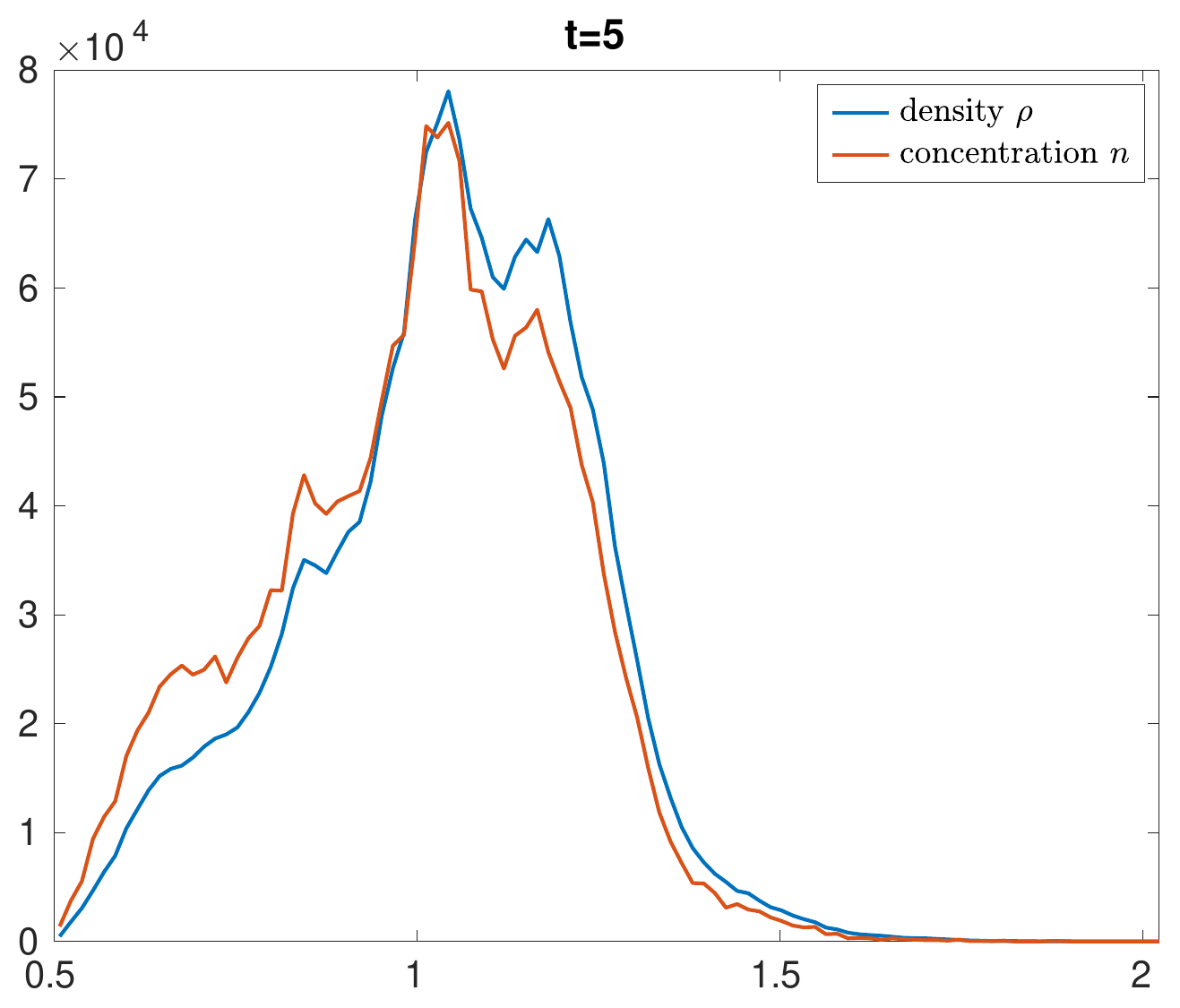}}
    \subfloat[T=10]{\includegraphics[width=0.43\textwidth]{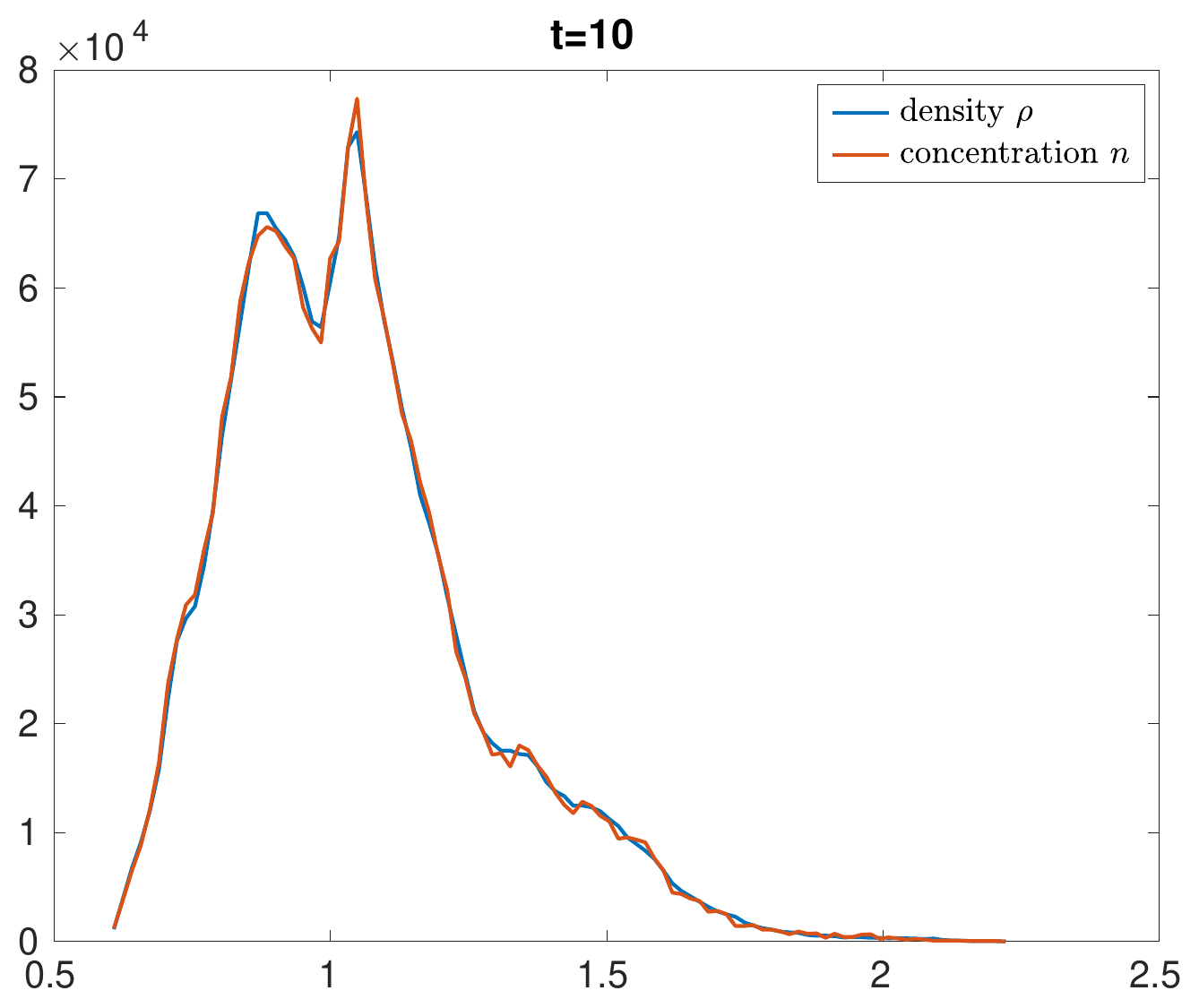}}
    \caption{\label{fig:rhon} Direct comparison of $\rho$ and $n$: Mass in different density $\rho$ regions.}
\end{figure}

We first divide the range of $\rho$:
$[\rho_{\min}, \rho_{\max}]$ into 100 regions $[\rho_i, \rho_{i+1}], i=0,\ldots,99$. The volume of the domain has a density $\rho(\boldsymbol{x})$ in the range $[\rho_i, \rho_{i+1}]$ is given by $V_i=\text{Vol}\bigl(\{\boldsymbol{x} \mid \rho_i < \rho(\boldsymbol{x}) \le \rho_{i+1}\}\bigr)$.
Then the fluid mass in $V_i$ is approximately given by $\bar{\rho}_i V_i$, where $\bar{\rho}_i = \frac{\rho_i+\rho_{i+1}}{2}$. Then we count the number of particles in $V_i$ to make the comparison. Such an approach can get results with smaller fluctuations.
We introduce $10^5$ particles into the system at $t=5$ with a uniform initial distribution. The results are presented in Fig.~\ref{fig:rhon}, from which
we see that even though the distributions of $\rho$ and $n$ at $t=5$ (the starting time) are different, they evolve into almost exactly the same distribution at $t=10$. This result is consistent with the Birkhoff ergodic theorem \cite{birkhoff_proof_1931}.

\begin{figure}[!ht]
    \centering
    \subfloat[T=5]{\includegraphics[width=0.42\textwidth]{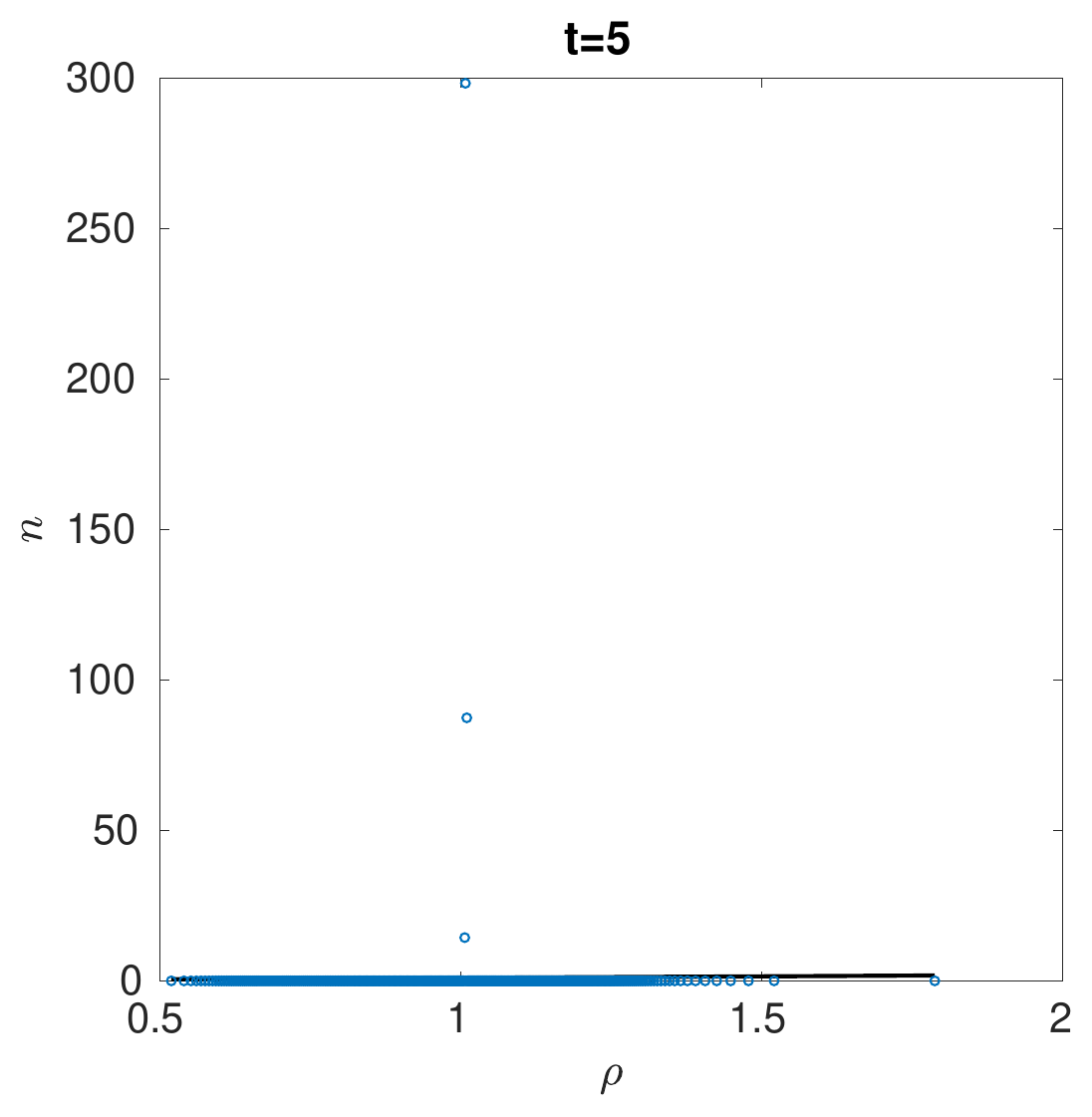}}
    \subfloat[T=10]{\includegraphics[width=0.42\textwidth]{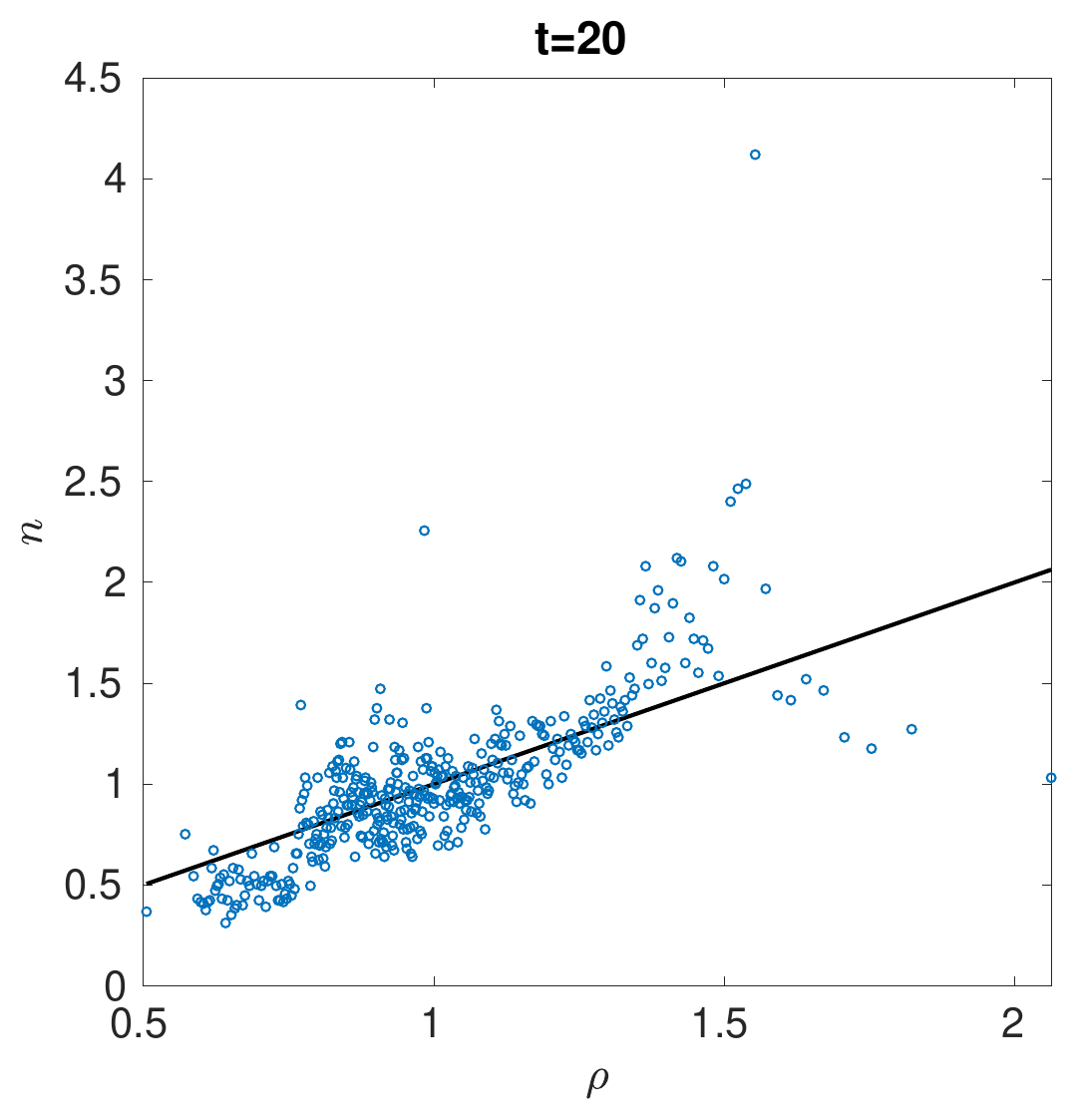}}
    \caption{\label{fig:rhon2} Direct comparison of the averaged particle density and fluid density in $V_i$ in an extreme case.}
\end{figure}

To further verify that the fluid density $\rho$ and particle number density $n$ have the same invariant distribution, we tested an extreme case, where the particles are
initialized in a small corner of the computational box: $\{0\le x<L/10, 0\le y<L/10, 0\le z<L/10\}$. The average particle density $n$ versus fluid density $\rho$ in $V_i$ are plotted in Fig.~\ref{fig:rhon2}, from which we see, at the very beginning ($t=5$), the two average densities are very different, but at a later time ($t=20$), they are linearly correlated.

\subsection{Spectrum of LEs}

\begin{figure}[!htb]
 \centering
 \includegraphics[width=0.8\textwidth]{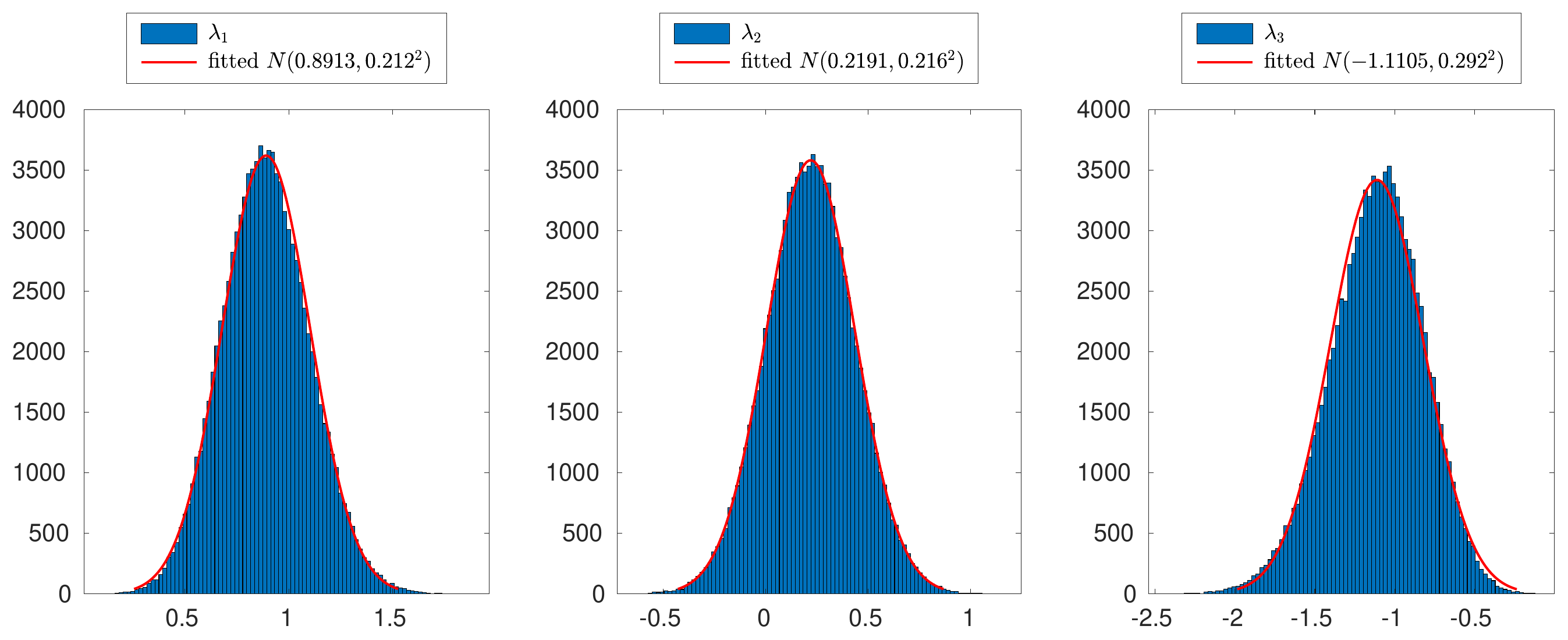}\\
 \includegraphics[width=0.8\textwidth]{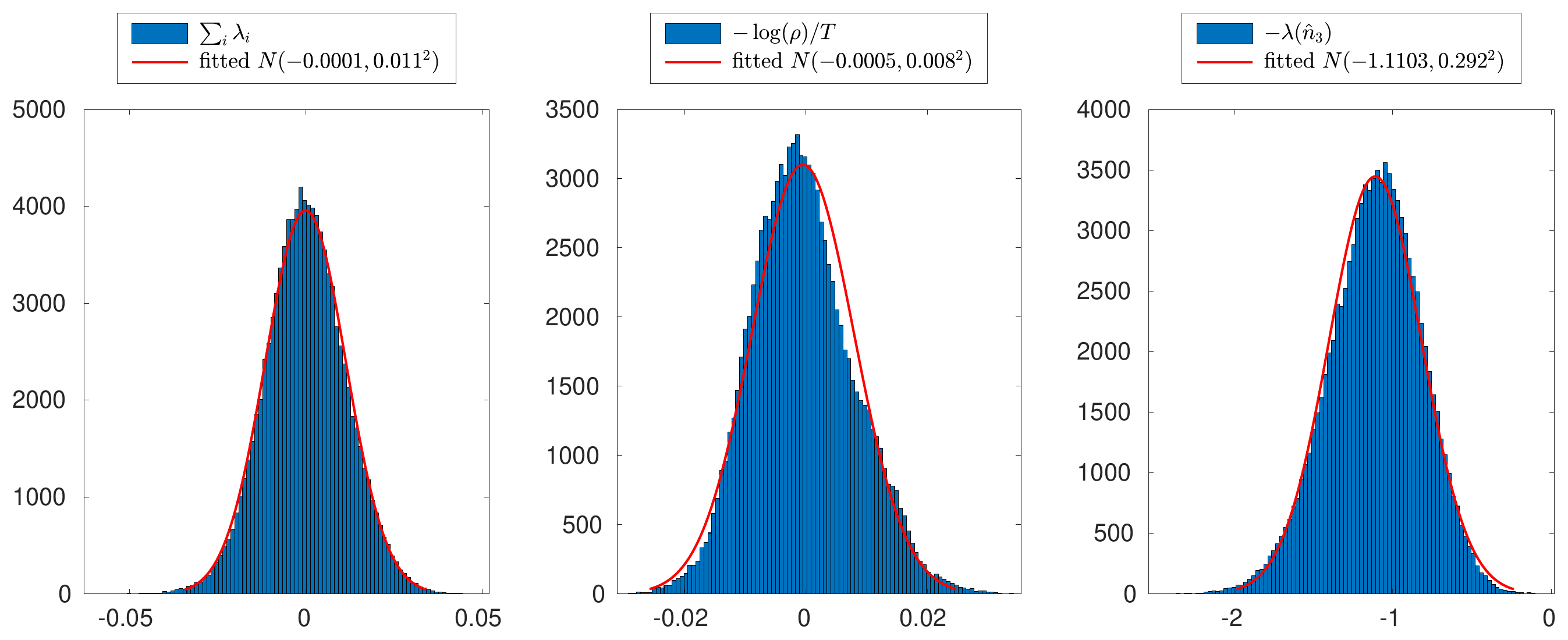}
 \caption{The distributions of calculated LEs and sum of LEs for the case ST-Re80M06 with solenoidal forcing. The last plot is the $\lambda_3$ calculated using approach \eqref{eq:LE3}.}
 \label{fig:LEpdf:STRe80M06}
 \end{figure}

\begin{figure}[!htb]
 \centering
 \includegraphics[width=0.8\textwidth]{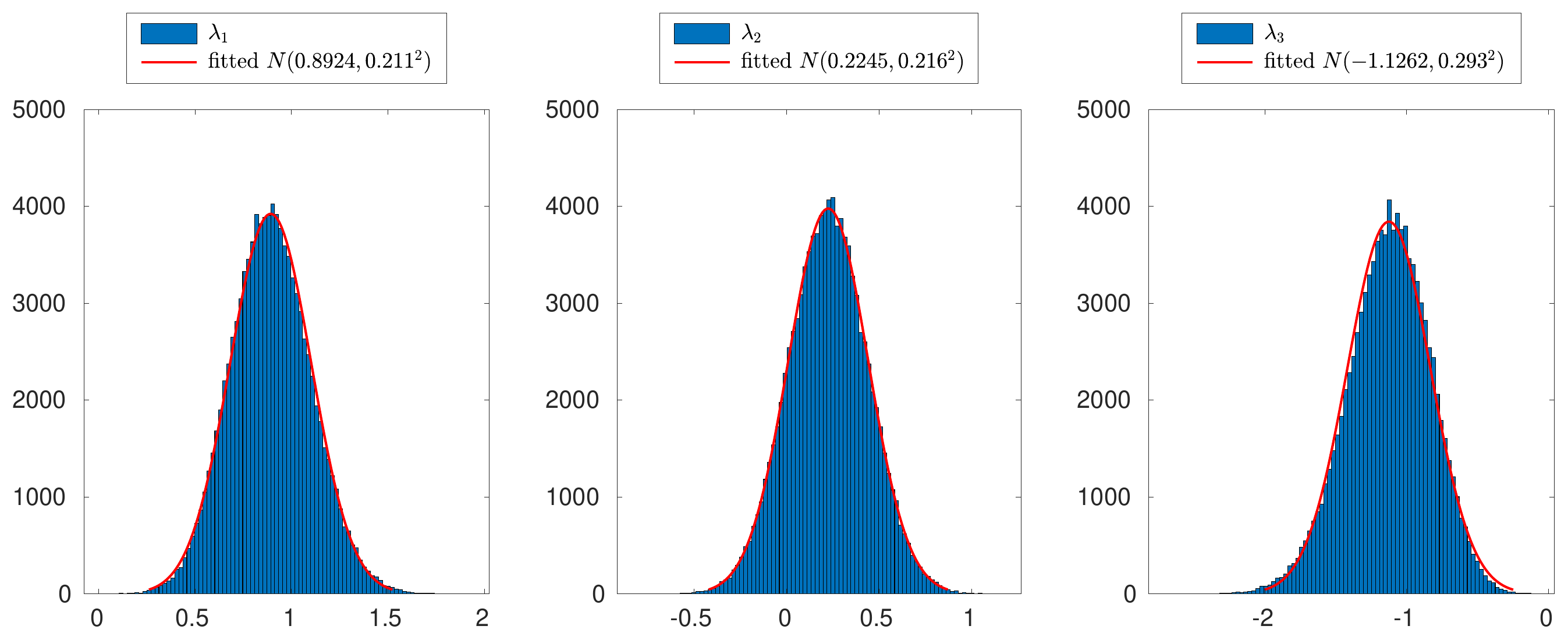}\\
 \includegraphics[width=0.8\textwidth]{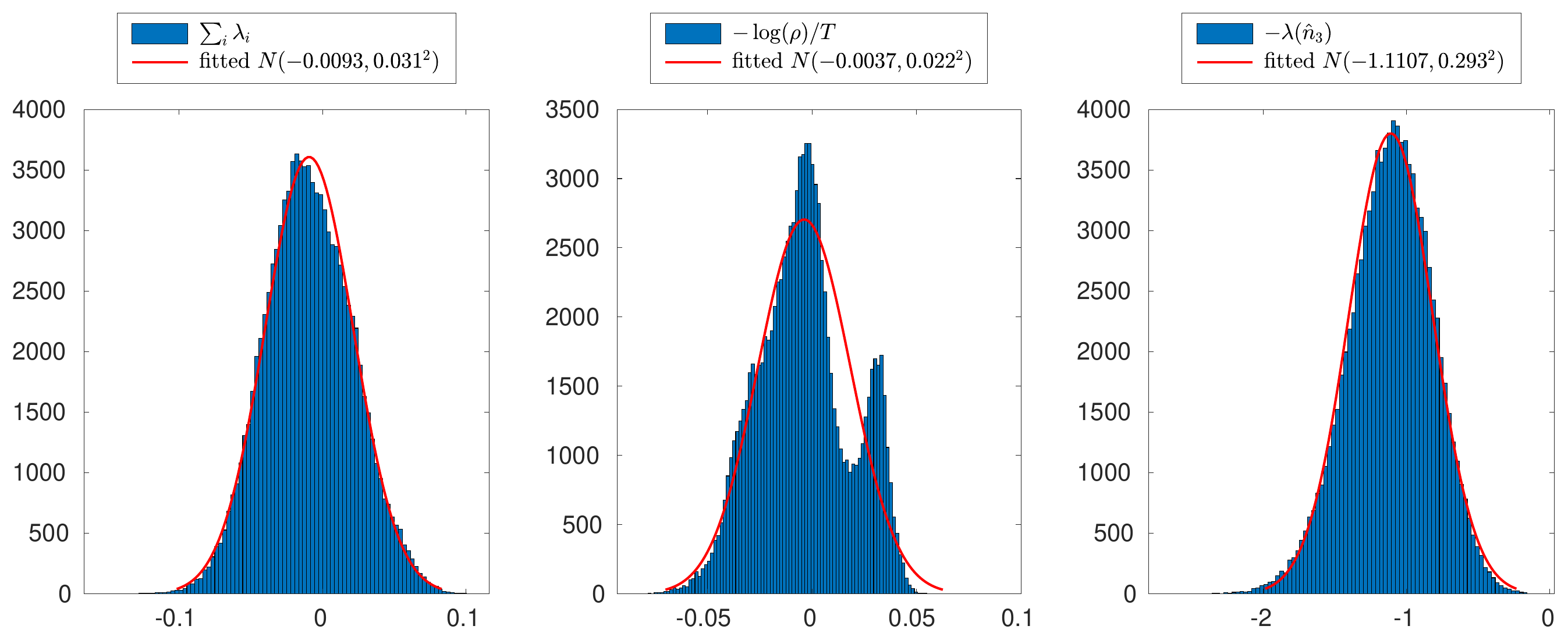}
 \caption{The distributions of LEs and the sum of LEs for the case C1-Re80M06.}
 \label{fig:LEpdf:C1Re80M06}
 \end{figure}

 Now, we present the results of the probability density functions (PDFs) of calculated LEs and the sum of all three LEs in Fig.~\ref{fig:LEpdf:STRe80M06} and \ref{fig:LEpdf:C1Re80M06}, where histogram and fitted normal distribution of calculated LEs and the sum of LEs using different methods are plotted. The simulation time period is $t=20 \approx 15 T_e \approx 150 \tau_\eta $. $N=10^5$ particles are tracked to generate the histogram. We tested taking $N=10^4, 10^5, 10^6$ particles, and the difference is very small. We see the PDFs of all three LEs are close to normal distributions. The results of using two different methods for $\lambda_3$ agree with each other very well.
 We notice that, even though $\lambda_2$ has smaller average values than $\lambda_1$, its variance is larger. The sums of LEs obtained using two different methods all have much smaller average values and variance than $\lambda_i, i=1,2,3$, which confirms the fact that the sum of all LEs should be 0.

 With similar other parameters used, we see that the ST forcing case has a smaller absolute value of $\sum \lambda_i$ than the C1 forcing case. This is due to the fact the density variations are larger in the C1 case, which can be verified by the corresponding PDF plots of $-\ln \rho /t$ in Fig.~\ref{fig:LEpdf:STRe80M06} and \ref{fig:LEpdf:C1Re80M06}. More passive particles will stay near high-density regions according to the invariant measure even though the initial distribution is uniform.

 \subsection{Dependence of LEs on turbulent Mach number and Taylor Reynolds number}
\begin{figure}[!htb]
    \centering
    \subfloat[Solenoidal forcing (ST)]{\includegraphics[width=0.48\textwidth]{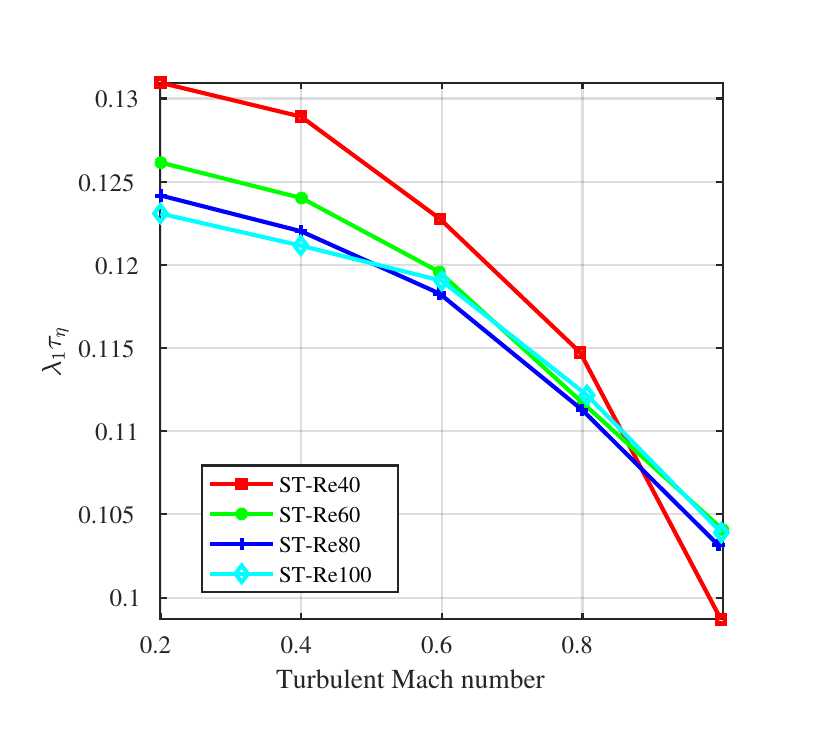}}
    \subfloat[Mixed forcing (C1)]{\includegraphics[width=0.48\textwidth]{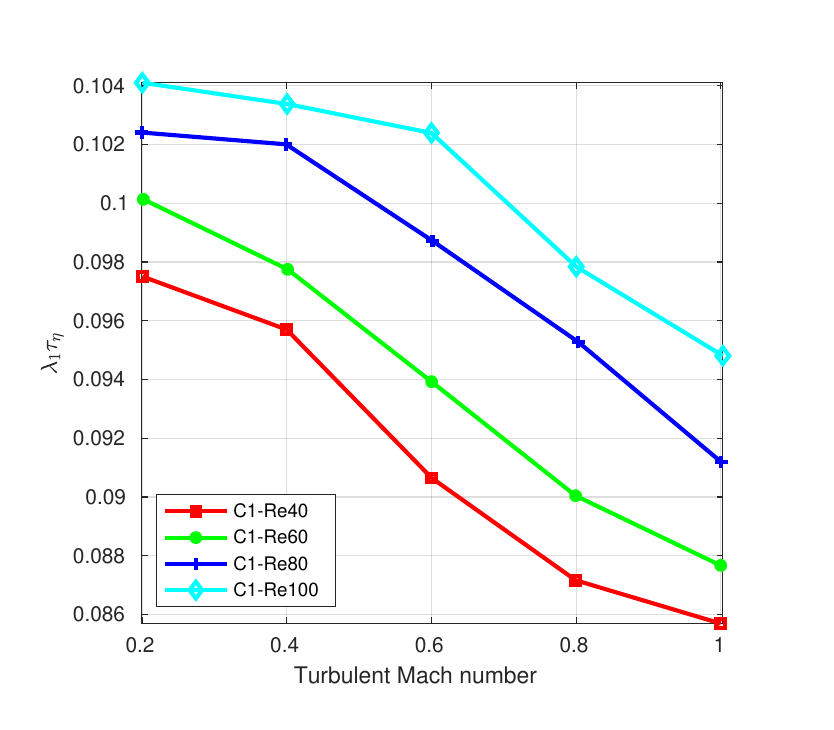}}
    \caption{The relation between $\lambda_1\tau_\eta$ and Taylor scale Reynolds number $Re_\lambda$ and turbulent Mach number.}
    \label{fig:LE1Re}
\end{figure}

\begin{figure}[!htb]
    \centering
    \subfloat[Solenoidal forcing (ST)]{\includegraphics[trim=0 10 40 30, clip,width=0.43\textwidth]{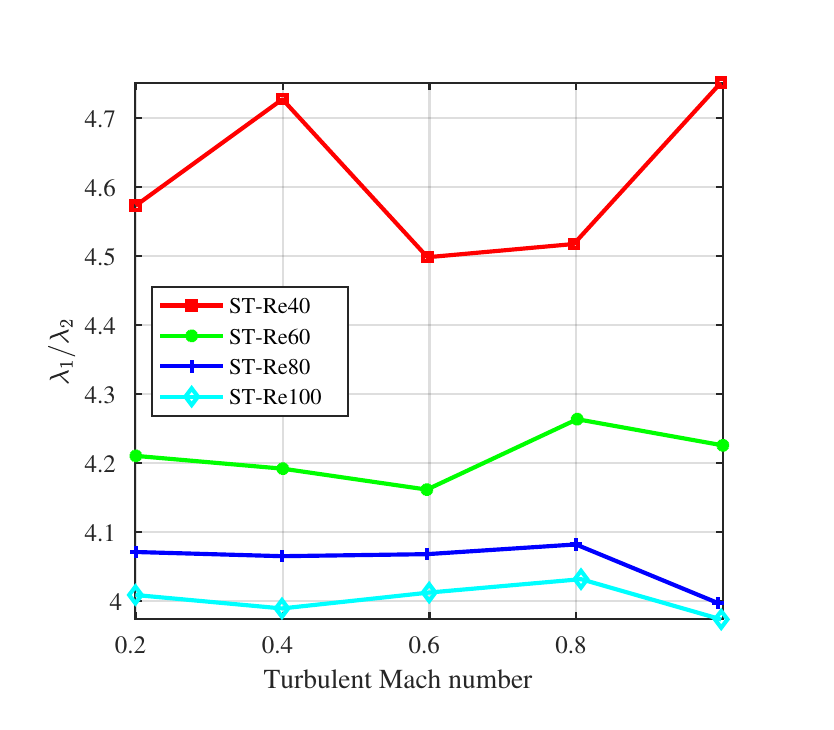}\ }
    \subfloat[Mixed forcing (C1)]{\includegraphics[trim=0 10 40 30, clip,width=0.43\textwidth]{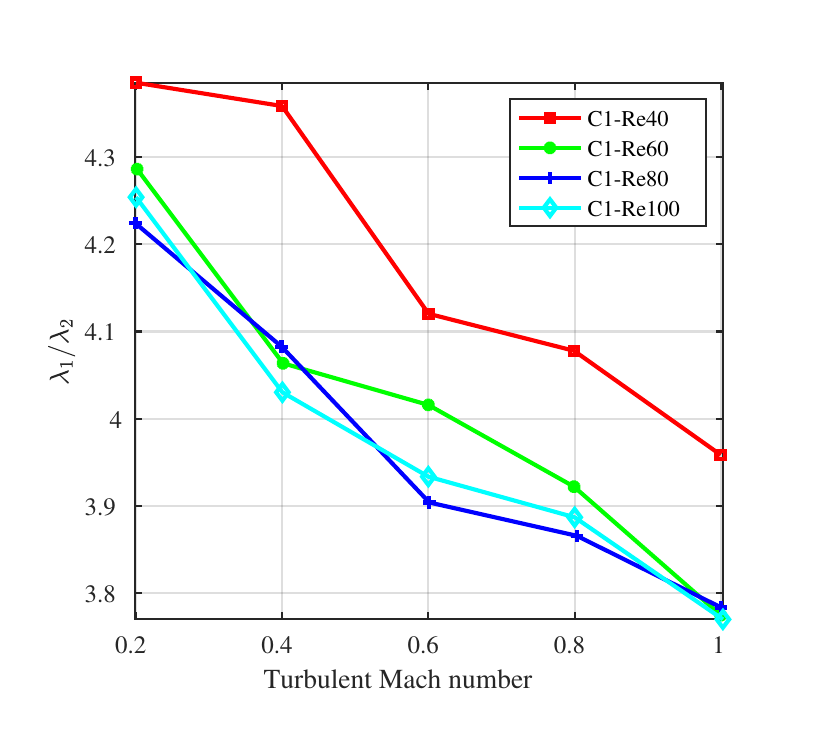}}
    \caption{ $\lambda_1/\lambda_2$ versus Taylor scale Reynolds number and turbulent Mach number.}
    \label{fig:LE1le2Ma}
\end{figure}

We show the dependence of $\lambda_1$ and $\lambda_2$ on turbulent Mach number $M_t$ and Taylor Reynolds number $Re_\lambda$ in Fig.~\ref{fig:LE1Re} and \ref{fig:LE1le2Ma}.

From Fig.~\ref{fig:LE1Re}, we see that $\lambda_1\tau_\eta$ is a decreasing function of $M_t$ for both ST and C1 forcing cases.
We attribute this to the fact that particles in velocity field with larger compressible components tend to steady clustered near shocks \cite{gawedzki_phase_2000, yang_interactions_2014}, which decreases the separation rate, since larger $M_t$ and mixed forcing leads to larger $\delta$, the ratio of the compressible(dilational) r.m.s. velocity to solenoidal r.m.s. velocity $u^c_{rms}/u^s_{rms}$.
The quantity $\delta$ has been studied as an additional non-dimensional parameter to get some universal scaling laws of compressible HIT by \citet{donzis_universality_2020}.

We also observe that the quantity $\lambda_1\tau_\eta$ is almost independent of $Re_\lambda$ (except for $Re_\lambda=40$) in the ST forcing case for $1 \ge M_t \ge 0.2$, but has a nearly linear dependence for the C1 forcing case when $M_t \le 0.8$.
For $M_t=0.2$, the results of the ST forcing case shown in Fig. \ref{fig:LE1Re} that $\lambda_1 \tau_\eta$ decreases slowly as $Re_\lambda$ increases, which is consistent with the incompressible flow case (see e.g. Fig.~7 in \citet{donzis_batchelor_2010}, and the results in \citet{fm}).
We numerically experience that when $Re_\lambda$ is small, simulations give large variations in $\lambda_1\tau_\eta$. This is because the flow is not fully developed turbulence, and the long time average is required to get rid of the chaotic oscillations in the dynamics of Navier-Stokes equations. Therefore the numerical results reported for $Re_\lambda \approx 40$ subject to larger fluctuation errors.
Note that \citet{bec_lyapunov_2006} report $\lambda_1 \tau_\eta \approx 0.15 $ for incompressible turbulence with $Re_\lambda =65, 105, 185$, which is slightly higher than our result interpolated at $M_t=0$.

In Fig \ref{fig:LE1le2Ma}, the dependence of $\lambda_1/\lambda_2$ on Taylor Reynolds number $Re_\lambda$ and turbulent Mach number $M_t$ are plotted for the ST and C1 forcing. We see that in ST cases, $\lambda_1/\lambda_2$ is a decreasing function of $Re_\lambda$. In the C1 cases, $\lambda_1/\lambda_2$ is also a decreasing function of $Re_\lambda$, but when $Re_\lambda$ gets larger,  the dependence $\lambda_1/\lambda_2$  on $Re_\lambda$ weakens quicker.
Due to the fact $\lambda_2$ is an indicator of time reversibility, smaller values of $\lambda_1/\lambda_2$ suggest a stronger irreversibility, thus suggesting the turbulent attractor gets more and more strange as $Re_\lambda$ increases. For $Re_\lambda>80$, the value $\lambda_1/\lambda_2$ seems to reach fixed values, which suggests that the flow is close to fully-developed turbulence. In particular, for the ST forcing, when $Re_\lambda \approx 100$, we observe that $\lambda_1 : \lambda_2 :  \lambda_3 \approx 4:1:-5$, which is similar to the known result for incompressible Navier-Stokes system\cite{johnson_largedeviation_2015}.

In the ST forcing case, the dependence on $M_t$ is very weak, but in the C1 case, higher $M_t$ leads to smaller values of $\lambda_1/\lambda_2$.  When $M_t$ is close to $1$, the ratio $\lambda_1/\lambda_2$ is smaller than 4.

From the above results, we see the dimensionless LEs depend on turbulent Mach number and Taylor Reynolds number nonlinearly. The scaling law is not universal, in other words, it depends on the driving force as well. We note that \citet{donzis_universality_2020} have introduced $\delta = u^c_{rms}/u^d_{rms}$ to derive universal scaling laws for compressible turbulence. However, we checked that the relationship between dimensionless LEs and $\delta$ is not linear either. Since the LEs depend on the dissipation $\sqrt{\langle |\nabla \boldsymbol{u}|^2 \rangle}$ rather than energy $\langle |\boldsymbol{u}|^2 \rangle$, we divide $\sqrt{\langle |\nabla \boldsymbol{u}|^2 \rangle}$ into the compressible part and solenoidal part, and define the ratio of dilation-to-vorticity (in magnitude) as
\begin{equation}\label{eq:r-cs}
    r_{dv} = \sqrt{\langle |\nabla\cdot\boldsymbol{u}|^2\rangle / \langle |\nabla\times\boldsymbol{u}|^2\rangle}.
\end{equation}
\begin{figure}[!htb]
    \centering
    \subfloat[ST forcing]{\includegraphics[width=0.45\textwidth]{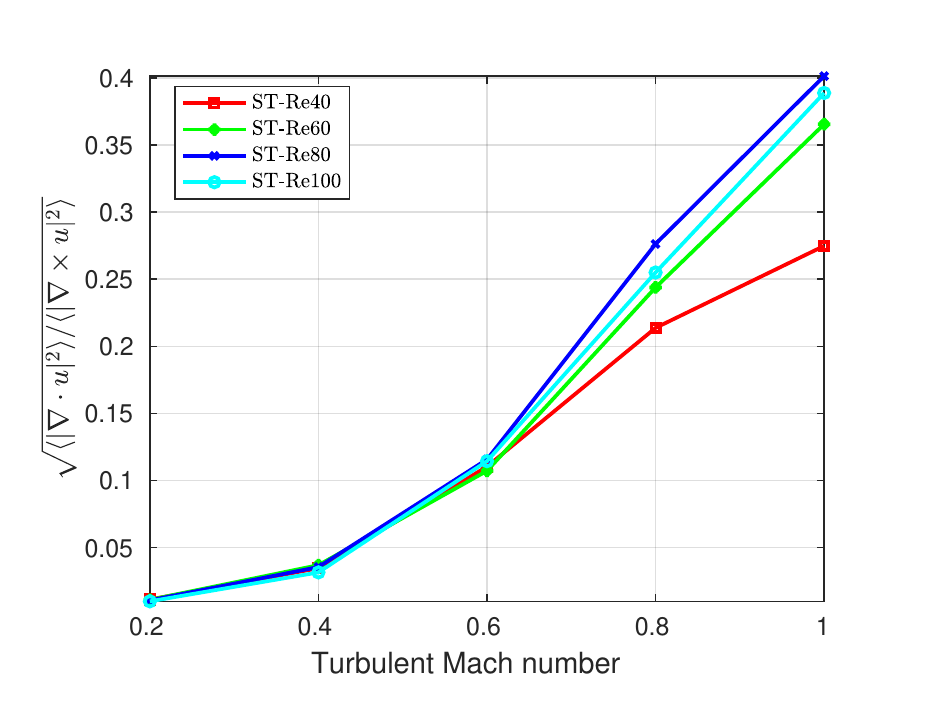}}
    \subfloat[C1 forcing]{\includegraphics[width=0.45\textwidth]{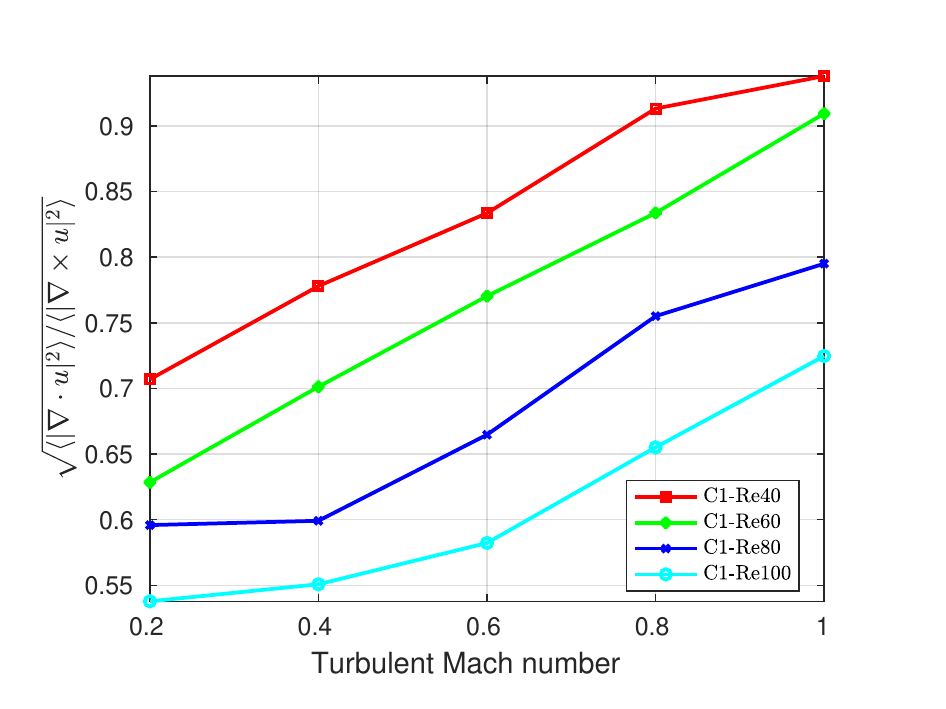}}
    \caption{\label{fig:Dcs} The ratio of dilation-to-vorticity in magnitude $r_{dv}$ defined in \eqref{eq:r-cs} as a function of turbulent Mach number and Taylor scale Reynolds number.}
\end{figure}
Fig.~\ref{fig:Dcs} shows how $r_{dv}$ depends on turbulent Mach number and Taylor scale Reynolds number. Again, we see a nonlinear relation. The $Re_\lambda=40$ in ST forcing case has a quite different scaling because the Reynolds number is very small, and the system is too dissipative to
allow small-scale compressible velocity to be excited.

\begin{figure}[!htb]
    \centering
    \subfloat[Fit $\lambda_1\tau_\eta$ for ST]{\includegraphics[trim=0 10 40 30, clip, width=0.45\textwidth]{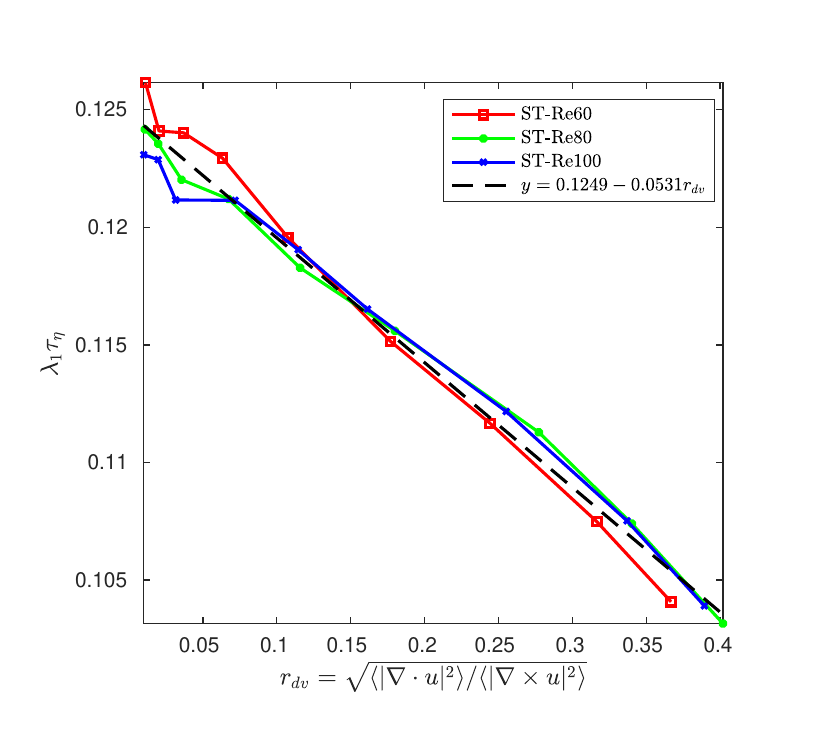}}
    \subfloat[Fit $\lambda_3\tau_\eta$ for ST]{\includegraphics[trim=0 10 40 30, clip, width=0.45\textwidth]{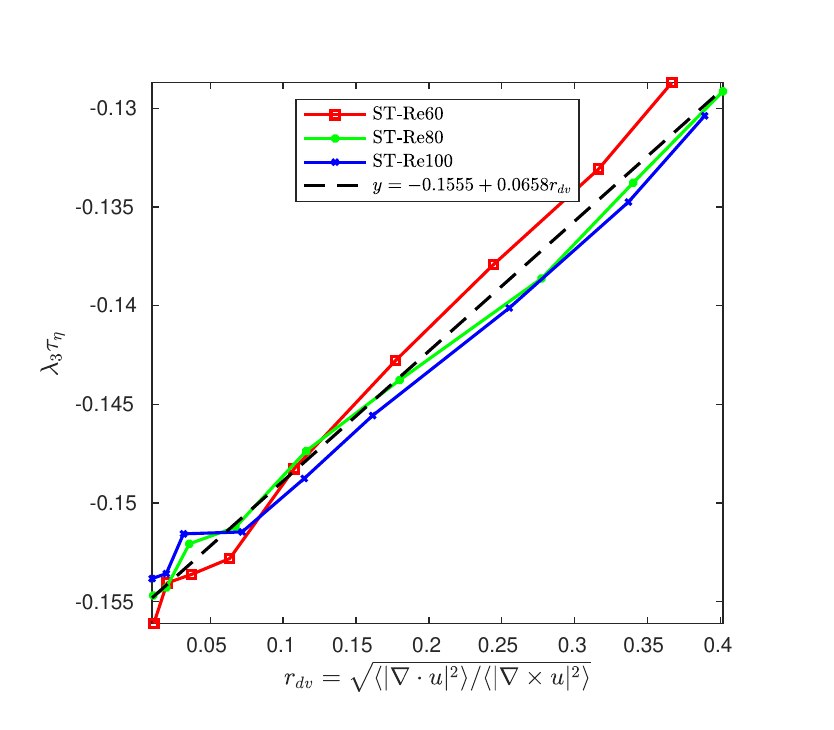}}\\
    \subfloat[Fit $\lambda_1\tau_\eta$ for C1]{\includegraphics[trim=0 10 40 30, clip, width=0.45\textwidth]{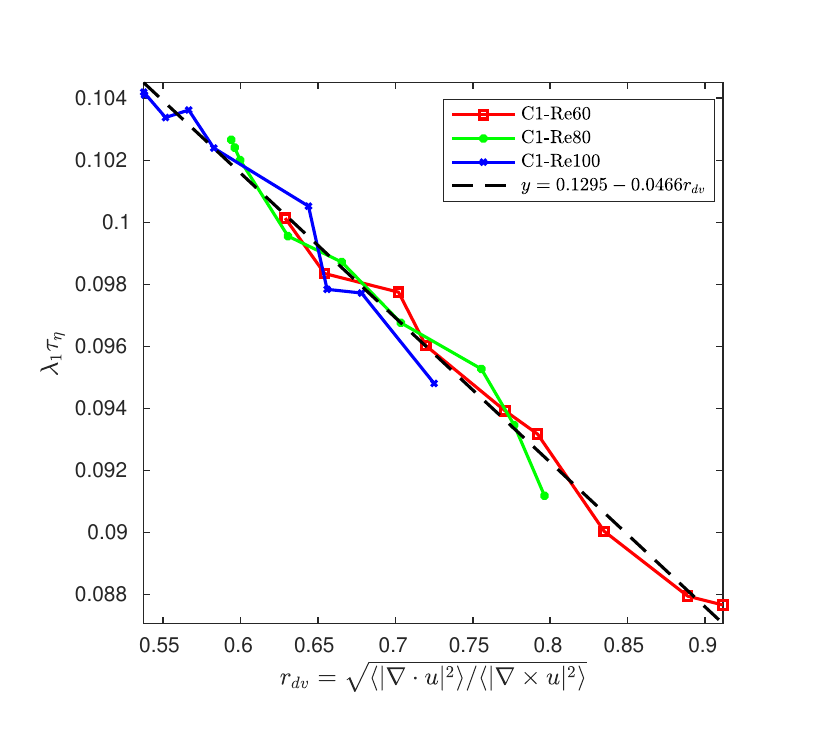}}
    \subfloat[Fit $\lambda_3\tau_\eta$ for C1]{\includegraphics[trim=0 10 40 30, clip, width=0.45\textwidth]{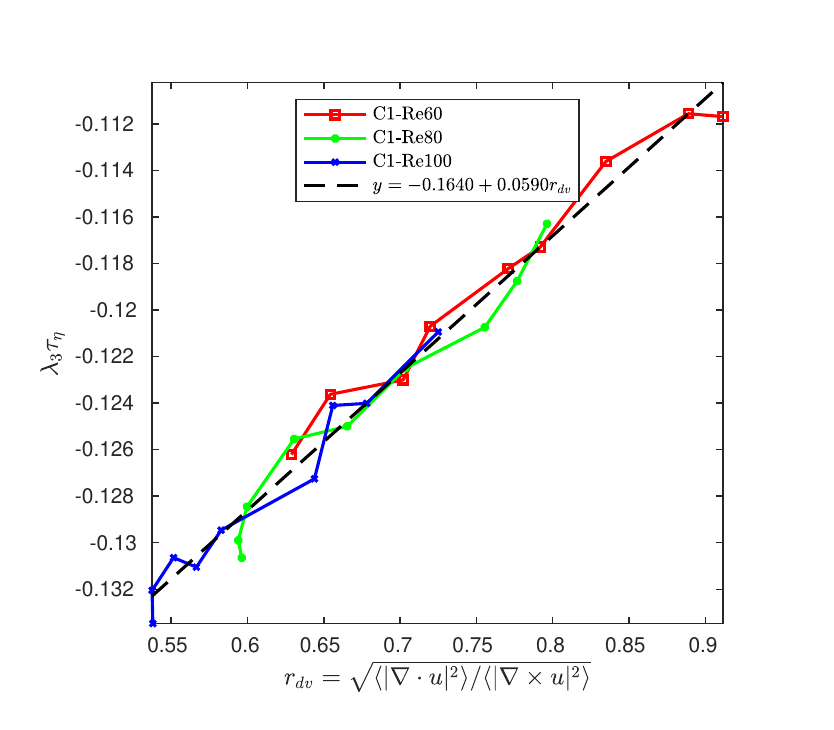}}\\
    \caption{\label{fig:LEsR-fitall} The linear fitting of LEs as functions of  $r_{dv}$ defined in \eqref{eq:r-cs} for different Taylor scale Reynolds number.}
\end{figure}

Next, we show that the dimensionless LEs depend on $r_{dv}$ linearly for both ST and C1 forcing cases in Fig.~\ref{fig:LEsR-fitall}, where nine $M_t$ values ($0.2, 0.3, \ldots, 0.9, 1$) are simulated for each $Re_\lambda$. From this figure, we see that the linear fitting presents very good results for all Taylor scale Reynolds numbers except for $Re_\lambda \approx 40$, and the slops in ST and C1 forcing cases are very close.

\begin{figure}[!htb]
    \centering
    \subfloat[Fit $\lambda_1\tau_\eta$ for ST, $Re_\lambda=60$]{\includegraphics[trim=0 10 40 20, clip, width=0.32\textwidth]{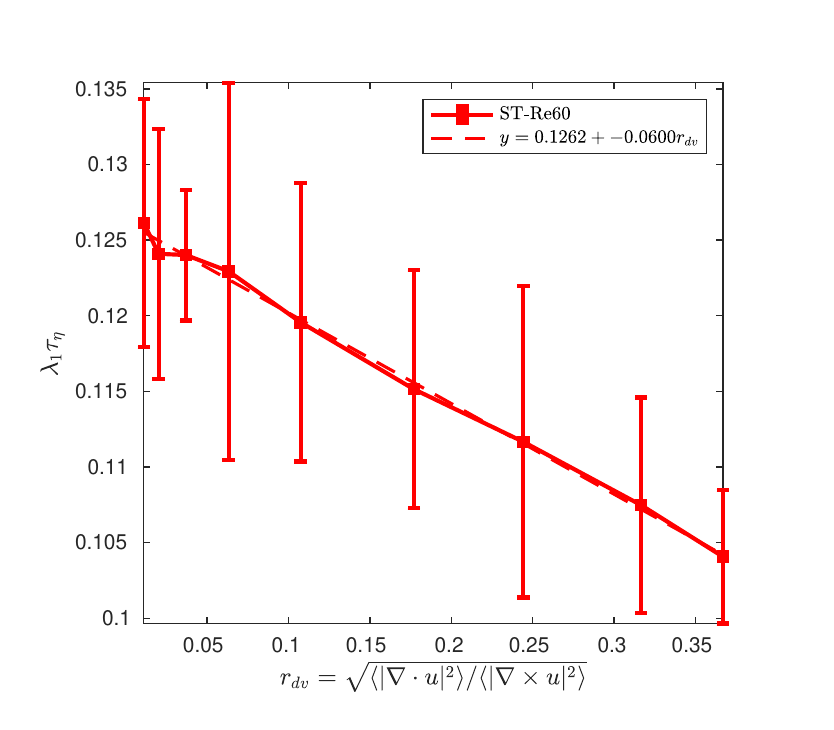}}
    \subfloat[Fit $\lambda_1\tau_\eta$ for ST, $Re_\lambda=80$]{\includegraphics[trim=0 10 40 20, clip, width=0.32\textwidth]{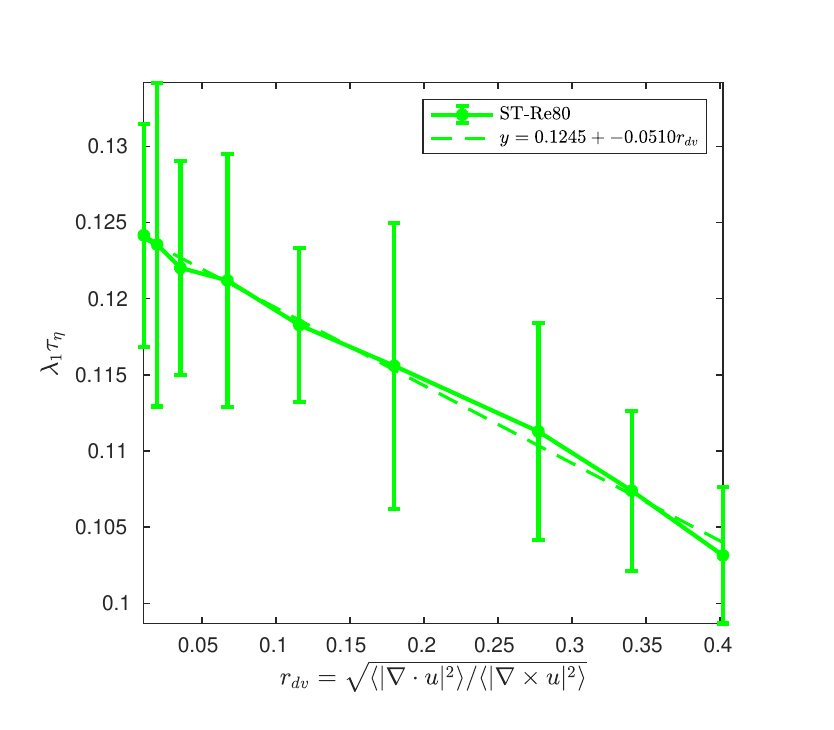}}
    \subfloat[Fit $\lambda_1\tau_\eta$ for ST, $Re_\lambda=100$]{\includegraphics[trim=0 10 40 20, clip, width=0.32\textwidth]{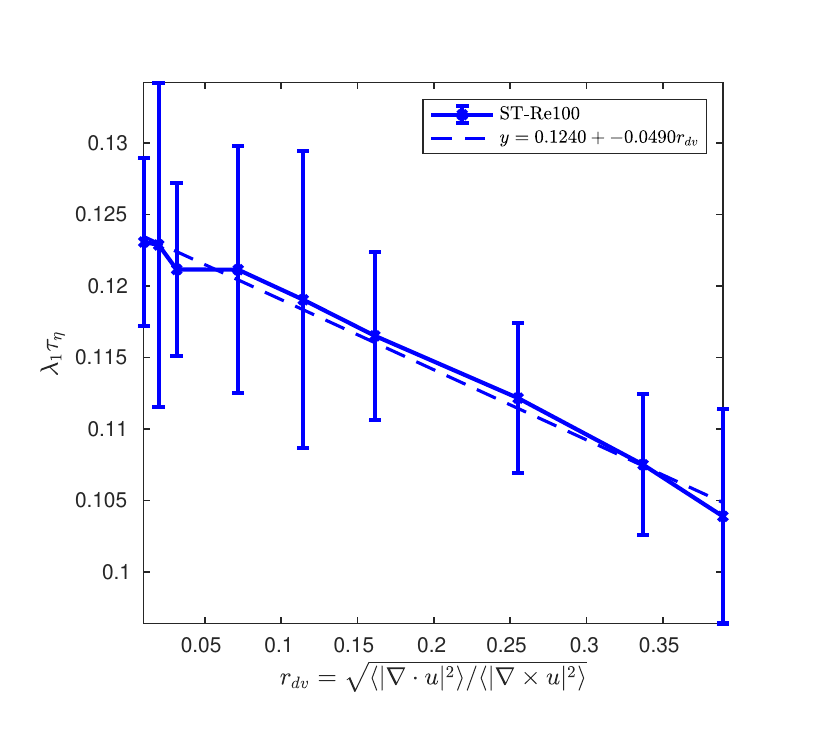}}\\
    \subfloat[Fit $\lambda_1\tau_\eta$ for C1, $Re_\lambda=60$]{\includegraphics[trim=0 10 40 20, clip, width=0.32\textwidth]{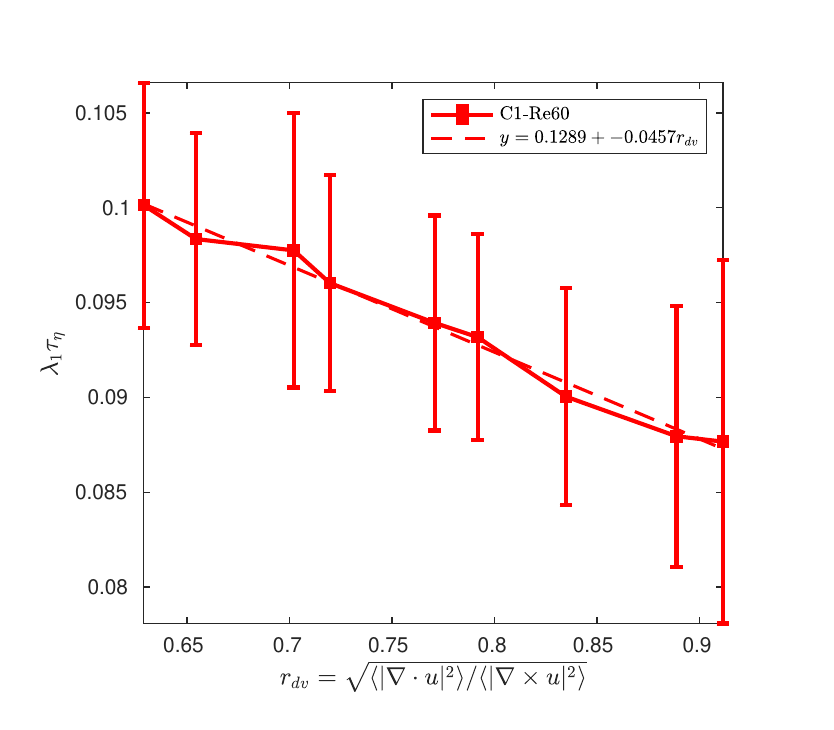}}
    \subfloat[Fit $\lambda_1\tau_\eta$ for C1, $Re_\lambda=80$]{\includegraphics[trim=0 10 40 20, clip,width=0.32\textwidth]{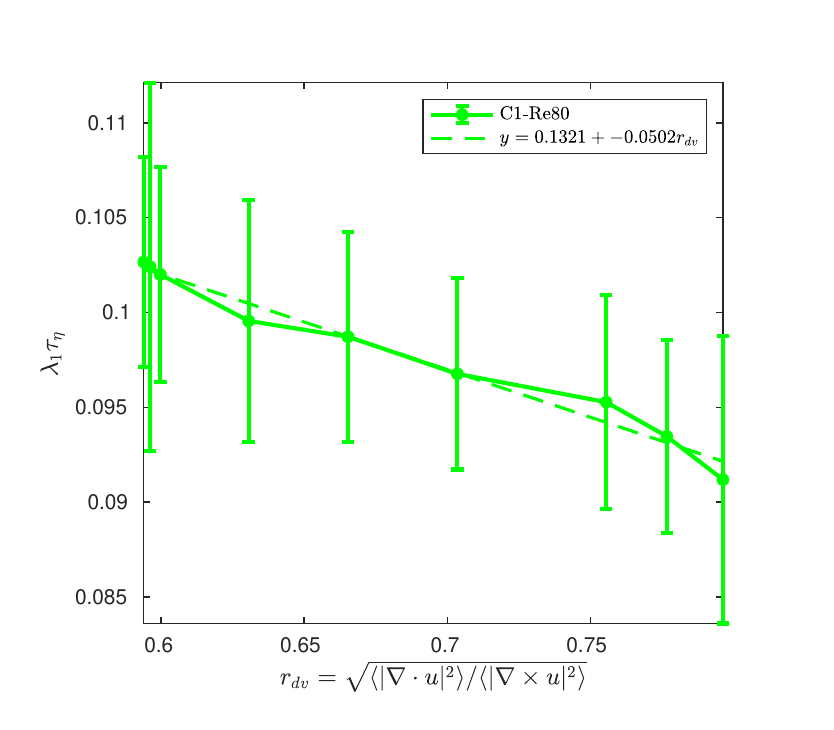}}
    \subfloat[Fit $\lambda_1\tau_\eta$ for C1, $Re_\lambda=100$]{\includegraphics[trim=0 10 40 20, clip,width=0.32\textwidth]{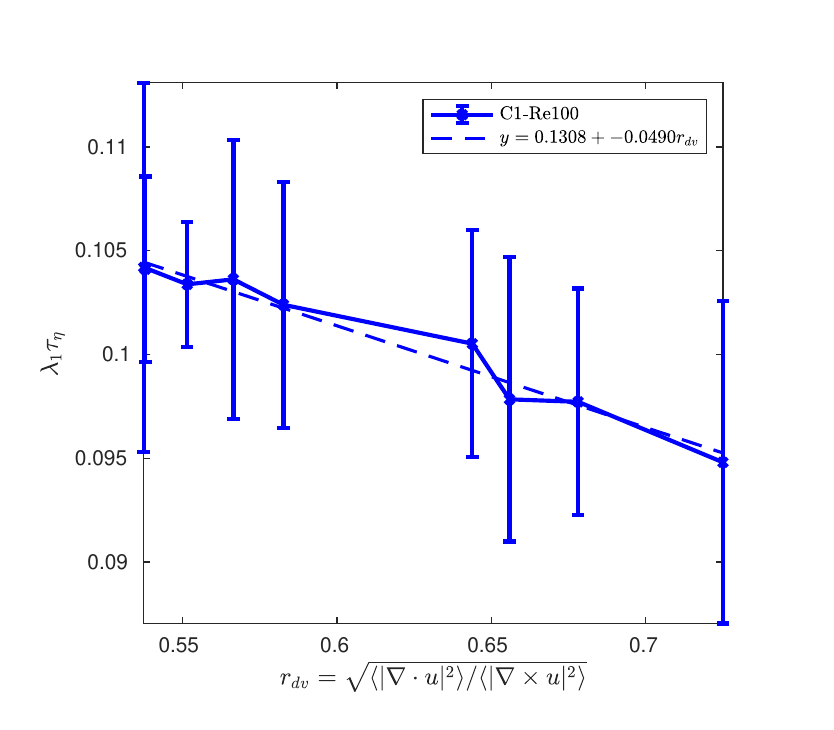}}\\
    \caption{\label{fig:LE1R-fit} The linear fitting of $\lambda_1\tau_\eta$ as functions of  $r_{dv}$ defined in \eqref{eq:r-cs} for different Taylor scale Reynolds number. Error bars reflect fluctuations in calculating $\tau_\eta$.}
\end{figure}

\begin{figure}[!htb]
    \centering
    \subfloat[Solenoidal forcing (ST)]{\includegraphics[width=0.49\textwidth]{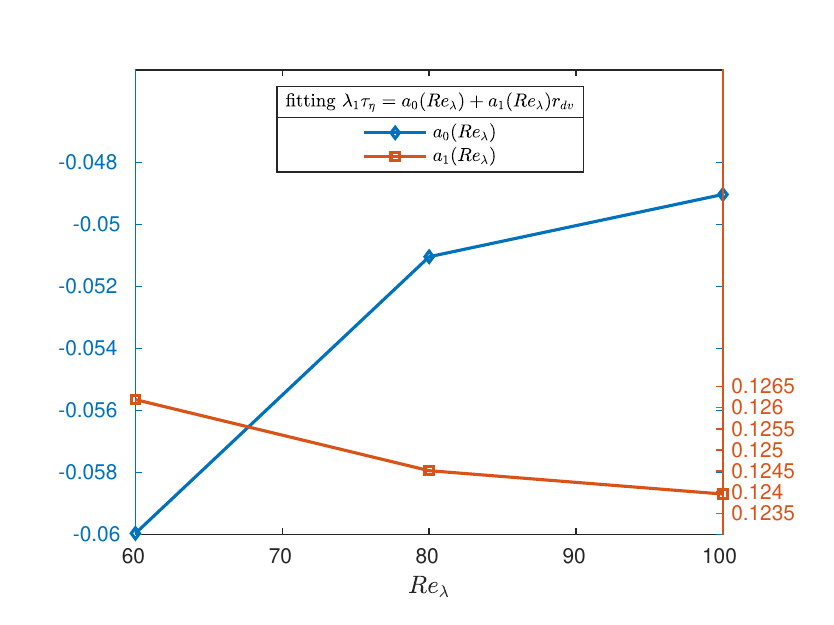}}
    \subfloat[Mixed forcing (C1)]{\includegraphics[width=0.49\textwidth]{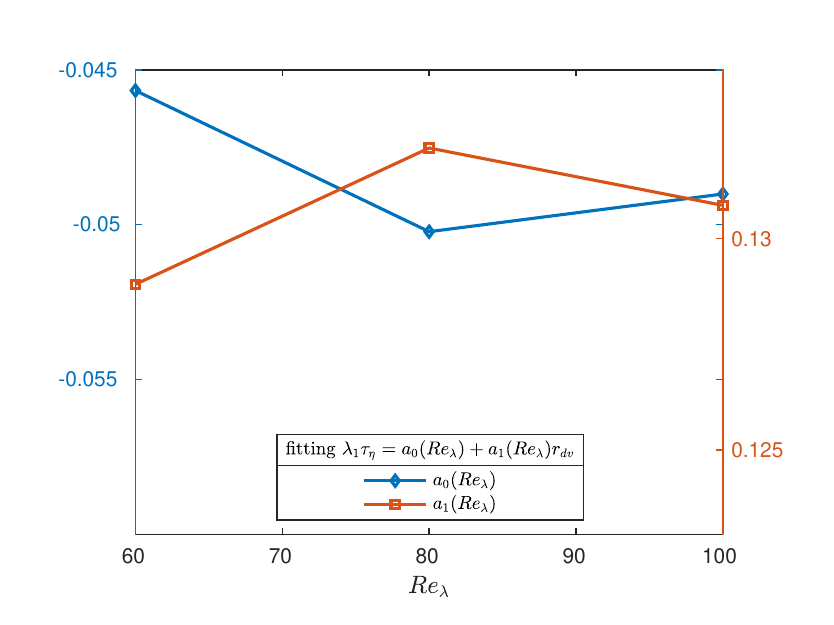}}
    \caption{The results of fitting $\lambda_1\tau_\eta = \alpha_0(Re_\lambda) + \alpha_1(Re_\lambda) r_{dv}$}
    \label{fig:LE1TauD2Fit}
\end{figure}

To see clearly the dependence on $Re_\lambda$, we carry out a linear fitting for $Re_\lambda=60,80,100$ separately, and present the results of $\lambda_1\tau_\eta$ in Fig.~\ref{fig:LE1R-fit}, where
temporal fluctuations of $\tau_\eta$ in the numerical simulation is used to plot error bars.
From this figure, we observe that the slopes and intercepts at different Taylor scale Reynolds numbers and forcing cases are very close (see Fig. \ref{fig:LE1TauD2Fit}), which suggests that $r_{dv}$ is a universal nondimensional parameter for LEs. The results of linear fitting for $\lambda_3 \tau_\eta$ are similar and not presented.

\begin{figure}[!htb]
    \centering
    \subfloat[ST-Re100M08, dilation]{\includegraphics[width=0.35\textwidth]{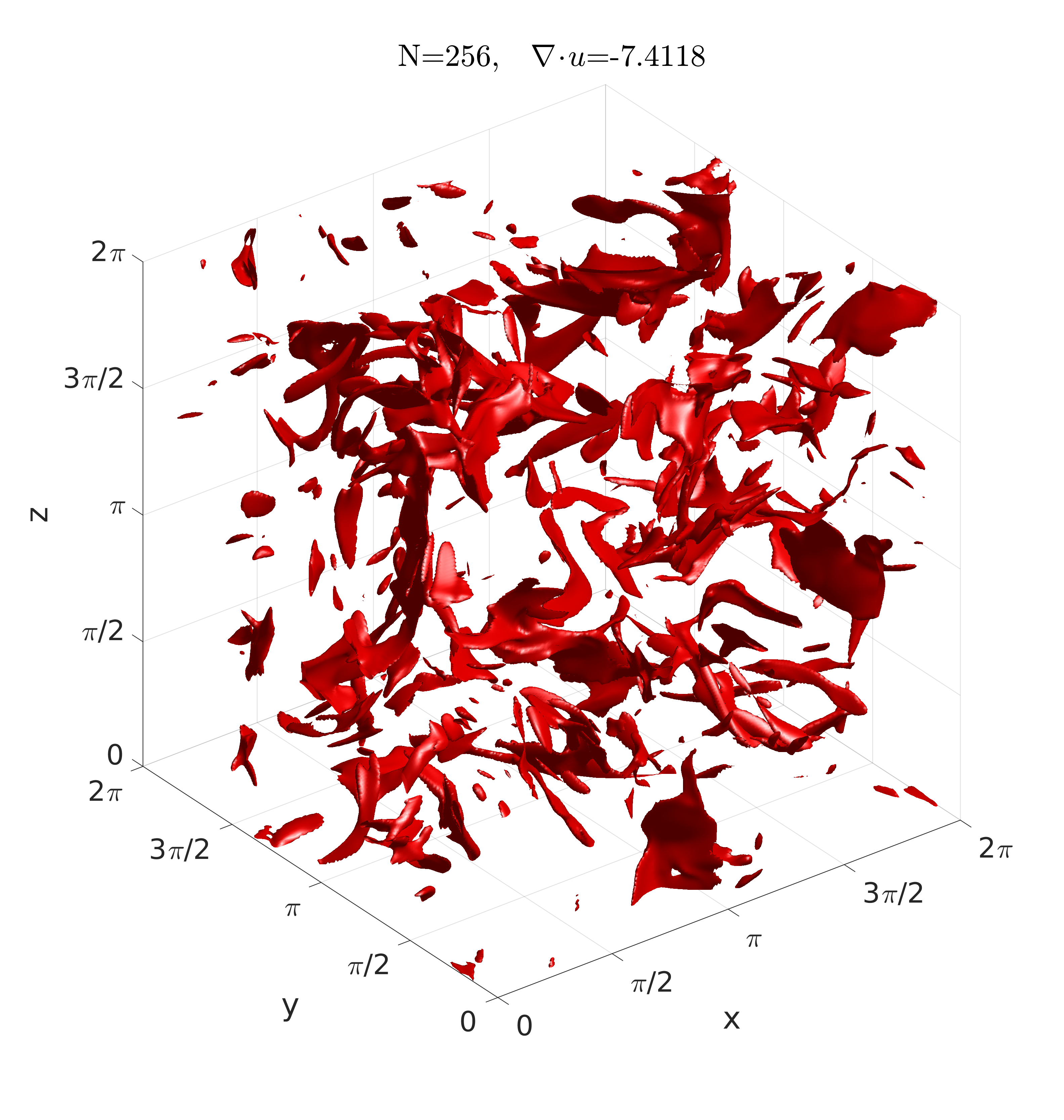}}
    \subfloat[ST-Re100M08, $|\nabla u|$]{\includegraphics[width=0.35\textwidth]{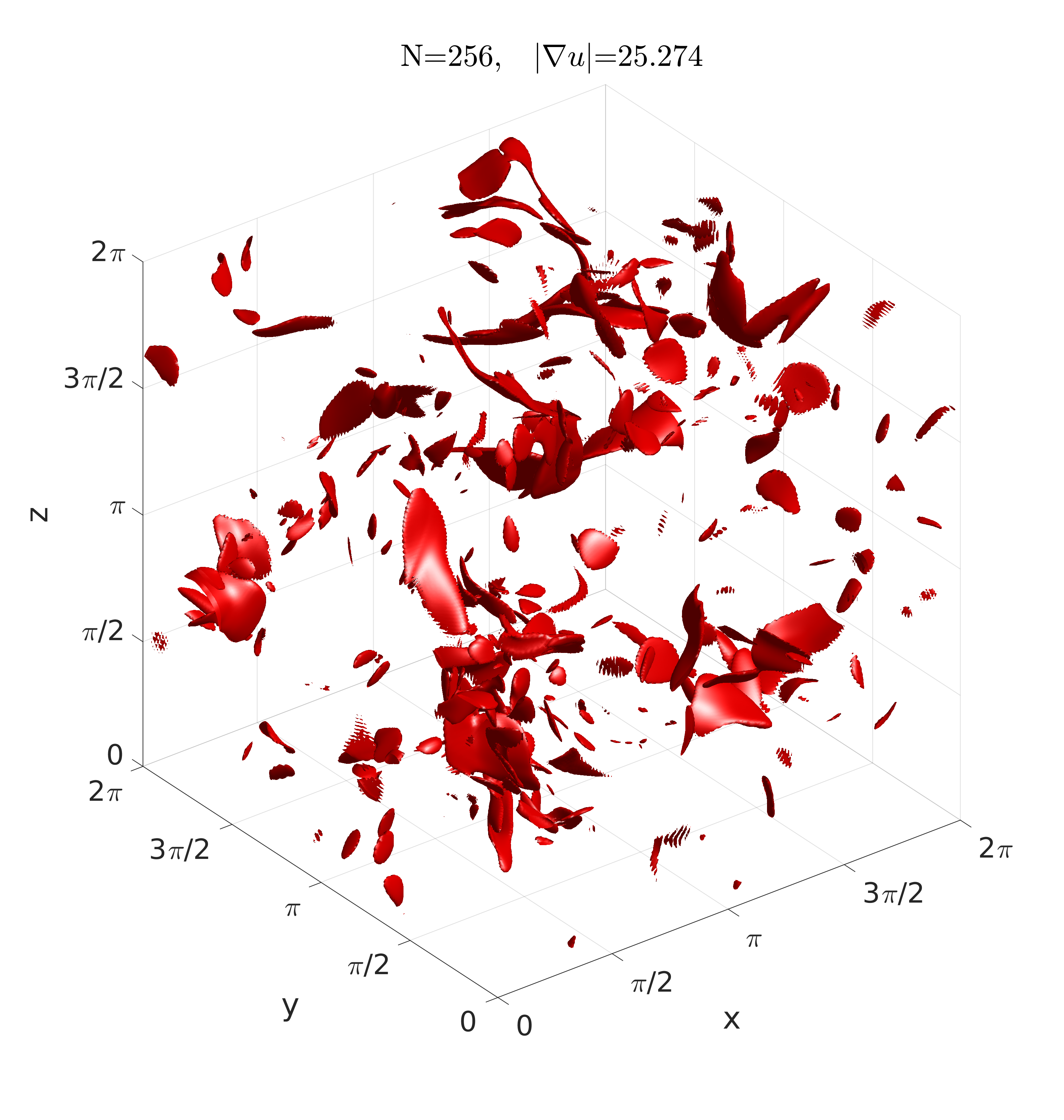}}
    \\
    \subfloat[C1-Re40M04, dilation]{\includegraphics[width=0.35\textwidth]{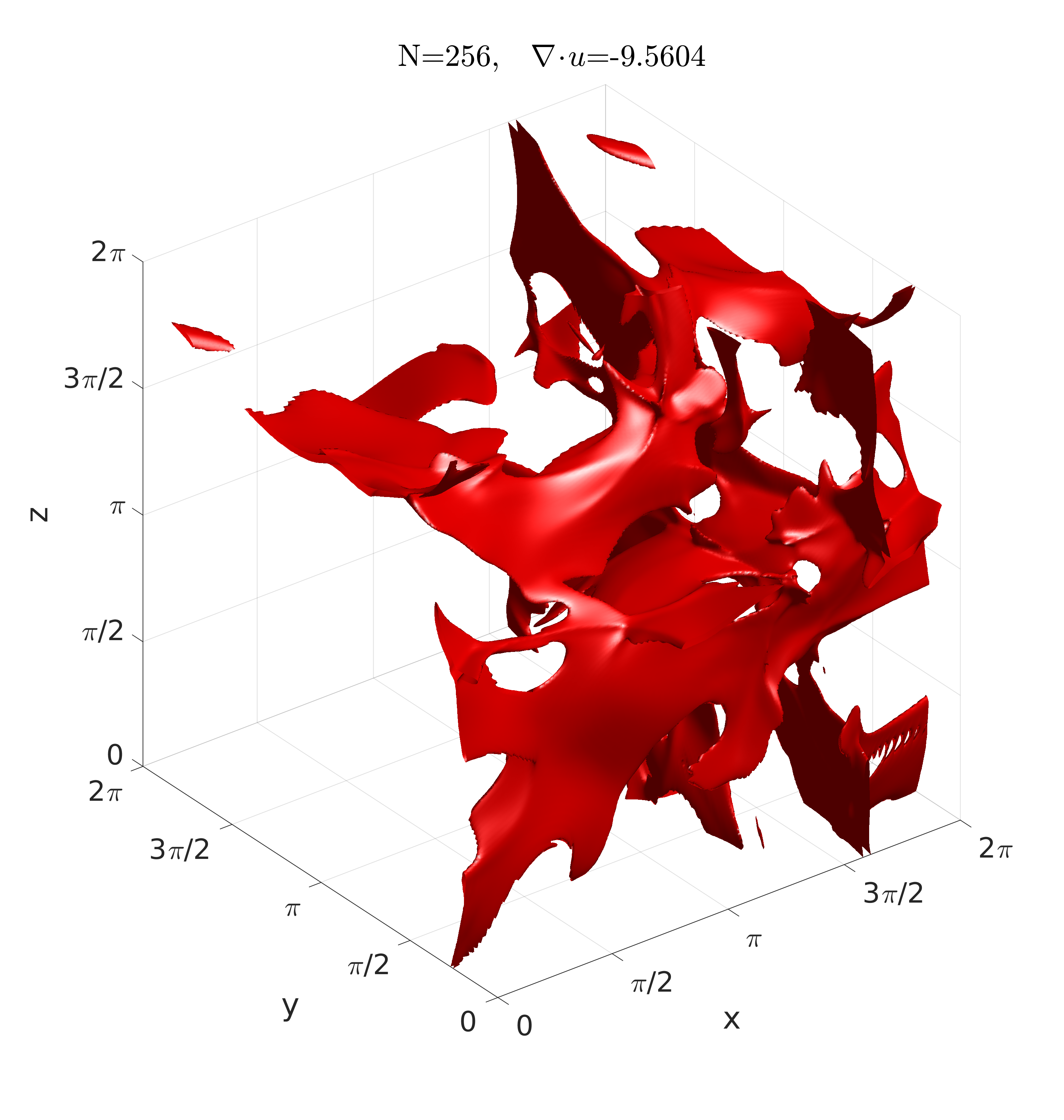}}
    \subfloat[C1-Re40M04, $|\nabla u|$]{\includegraphics[width=0.35\textwidth]{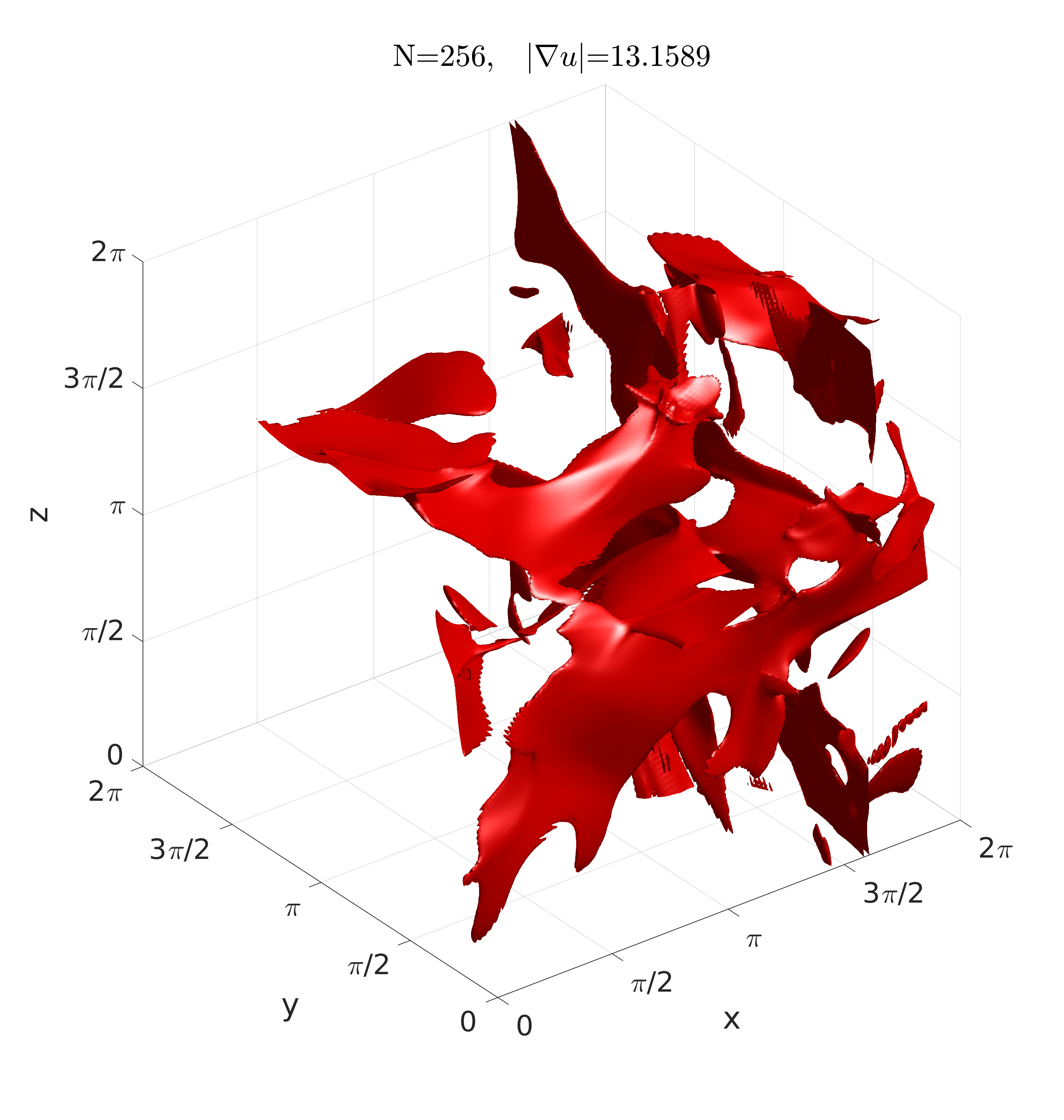}}
    \caption{Different correspondence between $|\nabla\cdot u|$ and $|\nabla u|$ in different driving forces. }
    \label{fig:isosurf}
\end{figure}

To give a better understanding of the dilation parts and vorticity parts in $\nabla \boldsymbol{u}$, we give the isosurfaces of the dilation $\nabla\cdot u$ and the Frobenius norm of $|\nabla \boldsymbol{u}|$ in Fig. \ref{fig:isosurf}.
If the driving force contains no compressible part, the dilation component in the velocity gradient is not easy to be excited, even at relatively high $M_t$, cf. Fig.~\ref{fig:isosurf}(a)(b), the correlation between high dilation region and high $|\nabla \boldsymbol{u}|$ region is weak. This explains why the dependence of $\lambda_1/\lambda_2$ on $M_t$ is weak in solenoidal forcing cases, since the linear fittings of $\lambda_1\tau_\eta$ and $\lambda_2\tau_\eta$ with respect to $r_{dv}$ have intercepts much larger than the slopes, meanwhile, $r_{dv}$ is small.
But in the C1 forcing case, even $M_t$ is as small as $0.4$, the dilation component is dominant in the high-frequency (i.e. small scale) parts, cf. Fig.~\ref{fig:isosurf}(c)(d).
In other words, the dilation component is responsible for the small-scale variations, which is the major thing to define the LEs. We note that when dilation components are dominant locally, sheet-like structures (thin but broader) corresponding to large-scale shock waves are observed. When vorticity components are dominant, small-scale vortex structures and shocklets are observed.

\section{Conclusions\label{sec:6}}
We developed the theory of statistics of finite-time Lyapunov exponents of compressible turbulence and have carried out a thorough numerical study of Lyapunov exponents of passive particles in compressible homogeneous isotropic turbulence, with emphasis on their dependence on turbulent Mach number and Taylor scale Reynolds number, as well as external forcing.

Our main theoretical result is the general form of the distribution of the finite-time Lyapunov exponents given by Eqs.~(\ref{for})-(\ref{fro}). For most purposes, the distribution has the same general form as in incompressible flow \cite{balkovsky_universal_1999}. This conclusion generalizes the previous observation that the sum of Lyapunov exponents vanishes in compressible turbulence \cite{falkovich_particles_2001,fouxon_density_2019}. It is remarkable that despite that the flow can have a possibly large Mach number, the small-scale mixing is effectively incompressible. The role of the Mach number seems mainly to suppress Lagrangian chaos and, consequently, mixing. We observed numerically that both the principal exponent of the Lagrangian flow $\lambda_1$ and its counterpart for the time-reversed flow $|\lambda_3|$ decrease with Mach number.

To the best of our knowledge no numerical study of the Lyapunov exponents of compressible turbulence has been previously published besides Ref. \onlinecite{schwarz_lyapunov_2010}. This reference, however, provided estimates of the Lyapunov exponents that do not sum to zero as they should. We believe that the discrepancy is caused by using velocity gradients that are not fully resolved and are effectively coarse-grained. Thus in our numerical simulations, we paid great attention to achieving the necessary resolution. In contrast to Ref. \onlinecite{schwarz_lyapunov_2010} whose simulations are performed at zero viscosity, our numerical scheme has a finite viscosity. To resolve the shocks, we focused on the middle range of Taylor scale Reynolds numbers and adopted high-order numerical schemes.  Our findings can be summarized as follows.
\begin{itemize}
 \item We numerically verified that the sum of LEs of passive particles in compressible turbulence is zero. To get this result, accurate numerical methods and long-time simulation were employed.
 \item For the solenoidal driving force with $Re_\lambda > 80$, we found $\lambda_1 : \lambda_2 : \lambda_3
 \approx 4:1:-5$ which is similar to the incompressible Navier-Stokes system
 \citep*{johnson_largedeviation_2015}.
 \item The dependence of $\lambda_1\tau_\eta$, $\lambda_1/\lambda_2$ on turbulent Mach number, Taylor Reynolds number is heavily affected by the type of external force.
 \item We find that the dilation-to-vorticity ratio $r_{dv} = \sqrt{\langle |\nabla\cdot\boldsymbol{u}|^2\rangle / \langle |\nabla\times\boldsymbol{u}|^2\rangle}$ is directly responsible for the behavior of dimensionless LEs, which can be used as a control parameter.
\end{itemize}

We note that there are some limitations to the current work. Firstly, due to the limitation of computational resources, we only investigated a special type of external forcing, although it has been extensively used in the literature.  Further investigations with other types of driving force should be performed in order to find out whether $r_{dv}$ is indeed a universal parameter. Secondly, the numerical error of small Reynolds number $Re_\lambda \le 40$ and small Mach number $M_t\le 0.2$ may have an undesirable effect, since the numerical scheme we used is not designed for small $Re_\lambda$ and small $M_t$.  Thirdly, this study is limited to $M_t \le 1$.  It would also be of interest to study the large deviations function that governs the distribution of the finite-time Lyapunov exponents. Performing this task and going beyond the considered special cases deserves further study.

\begin{acknowledgments}
    The computations were partially done on the
    high-performance computers of the State Key Laboratory of Scientific and Engineering
    Computing, Chinese Academy of Sciences. This work was financially supported by the National Natural Science Foundation of China under Grant No.~12161141017 and the Israel Science Foundation under Grant No.~3557/21.
    HY was also partially supported by NSFC under grant No. 12171467. LY was partially supported by NSFC under grant No. 12071470.
    SM was partially supported by NSFC under grant No. 12271514.
\end{acknowledgments}

\section*{Data availability}
The data that support the findings of this study are available from the corresponding author upon reasonable request.

\bibliography{main-PoF}


\end{document}